\documentclass[twocolumn]{aastex63}

\usepackage{longtable}

\def\simlt{\mathrel{\rlap{\lower 3pt\hbox{$\sim$}}
        \raise 2.0pt\hbox{$<$}}}
\def\simgt{\mathrel{\rlap{\lower 3pt\hbox{$\sim$}}
        \raise 2.0pt\hbox{$>$}}}

\received{March 14, 2022}
\revised{April 28, 2022}
\accepted{May 6, 2022}

\shorttitle{A multi-wavelength study of GRS~1716$-$249}
\shortauthors{Saikia et al. 2022}

\usepackage{multirow}
\begin{document}

\title{A multi-wavelength study of GRS~1716$-$249 in outburst : constraints on its system parameters}

\correspondingauthor{Payaswini Saikia}
\email{ps164@nyu.edu}

\author[0000-0002-5319-6620]{Payaswini Saikia}
\affiliation{Center for Astro, Particle and Planetary Physics, New York University Abu Dhabi, PO Box 129188, Abu Dhabi, UAE \\}
\affiliation{New York University Abu Dhabi, PO Box 129188, Abu Dhabi, United Arab Emirates\\}

\author[0000-0002-3500-631X]{David M. Russell}
\affiliation{Center for Astro, Particle and Planetary Physics, New York University Abu Dhabi, PO Box 129188, Abu Dhabi, UAE \\}
\affiliation{New York University Abu Dhabi, PO Box 129188, Abu Dhabi, United Arab Emirates\\}

\author[0000-0003-1285-4057]{M. C. Baglio}
\affiliation{Center for Astro, Particle and Planetary Physics, New York University Abu Dhabi, PO Box 129188, Abu Dhabi, UAE \\}
\affiliation{New York University Abu Dhabi, PO Box 129188, Abu Dhabi, United Arab Emirates\\}
\affiliation{INAF, Osservatorio Astronomico di Brera, Via E. Bianchi 46, I-23807 Merate (LC), Italy\\}

\author[0000-0002-1583-6519]{D. M. Bramich}
\affiliation{Center for Astro, Particle and Planetary Physics, New York University Abu Dhabi, PO Box 129188, Abu Dhabi, UAE \\}
\affiliation{Division of Engineering, New York University Abu Dhabi, PO Box 129188, Abu Dhabi, United Arab Emirates\\}

\author[0000-0002-0752-3301]{Piergiorgio Casella}
\affiliation{INAF, Osservatorio Astronomico di Roma Via Frascati 33 - Monte Porzio Catone, Italy\\}

\author[0000-0001-7796-4279]{Maria Diaz Trigo}
\affiliation{ESO, Karl-Schwarzschild-Strasse 2, D-85748 Garching bei M\"unchen, Germany\\}

\author[0000-0003-3105-2615]{Poshak Gandhi}
\affiliation{School of Physics \& Astronomy, University of Southampton, Highfield, Southampton SO17 \,1BJ, UK \\}

\author[0000-0002-9639-4352]{Jiachen Jiang}
\affiliation{Institute of Astronomy, University of Cambridge, Madingley Road, Cambridge CB3 0HA, UK
\\}

\author[0000-0003-0976-4755]{Thomas Maccarone}
\affiliation{Department of Physics \& Astronomy, Texas Tech University, Box 41051, Lubbock TX 79409-1051, USA\\}

\author[0000-0002-4622-796X]{Roberto Soria}
\affiliation{College of Astronomy and Space Sciences, University of the Chinese Academy of Sciences, Beijing 100049, China\\}
\affiliation{Sydney Institute for Astronomy, School of Physics A28, The University of Sydney, Sydney, NSW 2006, Australia\\}

\author[0000-0002-4187-4981]{Hind Al Noori}
\affiliation{Department of Physics, University of California, Santa Barbara, CA 93106, USA\\}

\author[0000-0003-2577-6799]{
Aisha Al Yazeedi}
\affiliation{Center for Astro, Particle and Planetary Physics, New York University Abu Dhabi, PO Box 129188, Abu Dhabi, UAE \\}
\affiliation{New York University Abu Dhabi, PO Box 129188, Abu Dhabi, United Arab Emirates\\}

\author[0000-0003-0168-9906]{Kevin Alabarta}
\affiliation{Center for Astro, Particle and Planetary Physics, New York University Abu Dhabi, PO Box 129188, Abu Dhabi, UAE \\}
\affiliation{New York University Abu Dhabi, PO Box 129188, Abu Dhabi, United Arab Emirates\\}
\affiliation{Kapteyn Astronomical Institute, University of Groningen, PO Box 800, NL-9700 AV Groningen, the Netherlands \\}
\affiliation{School of Physics and Astronomy, University of Southampton, Southampton, SO17 1BJ, UK \\}

\author[0000-0001-9621-3796]{Tomaso Belloni}
\affiliation{INAF, Osservatorio Astronomico di Brera, Via E. Bianchi 46, I-23807 Merate (LC), Italy\\}

\author{Marion Cadolle Bel}
\affiliation{Allane SE, Dr.-Carl-von-Linde-Str. 2, 82049 Pullach, Munich, Germany\\}

\author[0000-0002-4767-9925]{Chiara Ceccobello}
\affiliation{Department of Space, Earth and Environment, Chalmers University of Technology, Onsala Space Observatory, 439 92 Onsala, Sweden\\}

\author[0000-0001-5538-5831]{St\'ephane Corbel}
\affiliation{AIM, CEA, CNRS, Universit\'e de Paris, Universit\'e Paris-Saclay, F-91191 Gif-sur-Yvette, France\\}
\affiliation{Station de Radioastronomie de Nançay, Observatoire de Paris, PSL Research University, CNRS, Univ. Orl\'eans, 18330 Nan\c cay, France\\}

\author[0000-0002-5654-2744]{Rob Fender}
\affiliation{Department of Physics, University of Oxford, Denys Wilkinson Building, Keble Road, Oxford OX1 3RH, UK\\}
\affiliation{Department of Astronomy, University of Cape Town, Private Bag X3, Rondebosch 7701, South Africa\\}

\author[0000-0001-5802-6041]{Elena Gallo}
\affiliation{Department of Astronomy, University of Michigan, 1085 S University, Ann Arbor, Michigan 48109, USA\\}

\author[0000-0001-8371-2713]{Jeroen Homan}
\affiliation{Eureka Scientific, Inc., Oakland, CA 94602, USA \\}

\author[0000-0002-9677-1533]{Karri Koljonen}
\affiliation{Finnish Centre for Astronomy with ESO (FINCA), Vesilinnantie 5, FI-20014 University of Turku, Finland\\}
\affiliation{Institutt for Fysikk, Norwegian University of Science and Technology, Trondheim, Norway\\}
\affiliation{Aalto University Metsahovi Radio Observatory, Metsahovintie 114, 02540 Kylmala, Finland\\}

\author[0000-0003-3352-2334]{Fraser Lewis}
\affiliation{Faulkes Telescope Project, School of Physics and Astronomy, Cardiff University, The Parade, Cardiff, CF24 3AA, Wales, UK\\}
\affiliation{Astrophysics Research Institute, Liverpool John Moores University, 146 Brownlow Hill, Liverpool L3 5RF, UK\\}

\author[0000-0001-9564-0876]{Sera B. Markoff}
\affiliation{Anton Pannekoek Institute for Astronomy, University of Amsterdam, Science Park 904, 1098 XH Amsterdam, The Netherlands \\}
\affiliation{Gravitation \& AstroParticle Physics Amsterdam (GRAPPA), University of Amsterdam, Science Park 904, 1098 XH Amsterdam, The Netherlands \\}

\author[0000-0003-3124-2814]{James C. A. Miller-Jones}
\affiliation{International Centre for Radio Astronomy Research, Curtin University, GPO Box U1987, Perth, WA 6845, Australia \\}

\author[0000-0002-4151-4468]{Jerome Rodriguez}
\affiliation{AIM, CEA, CNRS, Universit\'e de Paris, Universit\'e Paris-Saclay, F-91191 Gif-sur-Yvette, France\\}

\author[0000-0002-7930-2276]{Thomas D. Russell}
\affiliation{INAF, Istituto di Astrofisica Spaziale e Fisica Cosmica, Via U. La Malfa 153, I-90146 Palermo, Italy\\}

\author[0000-0003-1331-5442]{Tariq Shahbaz}
\affiliation{Instituto de Astrofisica de Canarias (IAC), E-38205 La Laguna,  Tenerife, Spain\\}
\affiliation{Departamento de  Astrofisica, Universidad de La Laguna (ULL),  E-38206 La Laguna, Tenerife, Spain \\}

\author[0000-0001-6682-916X]{Gregory R. Sivakoff}
\affiliation{Department of Physics, University of Alberta, CCIS 4-181, Edmonton, AB, T6G 2E1, Canada\\}

\author[0000-0003-1033-1340]{Vincenzo Testa}
\affiliation{INAF, Osservatorio Astronomico di Roma Via Frascati 33 - Monte Porzio Catone, Italy\\}

\author[0000-0003-3906-4354]{Alexandra J. Tetarenko}
\altaffiliation{NASA Einstein Fellow}
\affiliation{Department of Physics \& Astronomy, Texas Tech University, Box 41051, Lubbock TX 79409-1051, USA\\}

\begin{abstract}

We present a detailed study of the evolution of the Galactic black hole transient GRS~1716$-$249 during its 2016--2017 outburst at optical (Las Cumbres Observatory), mid-infrared (Very Large Telescope), near-infrared (Rapid Eye Mount telescope), and ultraviolet (the Neil Gehrels Swift Observatory Ultraviolet/Optical Telescope) wavelengths, along with archival radio and X-ray data. We show that the optical/near-infrared and UV emission of the source mainly originates from a multi-temperature accretion disk, while the mid-infrared and radio emission are dominated by synchrotron emission from a compact jet. The optical/UV flux density is correlated with the X-ray emission when the source is in the hard state, consistent with an X-ray irradiated accretion disk with an additional contribution from the viscous disk during the outburst fade. We find evidence for a weak, but highly variable jet component at mid-infrared wavelengths. We also report the long-term optical light curve of the source and find that the quiescent ${i}^{\prime }$-band magnitude is 21.39$\pm$0.15 mag. Furthermore, we discuss how previous estimates of the system parameters of the source are based on various incorrect assumptions, and so are likely to be inaccurate. By comparing our GRS~1716$-$249 data-set to those of other outbursting black hole X-ray binaries, we find that while GRS~1716$-$249 shows similar X-ray behaviour, it is noticeably optically fainter, if the literature distance of 2.4 kpc is adopted. Using several lines of reasoning, we argue that the source distance is further than previously assumed in the literature, likely within 4--17 kpc, with a most likely range of $\sim4$--8 kpc.

\end{abstract}


\section{Introduction} \label{sec:intro}

Black hole X-ray binaries (BHXBs) are interacting binary systems composed of a black hole (BH) accreting matter from a secondary companion star. The accreted matter forms a differentially rotating disk around the BH known as an accretion disk \citep{ss}. A large fraction of the accretion energy is often channeled into relativistic, collimated outflows known as jets \citep[e.g.][]{bk,fender04}. Many BHXBs are transient in nature, alternating between periods of quiescence (typically lasting years to decades, with the X-ray luminosities in the range of $10^{30-33}$ erg s$^{-1}$) and outburst \citep[typically lasting weeks to months, with X-ray luminosities reaching $10^{36-39}$ erg s$^{-1}$, e.g.][]{bc,tt}. 

During an outburst, many BHXBs undergo hysteresis in the spectral state transitions following a q$-$shaped evolutionary pattern in the hardness-intensity diagram \citep[HID;][]{my,homan2001,hb,b10}. The rise of the outburst is generally dominated by a hard, power law-like spectral component (with photon index $\Gamma <$2) with an high-energy cut-off at 50–100 keV. This is known as the hard state (HS), which is usually associated with thermal Comptonization due to Compton up-scattering of soft disk photons by a corona of hot electrons \citep[e.g.][]{thorne,sunyaev,done}. During the HS, collimated compact jets are launched, emitting self-absorbed synchrotron emission that dominate radio though infrared (IR) wavelengths \citep[e.g.][]{Corbel2000,fender04}, in analogy with those observed in active galactic nuclei \citep{bk,hjellming88}. Many BHXBs in the HS follow a non-linear radio/X-ray luminosity correlation, where $L_{R} \propto L_{X}^{\beta}$ with $\beta \sim$0.5–0.7 \citep[e.g.][]{Corbel2003,Corbel2013,gallo2018}, which extends to active galactic nuclei through the fundamental plane of BH activity \citep{merloni,falcke,saikia15,saikia18}, suggesting scale invariance of compact jets.

During the peak and decay of an outburst, when the system is said to be in the soft state (SS), the spectra are dominated by a soft, blackbody-like spectral component due to an optically thick, geometrically thin accretion disk \citep{ss}. The jets are suppressed in this state \citep[e.g.][]{oldjet,Fender1999,Coriat2011,Russell2011b,Koljonen2018,r2019,c2021}. During the transition between these two states, the system enters the intermediate state (IS), dominated by a thermal disk component with a color temperature of 0.1--1 keV, which is further classified based on the X-ray timing properties into hard--intermediate and soft--intermediate states  \citep[e.g.][]{hb,b10}. Depending on the source state, fast variability can be observed, including quasi-periodic oscillations (QPOs), that have been classified into three types: A, B and C \citep[e.g.][]{ingram}. A number of BHXBs remain in the HS for the entire duration of the outburst (or only transition to the hard--intermediate state). These are referred to as ‘hard-only state outbursts’ \citep{tt}, ‘low/hard state outbursts’ \citep{bb02}, ‘failed outbursts’ \citep[e.g.][]{cap,curran2013}, ‘failed state transition outbursts’ \citep{bassi} or ‘failed-transition outbursts’ \citep{alabarta}.

\subsection{GRS~1716$-$249}

In 1994 September, GRS~1716$-$249 had a series of several X-ray re-flares or mini-outbursts, as observed by both SIGMA and BATSE at the level of $\sim$10$\%$ of its peak value in 1993 \citep[][]{rr98}. During this period, the X-ray light curve was dominated by at least four sawtooth-like rebrightening events with slow rise ($\sim$30-70 days) and dramatic decay ($\sim$10 days), accompanied by simultaneous radio flares following the onset of decays \citep{hjellming96}. This re-brightening event lasted $\sim$400 days and had at least 4 separate peaks in hard X-rays.

The source had another outburst after almost 21 years in quiescence, and was detected by the Monitor of All-sky X-ray Image (MAXI) on 2016 December 18 \citep[MJD 57740,][]{negoro}, with a photon index of $\Gamma$= 1.62$\pm$0.06 on 2016 December 21 \citep[MJD 57743,][]{mashu}. It was found to be in the hard spectral state with {\it Chandra X-ray Observatory} observations on 2017 February 6 \citep[MJD 57790,][]{milleratel} and International Gamma-Ray Astrophysics Laboratory (INTEGRAL) observations on 2017 February 10 \citep[MJD 57794,][]{delsanto}. The source was then seen transitioning to the hard--intermediate state for some time with Neil Gehrels Swift Observatory (Swift) observations on 2017 March 27 and April 2 \citep[MJD 57839 and MJD 57845,][]{armas}, and then returning to the hard state after a failed-transition outburst to the soft state on 2017 May 5 and 11 \citep[MJD 57878 and MJD 57884,][]{bassi17}, as was also the case in the 1993 event \citep{rr98}. \cite{bassi} studied the HID of the source and found that it had three softening events when the source transitioned from the hard to the hard-intermediate state. Along with the three softest points (MJD 57854.2, 57895.9 and 57960.7), we consider all the dates with hardness ratio $\lesssim$ 0.7, which lies in the range of 2017 July 6 and August 13 (MJD 57940--57978) as the hard-intermediate state. The source was found to be one of the ‘outlier’ BHXBs \citep{bassi} in the radio/X-ray correlation plane \citep[which are radio fainter by 1--2 orders of magnitude, and tend to have a steeper correlation index, with $\beta \sim$ 1.4, e.g.,][]{Corbel2004,Coriat2011,gallo2012}. A type-C quasi-periodic oscillation (QPO) was also detected in the hard state \citep{bharali}, and signatures of a hot and dense accretion disk wind (with terminal velocity $\sim$2000 km s$^{-1}$) were observed \citep{cuneo}. From the broadband spectral fitting of the source, the irradiated accretion disk was found to dominate the optical emission, while a hint of an excess near-IR emission above the prediction of the irradiated disk model was observed, likely due to synchrotron emission originated in the jet \citep{rout}.

\subsection{System parameters of GRS~1716$-$249}

The system parameters of GRS~1716$-$249 are not well constrained. From the 1993 outburst, \cite{della} proposed that the system contains a low-mass main-sequence star with spectral type K (or later), at a possible distance between 2.2 kpc (lower limit obtained from the equivalent width of the NaD absorption lines) and 2.8 kpc (upper limit based on an incorrect maximum luminosity of an X-ray transient). But in light of several arguments we explore in Section 4.2.2., we find that the estimated upper limit of 2.8 kpc is not a reliable constraint for its distance. \cite{masetti} discovered superhumps in the lightcurve (although these could also be due to irradiation modulation, see Section 4.2.1 for a discussion). Assuming that the donor is a main sequence star, they estimated the companion star mass to be $\sim$1.6 M$_{\odot}$ and inferred an orbital period of $\sim$0.6127 days or $\sim$14.7 hrs for a Roche lobe filling star. Then they used the maximum mass ratio criterion for having superhumps, which is about 3:1, and proposed that the mass of the accreting compact object is $>$ 4.9 M$_{\odot}$, hence classifying it as a black hole. They also suggested that a 1.6 M$_{\odot}$ main sequence star at 2.4$\pm$0.4 kpc would exceed the quiescent luminosity of the binary substantially (although it is important to note that the quiescent luminosity limit of the source was not confidently known, see Section 3.6). Despite all the crude assumptions employed, these limits on the mass of the compact object and the distance to the source have been used for all subsequent studies on the source, until this paper.

\begin{figure*}[ht]
\center
\includegraphics[width=8.9cm,angle=0]{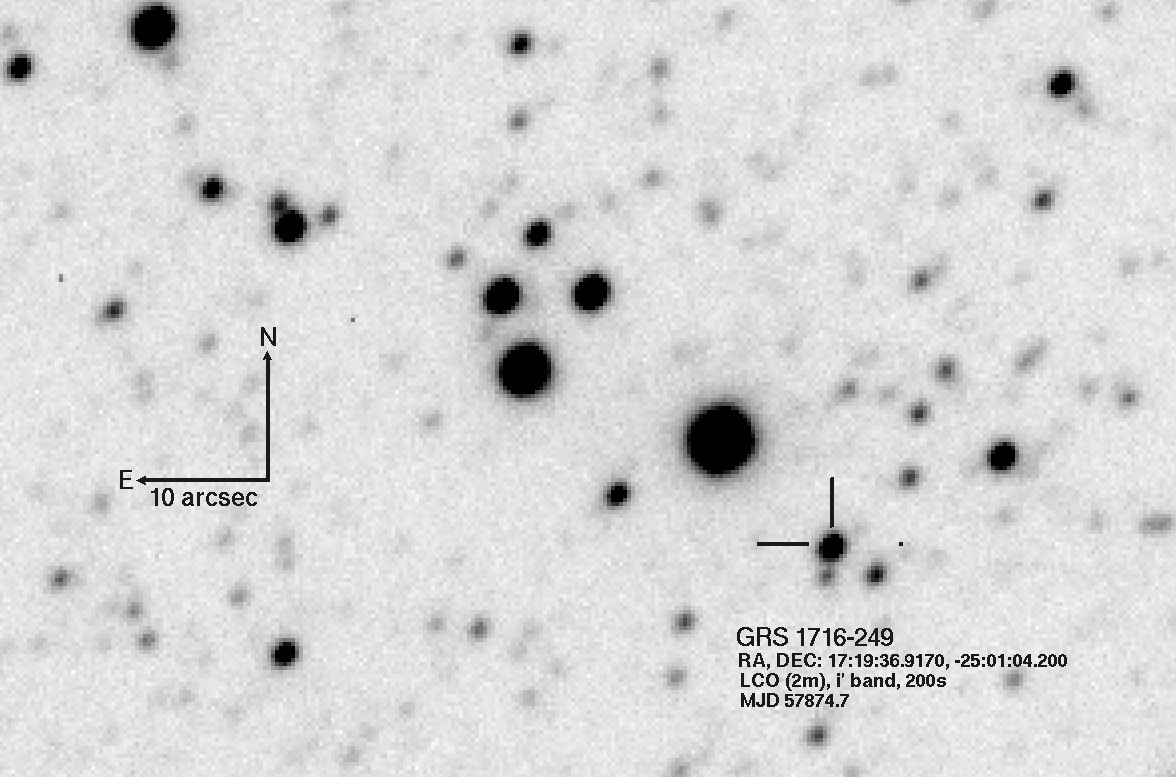}
\includegraphics[width=8.9cm,angle=0]{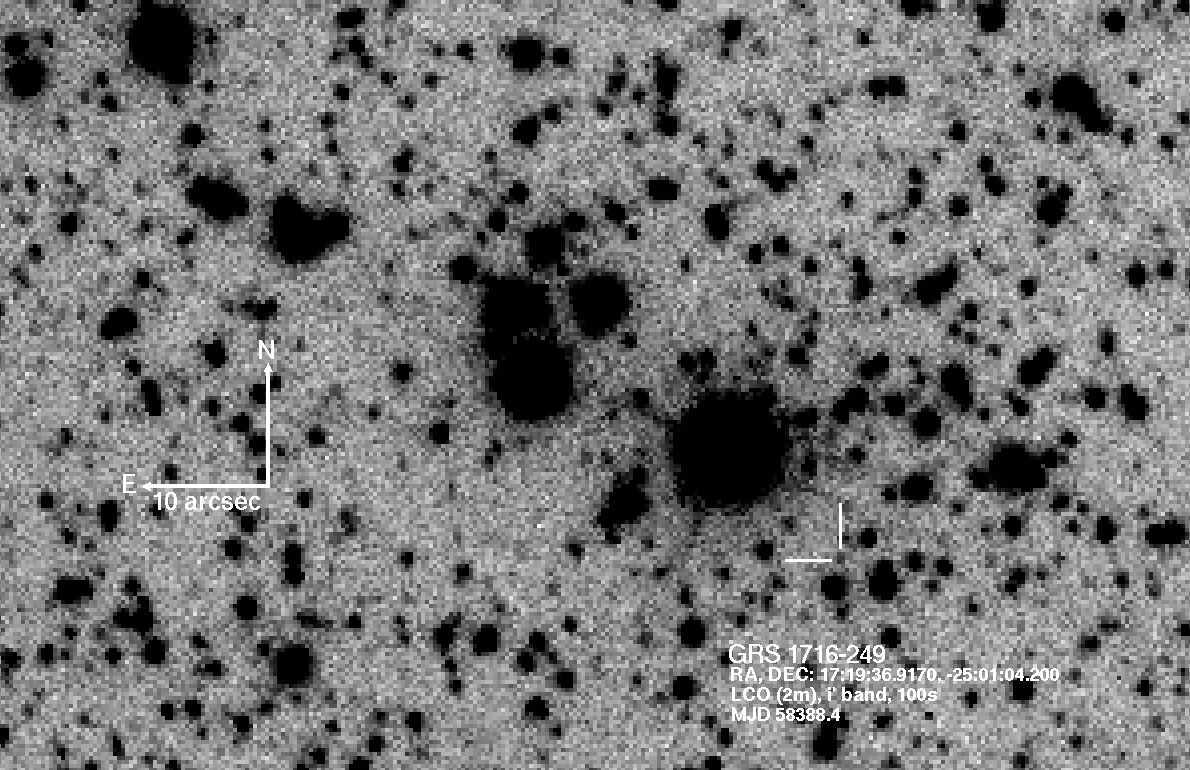}
\caption{The left panel shows the optical finding chart of GRS~1716$-$249 during outburst (MJD 57874.7) with the 2-m LCO telescope in the ${i}^{\prime }$-band with 200s exposure time. Previously, a lower-resolution optical finding chart during outburst is available in the V band \citep{masetti}. The target is indicated with hash mark in both the panels. The right panel shows the quiescent optical finding chart (MJD 58388.4) with the 2-m LCO telescope in the ${i}^{\prime }$-band with 100s exposure time (an image taken under excellent conditions, with seeing of 0.82 arcsec). The counterpart is just 1.6 arcsec from a star of similar magnitude to the north of GRS 1716--249.}
\label{fig:findingchart}
\end{figure*}

During the 2016--2017 outburst, GRS~1716$-$249 was extensively studied in the X-ray wavelengths. \cite{tao2019} used spectral fits of three NuSTAR and Swift datasets in its hard-intermediate state, and constrained the black hole mass to be $<$ 8 M$_{\odot}$ at a 90\% confidence level under the assumption that the distance to the source is 2.4$\pm$0.4 kpc. Using the same assumption, they also inferred the inclination angle of the inner disk to be in the range of 40$^{\circ}$--50$^{\circ}$ by performing joint modelling of the continuum and the reflection components. 
An analysis of the broadband (1--78 keV) X-ray spectra of the source taken by NuSTAR and Swift constrained the accretion disk density parameter of GRS~1716$-$249 to be in the range of 10$^{19}$--10$^{20}$ cm$^{-3}$ \citep{jiang}. Recently, the black hole mass was claimed to lie in the range of 4.5--5.9 M$_{\odot}$ according to a two-component advective flow (TCAF) model \citep{chatterjee}, although this method uses model-dependent spectral fitting of the source to obtain these values.

In this paper, we present a detailed multi-wavelength study of GRS~1716$-$249 during its 2016--2017 outburst, with particular focus on its UV/optical/IR emission to investigate the physical mechanisms contributing to the emission in these wavebands, and reveal the system parameters of the source. In Section 2, we describe in detail the observations and the analyses of the data used for this study. In Section 3, we present the characteristics of the outburst using various tools like the light curves, variability of the source during the peak of the outburst using fractional rms values, the optical/UV spectra of the source, the broadband spectral energy distributions (SEDs), the colour-magnitude diagrams to study the colour evolution of the source during the outburst, and the optical/UV/X-ray correlations to explore the various emission mechanisms. We also report long-term ($\sim$ 10 years) monitoring of the source and discuss its quiescent optical magnitude, which is important as the optical brightness of BHXBs in quiescence has minimal contribution from the accretion disk and is dominated by the companion star \citep{cheva}. In Section 4, we interpret and discuss our results, including the implications of our analyses on the system parameters of the source, and present new estimates for the distance to the source. Finally, we present our conclusions in Section 5.

\section{Observations and Data Reduction} \label{sec:obs}

\subsection{Optical observations}

\subsubsection{LCO optical data}

We monitored GRS~1716$-$249 during its 2016--2017 outburst extensively with the Las Cumbres Observatory (LCO) between 2017 January 28 and October 21 (MJD 57781--58046). Observations were made using the 1 m LCO telescopes at Siding Spring Observatory, Australia, Cerro Tololo Inter-American Observatory, Chile, and the South African Astronomical Observatory (SAAO), South Africa, as well as the 2 m Faulkes Telescopes at Haleakala Observatory, Maui, Hawai`i, USA and Siding Spring Observatory, Australia. The source was also monitored during quiescence, before and after the 2016--2017 outburst, for 11 years since 2006 February, as part of an on-going monitoring campaign of $\sim$50 low-mass X-ray binaries (LMXB) coordinated by the Faulkes Telescope Project \citep{Lewis2008,Lewis2018}.

Imaging data were primarily taken in the Sloan Digital Sky Survey (SDSS) ${g}^{\prime }$, ${r}^{\prime }$, ${i}^{\prime }$ and the Panoramic Survey Telescope and Rapid Response System (Pan-STARRS) $Y$-band filters, with some data also taken in Bessel $B$ and $V$-bands. The data were initially processed using the LCO BANZAI pipeline \citep{banzai}. The multi-aperture photometry on the reduced data was performed using ``X-ray Binary New Early Warning System (XB-NEWS)'', a real-time data analysis pipeline that aims to detect and announce new X-ray binary outbursts within a day of the first optical detection of an outburst \citep[e.g.][]{Russell2019,Pirbhoy2020,goodwin2020}. The XB-NEWS pipeline downloads new images of all targets of interest from the LCO archive along with their associated calibration data and performs several quality control steps to ensure that only good quality images are analysed. XB-NEWS then computes an astrometric solution for each image using Gaia DR2 positions\footnote{\url{https://www.cosmos.esa.int/web/gaia/dr2}}, performs aperture photometry of all the stars in the image, solves for zero-point calibrations between epochs \citep{dan}, and flux calibrates the photometry using the ATLAS All-Sky Stellar Reference Catalog \citep[ATLAS-REFCAT2,][]{tonry}. The pipeline also performs multi-aperture photometry \citep[azimuthally-averaged PSF profile fitting photometry,][]{Stetson}. Light curves are produced in near real-time. If the location of the source is well-known, but the source is fainter than the formal detection threshold, the pipeline performs forced photometry on the position. Magnitude errors larger than $\sim 0.25$ mag are considered as marginal detections and are not included in our study.

We detected the source during outburst in a total of 192 images between 2017 January 28 (MJD 57781) and 2017 October 21 (MJD 58046), generally at a cadence of every 2--3 days during the brighter phase of the outburst, and every $\sim$75 s for the high cadence images taken on 2017 May 9 (MJD 57882). A detailed observation log containing information about the LCO epochs, filters and magnitudes, is summarized in the Appendix (see Table A1). From our XB-NEWS optical analysis, the accurate optical position of the source was found to be RA: 17:19:36.917 and Dec: -25:01:04.20 (J2000), consistent with the VLBI coordinates to within $\lesssim$0.1 arcsec \citep{atri}. The optical finding charts in the ${i}^{\prime }$-band during both outburst and quiescence, are shown in Fig.~\ref{fig:findingchart}. The systematic error in the position measurement is small ($\lesssim$0.3 arcsec) and has better precision than previously reported optical measurements \citep[$\sim$1 arcsec in][]{della}. To convert the multi-aperture photometry magnitudes obtained from the XB-NEWS pipeline to the intrinsic de-reddened flux densities, we use the absorption column density N$_H$ = (0.70$\pm$0.01) $\times 10^{22}$ cm$^{-2}$, as reported in \cite{bassi2020}. Using the relation between optical extinction and hydrogen column density \citep{foight16}, the $V$-band absorption coefficient is inferred as A$_V$ = 2.44 $\pm$ 0.11 mag. There are different determinations of the relation between extinction and hydrogen column density in the literature \citep[see for e.g.][]{av1,av2,av3}, but we use the \cite{foight16} value as they provide the most recent estimates using updated abundances and hence are likely more reliable. We note that \cite{bahr15} also arrive at a similar relation using the updated abundances, while including X-ray binaries in their sample. This leads to a color excess of E($B-V$)$\sim$0.8 mag \citep[assuming a mean value of  A$_V$/E($B-V$)$\sim$3.1 for the diffuse interstellar medium,][]{fitz}, which is consistent with the historical value of E($B-V$)$\sim$0.9$\pm$0.2 mag \citep[][]{della} obtained based on multiple lines of reasoning. The wavelength-dependent extinction terms, used for de-reddening in other bands, are obtained from the extinction curve of \cite{cardelli}.

\begin{table*}
\centering
\caption{Observation log and results of the mid-IR photometry performed on GRS~1716$-$249 with the VISIR instrument in 2017 March and April, tabulating the start of observing time, filter used and the central wavelength, weather condition and airmass at mid-observation, the exposure time for each observation, flux density of the detections and the $3 \sigma$ upper limits for non-detections, and a list of the standard stars observed within a month of our observation.}
\begin{tabular}{ l l l l  l l l l}
\hline
\hline
Epoch (UT) & Filter  &  Wavelength & Weather & Airmass & Exposure & Flux density & Standard stars\\
 start time (MJD) &    &  ($\mu$m) &  conditions & & (s) & (mJy) & within a month \\ 
\hline
2017.03.25 06:29:22 (57837)	&  $\textit{B10.7}$	&   10.64 & Photometric &	 1.398 	&	1500 & $<$ 1.35 & 1-9 \\
2017.03.25 07:30:25 (57837)	&  $\textit{J8.9}$	&  8.70 & Photometric	& 1.165 	&	1500 & $<$ 1.37 & 10 \\
2017.03.25 08:18:47 (57837)	&  $\textit{PAH2\_2}$	&  11.68 & Photometric	& 1.066 	&	1500 & $<$ 3.16 & 10 \\
2017.04.21 08:59:26 (57864)	&  $\textit{J8.9}$	&  8.70 & Clear, humid	& 1.029	&	1500 & 3.22$\pm$0.59 & 11, 12\\
2017.04.22 08:24:43 (57865) &  $\textit{PAH2\_2}$ &  11.68 &	Photometric & 1.007	&	1200 & $<$ 2.36 & 11,13-16 \\
2017.04.22 09:03:19 (57865)  &  $\textit{M-band}$ &  4.67 & Photometric &	 1.035	&	1500 & $<$ 2.71 & 11,15,16\\
\hline
\end{tabular}
\\
\vspace{0.2cm}
\raggedright \textbf{Note.} The reported fluxes are not de-reddened. Standard stars used are: 1=HD039523; 2=HD046037; 3=HD047667; 4=HD061935; 5=HD075691; 6=HD097576; 7=HD099167; 8=HD111915; 9=HD133774; 10=HD145897; 11=HD151680; 12=HD178345; 13=HD082668; 14=HD108903; 15=HD123139 and 16=HD163376.\\
\par
\end{table*}

\subsubsection{Archival optical data}

We also use the archival data of the source obtained in the G spectral filter with the Gaia telescope\footnote{http://gsaweb.ast.cam.ac.uk/alerts/alert/Gaia17agz/} during the recent outburst. Gaia first detected the source on 2017 January 27 (MJD 57780.8) at $G$=16.44. Prior to that, the last observation it had on 29 October 2016 (MJD 57690) was a non-detection \citep[typical detection limit of Gaia is $\sim$ 20.7 mags,][]{gaia}. Gaia detected GRS~1716$-$249 on 13 days during the outburst, with the last detection on 2017 September 23 (MJD 58019). We use these public data in Fig.~\ref{fig:lightcurve}, while studying the optical light curve of the source.

\subsubsection{Archival historical optical data}

To compare the 2016--2017 outburst of the source with its discovery outburst from 1993, we include the simultaneous optical detections taken on 1993 October 8 (MJD 49268) in the $B, V$ and $R$ filters as 17.7$\pm$0.1, 16.7$\pm$0.1 and 16.0$\pm$0.1 mags, respectively \citep{della}. We also use the historical $B$ and $V$-band observations from \cite{masetti}. We use these data in the spectral energy distribution study (see Section 3.3) and the detailed analysis of the evolution of the source through the colour-magnitude diagram (see Section 3.5).

\subsection{Infrared observations}

\subsubsection{VISIR mid-IR observations}

We acquired targeted observations of GRS~1716$-$249 with the Very Large Telescope (VLT) in mid-IR wavelengths on three nights during the 2016--2017 outburst, using the VLT Imager and Spectrometer for the mid-IR \citep[VISIR;][]{Lagage2004} instrument on the VLT's UT3 (Melipal). The observations were made under the programmes 098.D-0893 and 099.D-0884 (PI : D. Russell) in $M$-band (4.15--5.19 $\mu$m), $J8.9$ (8.00--9.43 $\mu$m), $B10.7$ (9.28--12.02 $\mu$m) and $PAH2\_2$ (11.5--12.3 $\mu$m) filters on 2017 March 25 (MJD 57837), April 21 (MJD 57864) and April 22 (MJD 57865), for approximately 40--45 minutes total telescope time on each date. The integration time on source was usually 25 min, with additional substantial overheads due to the chopping and nodding pattern. The observing conditions were photometric during the observations of GRS~1716$-$249 on March 25 and April 22, and were clear with some humidity on April 21. Nevertheless, standard stars taken just before and after GRS~1716$-$249 were used to achieve accurate flux calibration on April 21. The detailed VISIR observing log and the photometric results are reported in Table 1.

The data were reduced using the VISIR pipeline in the \emph{gasgano} environment. We combined the raw images from the chop/nod cycle and performed aperture photometry in IRAF using a large enough aperture to minimize the effect of small seeing variations on the fraction of flux in the aperture \citep[the method is the same as that used in][]{baglio18}. To flux calibrate the photometry and estimate the flux density of the source, we used all the standard star observations taken within one month of the observation night in the same filter during clear sky conditions. All the standard stars used are listed in the final column of Table 1. At mid-IR wavelengths, the zero point corrections rarely vary much. In fact, we found that the ADU/flux conversion factor measured from different standard star observations within a month varied only by 5\%–10\% when the airmass is less than 1.5. For all the filters for which we had only one standard star available within one month of the observation, we use an error of 5\% in the ADU to calculate the uncertainty on the flux density of the source.

The source was only detected on one date, 21 April, in the J8.9 filter, with a magnitude of 18.24$\pm$0.19, or a flux density of 3.22$\pm$0.59 mJy (the detection has a signal-to-noise ratio of 6.2). Although the photometric error for the detection is small, the uncertainty in the flux density is increased due to systematic errors arising from the limited number of available standard stars within a month of the detection. On the other two dates when the source was not detected, we derive $3\sigma$ upper limits from the root mean square (rms) in a region centred on the position of GRS~1716$-$249. The closest WISE catalogue star is 12 arcseconds away from the position of GRS~1716$-$249 (outside the field-of-view of VISIR), with a flux density of 2.24 mJy at 12$\mu$m.

\subsubsection{REM near-IR observations}\label{REM_description}

We observed GRS~1716$-$249 in the near-IR wavelengths ($J,H$ and $K$ bands, one filter at a time) with the REMIR camera mounted on the Rapid Eye Mount (REM; La Silla, Chile) telescope between 2017 February 8 and October 1 (MJD 57792-58027). For each epoch, the reduction of the images was performed by subtracting the sky contribution; this was obtained as the median of 5 misaligned exposures of 60s and 30s in the $J$ and $H$ filter, respectively, and of 10$\times$15s exposures in $K$-band. Once the sky was subtracted, we registered and averaged the exposures to enhance the signal to noise.

We performed aperture photometry on each reduced image using IRAF. The magnitudes were then calibrated against a group of five 2MASS reference stars in the field.
We note that a 2MASS star is observed at a distance of $\sim 3.5''$ from the target in the REM images. Considering the spatial resolution of REMIR ($1.22''$/pixel) and the seeing, the two stars are therefore blended together in all images. Under the reasonable hypothesis that the 2MASS star is not variable, we subtract the contribution of the 2MASS star from the flux extracted with our analysis to build spectral energy distributions. The magnitudes of the 2MASS star are tabulated in the 2MASS catalogue ($J=15.33\pm0.06$; $H=14.46\pm0.06$; $K=14.16\pm0.08$). To double check, we found and downloaded archival $J$-band images of the field taken with the SOFI instrument at the New Technology Telescope (NTT; La Silla, Cile) during quiescence in 1999 (July 5 and 7; Program ID: 63.H-0232). The $J$-band magnitude of the 2MASS star, after calibration, is $J=15.38\pm0.05$, entirely consistent with the value reported in the 2MASS catalogue (which suggests that source has probably been stable over the years). We tabulate the REM epochs, filters and magnitudes in the Appendix (see Table A2).

\subsubsection{Archival near-IR observations}

We use the archival near-IR photometric observations of GRS~1716$-$249 during the outburst. \cite{bassi2020} reported near-IR detections of the source obtained with the Rapid Eye Mount telescope (REM) on 2017 February 9 (MJD 57793) of J=14.18$\pm$0.22, H=13.81$\pm$0.14 and K = 13.84$\pm$0.29 and 13.59$\pm$0.16, with exposure times of 300s, 150s, 75s and 75s, respectively (the same observation is also included in the dataset presented in Section \ref{REM_description}). Later, \cite{joshi17} observed the source with the Mount Abu 1.2 meter telescope and the Physical Research Laboratory (PRL) near-IR Imager/Spectrograph, and reported near-IR magnitudes on 2017 March 20 (MJD 57832) of J = 14.3, H = 14.0 and K$_S$ = 13.7, with typical errors of 0.1 magnitude, for a total integration time of 15, 15 and 17.5 minutes, respectively. These magnitudes are consistent with those derived in our analysis of the REM data during the 2016--2017 outburst, before the subtraction of the contribution from the nearby 2MASS star (see Section \ref{REM_description} for a detailed discussion). 

We also use the historical IR data of the source from its mini-outburst in 1994, taken on 1994 July 8 (MJD 49541) in the J and K filters of 16.2$\pm$0.3 and 15.5$\pm$0.3, respectively \citep{chaty}, especially for the spectral energy distribution study (see Section 3.3).

\subsection{Archival Swift/UVOT Observations}

We gathered publicly available Swift UV/Optical Telescope (UVOT) pointing observations of the source during its entire outburst from the NASA/HEASARC data center. We used the pipeline processed images and obtained the magnitude of the source using the {\tt uvotsource} HEASOFT routine, with an aperture of 5 arcsec centered on the source. An empty region close to GRS~1716$-$249 was chosen as the background region. We select only those 81 pointings where the source flux estimate is at least 5$\sigma$ above the sky background. Although the UVOT observations of this source were available in all the filters, most of the significant and usable detections were found to be in the $V$ (0.546\,$\mu$m), $B$ (0.439\,$\mu$m) and $U$ (0.346\,$\mu$m) bands, with a smaller amount of detections in the $UVW1$ (0.260\,$\mu$m) bands.

Similar to the optical flux values, we de-reddened the UV data. We use the absorption column density N$_H$ = (0.70$\pm$0.01) $\times 10^{22}$ cm$^{-2}$, reported in \cite{bassi2020}, the generic relation between optical extinction and hydrogen column density \citep{foight16}, and the wavelength-dependent extinction terms using the extinction curve of \cite{matthis}.

\subsection{Archival radio detections}

We use the radio observations of the source during its outburst with the Karl G. Jansky Very Large Array (VLA; 5.25, 7.45, 8.8 and 11.0 GHz), Australia Telescope Compact Array (ATCA; 5.5 and 9.0 GHz) and Australian Long Baseline Array (LBA; 8.4 GHz) as reported in \cite{bassi} and \cite{atri}. Radio detections of the source are available for 2017 February 9 and 11 (MJD 57993 and 57795), April 22 (MJD 57865), August 12 and 13 (MJD 57977 and 57978). We use all the radio detections for which more than one quasi-simultaneous (within 24 hours) optical/UV measurement is available, to study the broadband spectral energy distribution (SED) of the source (see Section 3.3).

\begin{figure*}[ht]
\center
\includegraphics[width=14.0cm,angle=0]{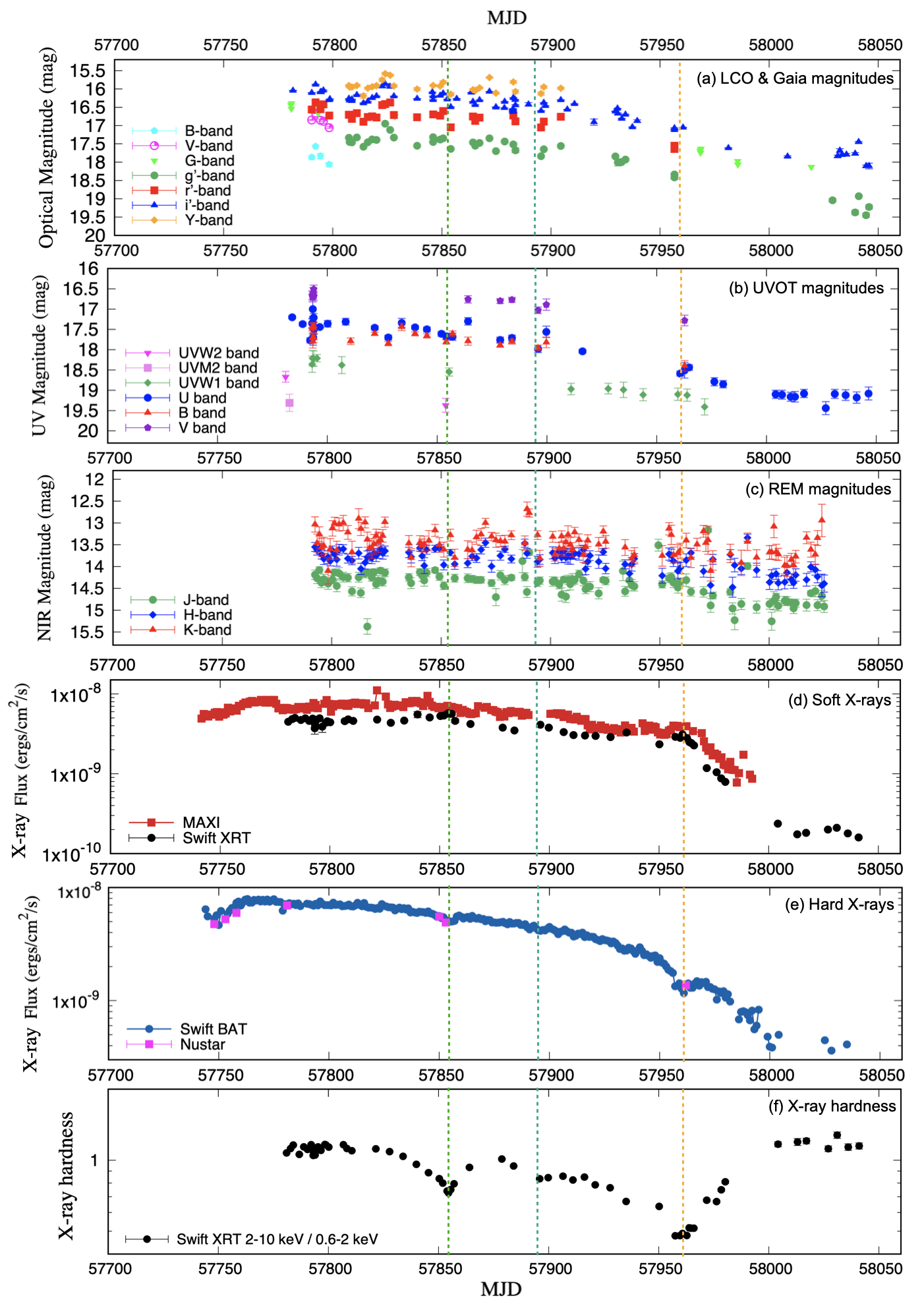}
\caption{Light curves of the 2016--2017 outburst of GRS~1716$-$249 in different wavelengths. The three vertical dashed lines depict the softest points in each of the three softening events when the source transitioned from the hard to the hard-intermediate state, as reported by \cite{bassi}, on MJD 57854.2 (green), MJD 57895.9 (blue) and MJD 57960.7 (yellow). \emph{(a)} The optical light curves from the LCO telescopes in $B$, $V$, ${g}^{\prime }$, ${r}^{\prime }$, ${i}^{\prime }$ and $Y$ bands, and from the Gaia telescope in $G$ band. \emph{(b)} The UV light curves represent Vega magnitudes from Swift UVOT in $UVW2$, $UVM2$, $UVW1$, $U$, $B$ and $V$ filters. \emph{(c)} The near-IR light curves with vega magnitudes obtained with the REM telescope; here we show the $J, H, K$ magnitudes of the source blended with a non-variable 2MASS star with magnitudes $J=15.33\pm0.06$, $H=14.46\pm0.06$ and $K=14.16\pm0.08$ (see text). \emph{(d)} The soft X-ray fluxes obtained from Swift/XRT in the 2-10 keV and MAXI/GSC in the 2-20 keV range, plotted in a logarithmic scale. \emph{(e)} The hard X-ray fluxes obtained with Swift/BAT and NuSTAR in the 15-50 keV telescopes, plotted in a logarithmic scale. We only plot the 5 sigma detections in the X-rays, converting the source count-rate to fluxes, assuming a photon index of $\Gamma$ = 1.68$\pm$0.01 and N$_H$ = (0.70$\pm$0.01) $\times 10^{22}$ cm$^{-2}$ \citep{bassi2020}. \emph{(f)} X-ray hardness from Swift/XRT showing the ratio of X-ray fluxes in the ranges of 2-10 keV and 0.6-2 keV.}
\label{fig:lightcurve}
\end{figure*}

\subsection{Archival data from X-ray telescopes}

We acquired X-ray monitoring data of GRS~1716$-$249 from the Swift/BAT and Swift/XRT telescopes. Swift/BAT has observed the source almost daily from 2016 December 1 (MJD 57723) in the 15$-$50 keV flux range. We extracted the daily average light curve data of this source from the Swift/BAT transient monitor\footnote{https://swift.gsfc.nasa.gov/results/transients} \citep{bat}.  To convert the count-rates to flux, we used the hydrogen column density N$_H$ = (0.70$\pm$0.01) $\times 10^{22}$ cm$^{-2}$ and a photon index of $\Gamma$ = 1.68$\pm$0.01, as reported in \citep{bassi2020}. Swift/XRT observations were made every few days between 2017 January 28 (MJD 57781) and 2017 October 20 (MJD 58046), mostly in the window timing mode, with target IDs 34924 and 88233 \citep[see Table 1, of][for a detailed observation log]{bassi}. We used the on-line Swift/XRT products generator\footnote{https://www.swift.ac.uk/user\textunderscore objects/} \citep{evans1,evans2} to extract the 2$-$10 keV count rate of GRS~1716$-$249 from each XRT observation, after correcting for instrumental artefacts.

GRS~1716$-$249 was also observed with Nuclear Spectroscopic Telescope Array (NuSTAR) during the 2016--2017 outburst. We calculate the NuSTAR flux density of the source using {\tt NUPIPELINE V0.4.6} in HEASOFT V6.25, with the calibration file version v20171002. Both Science Mode and Spacecraft Mode data were considered. The flux was calculated in the Swift BAT energy band (15-50 keV) using the best-fit spectral models provided in \cite{jiang}. A relativistic disk reflection model with a variable disk density parameter was used \citep{ross}. Detailed descriptions of spectral modelling can be found in \cite{jiang}.

We also gathered daily X-ray monitoring data of GRS~1716$-$249 from MAXI/GSC\footnote{\url{http://maxi.riken.jp/top/lc\textunderscore bh.html}} \citep{maxi} in the 2$-$20 keV range covering the complete outburst.

\section{Results}

\subsection{Multi-wavelength light curve}

The light curves of the entire outburst are plotted in Fig.~\ref{fig:lightcurve} including data in near-IR (REM), optical (LCO and Gaia), UV (UVOT) and X-ray (NuSTAR, Swift XRT and BAT, MAXI) wavelengths.

\begin{figure}[ht]
\center
\includegraphics[width=8.4cm,angle=0]{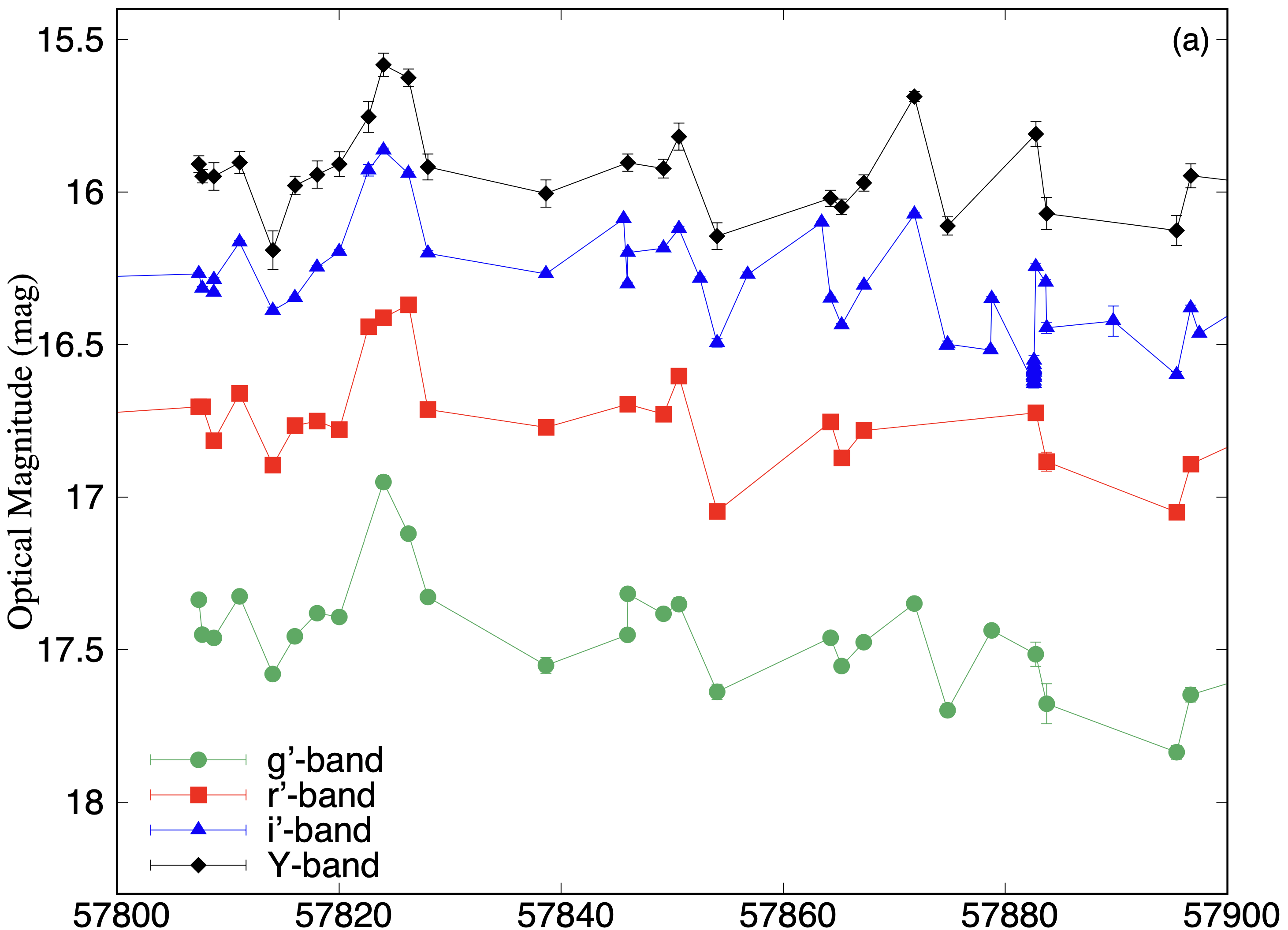}
\includegraphics[width=8.4cm,angle=0]{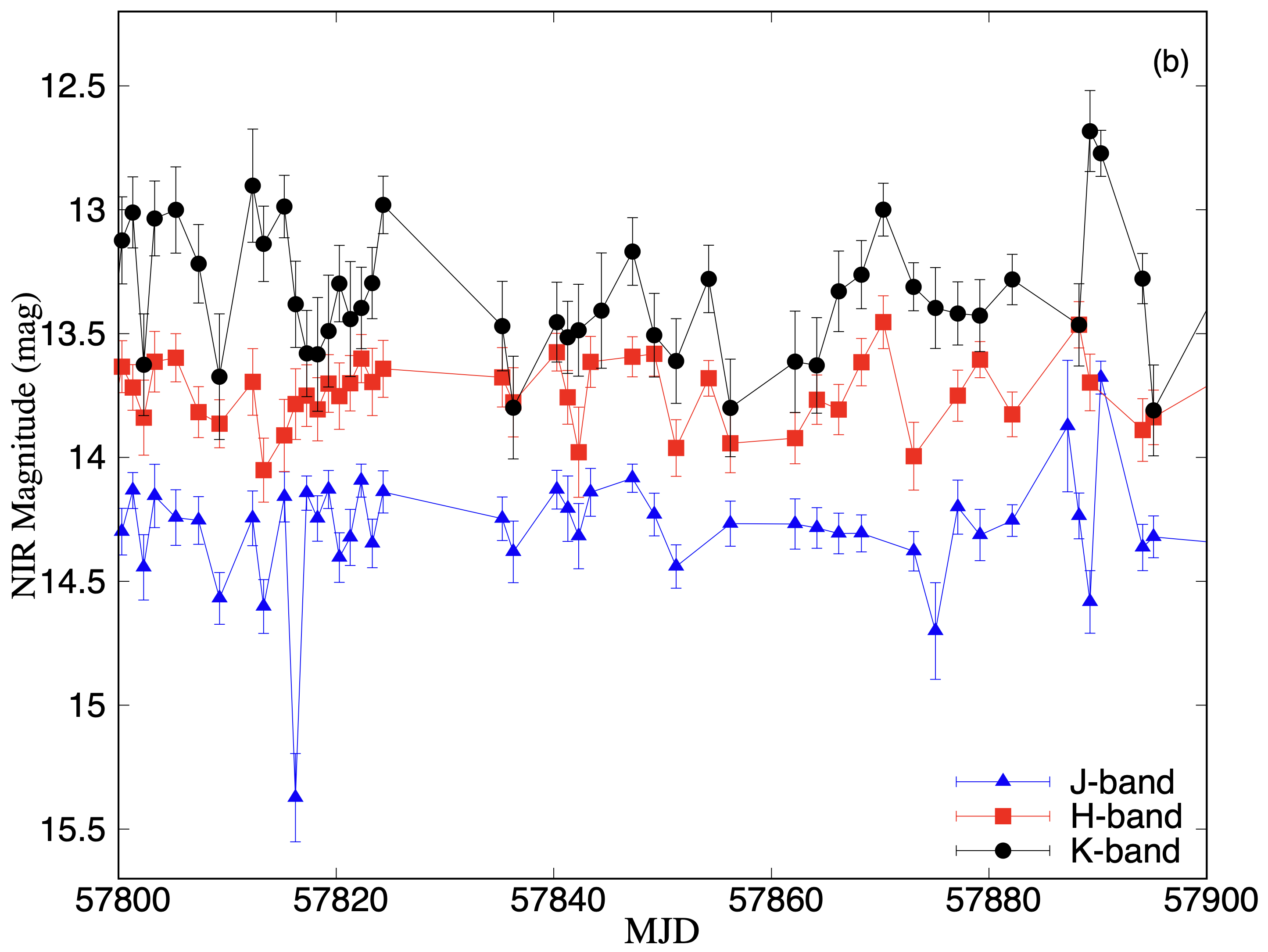}
\caption{Zoom-in of the \emph{(a)} optical and \emph{(b)} near-IR light curve of GRS~1716$-$249 during the peak of the 2016--2017 outburst to show the optical/IR variability.}
\label{fig:zoomin}
\end{figure}

The first optical detection of GRS~1716$-$249 during the 2016--2017 outburst was obtained by LCO on MJD 57781, when the optical magnitude was already bright, with ${i}^{\prime }$ = 15.97 $\pm$ 0.01. Since that time we regularly monitored the
source in ${i}^{\prime }$, ${g}^{\prime }$, ${r}^{\prime }$ and $Y$ bands until its optical emission faded back to quiescence. There are also some scattered observations taken in the $B$ and $V$ bands. A zoom-in of the optical light curve during the peak of the outburst between 2017 January 28 and May 27 (MJD 57800-57900) is shown in Fig.~\ref{fig:zoomin}a. The general trend of the LCO light curve is an almost constant plateau in all the optical bands during the outburst, until $\sim$ 2017 May 27 (MJD 57900), followed by a slow and steady decline to quiescence until 2017 October 20 (MJD 58046). The same behaviour was also observed in the Gaia optical light curve in $G$-band. At the end phase of the decline, we find a small amplitude optical brightening of the source, with ${i}^{\prime }$ magnitudes changing from 17.80 $\pm$ 0.01 on 2017 October 7 (MJD 58033), to 17.45 $\pm$ 0.01 on October 15 (MJD 58041), and then again back to 18.10 $\pm$ 0.01 on October 18 (MJD 58044). 

The optical/UV light curves obtained with Swift/UVOT in the $U$, $B$, $V$, $UVW1$, $UVW2$ and $UVM2$ bands, show a similar outburst profile as the LCO and Gaia light curves. The complete outburst, including the decay towards quiescence, is well-covered by the $U$-band data. The brightening observed in optical wavelengths during the decline of the outburst is not evident in the UV data.

The near-IR REM light curve is approximately constant, with some scatter and flickering, until 2017 June 21 (MJD 57925). A zoom-in of the near-IR light curve during the peak of the outburst is shown in Fig.~\ref{fig:zoomin}b. After MJD 57925, the flux starts to show a decreasing trend in all bands until $\sim$ 2017 September 4 (MJD 58000), when the flux experiences a plateau that lasts until the end of the observations. This behavior is similar to that of the higher energy light curves.

The X-ray light curves in both the hard (Swift/BAT and NuSTAR) and the soft (Swift/XRT and MAXI) energy ranges follow the morphology of the near-IR/optical light curve, indicating a correlated behaviour, which is explored in detail in Section 3.4.

\subsection{Optical/IR variability}

\subsubsection{Optical variability}

Fig.~\ref{fig:zoomin}a shows the zoom-in of the optical light curve during the peak of the outburst. On longer timescales, the four LCO optical filters (${g}^{\prime }$, ${r}^{\prime }$, ${i}^{\prime }$ and $Y$ band) are clearly correlated. We also took higher cadence optical observations on 2017 May 9 (MJD 57882; 15 detections in $\sim$17.5 mins with a time resolution of $\sim$ 75 seconds) of the source with LCO ${i}^{\prime }$-band. The optical fractional rms deviation in the flux on such short timescales (minutes; i.e. a frequency range of 0.0010--0.013 Hz) during the hard state, evaluated following the method described in \cite{vaughan} and \cite{gandhi2010}, is found to be 1.3$\pm0.4\%$, reflecting on a very weak variability.

The observed rms is substantially lower than the optical fractional rms of BHXBs like GX~339--4 and Swift~J1357.2--0933 in the hard state and V404~Cyg in the flaring state, which are $\sim 5$--20\% on similar and shorter timescales \citep{gandhi2009,gandhi2010,cadollebel2011,gandhi2016,paice2019}. The variability seen in GRS~1716$-$249 is similar to the lower optical fractional rms values of $\sim$3--5\% seen in the hard accretion states of Swift~J1753--0127, XTE~J1118+480 and MAXI~J1535--571 \citep{gandhi2009,hynes2009,baglio18}. Such variability is also observed in the soft accretion states of GX~339--4 and GRO~J1655--40 \citep{hynes1998,obrien2002,cadollebel2011}, when the accretion disk dominates the emission.

\begin{figure*}[t]
\center
\includegraphics[width=8.2cm,angle=0]{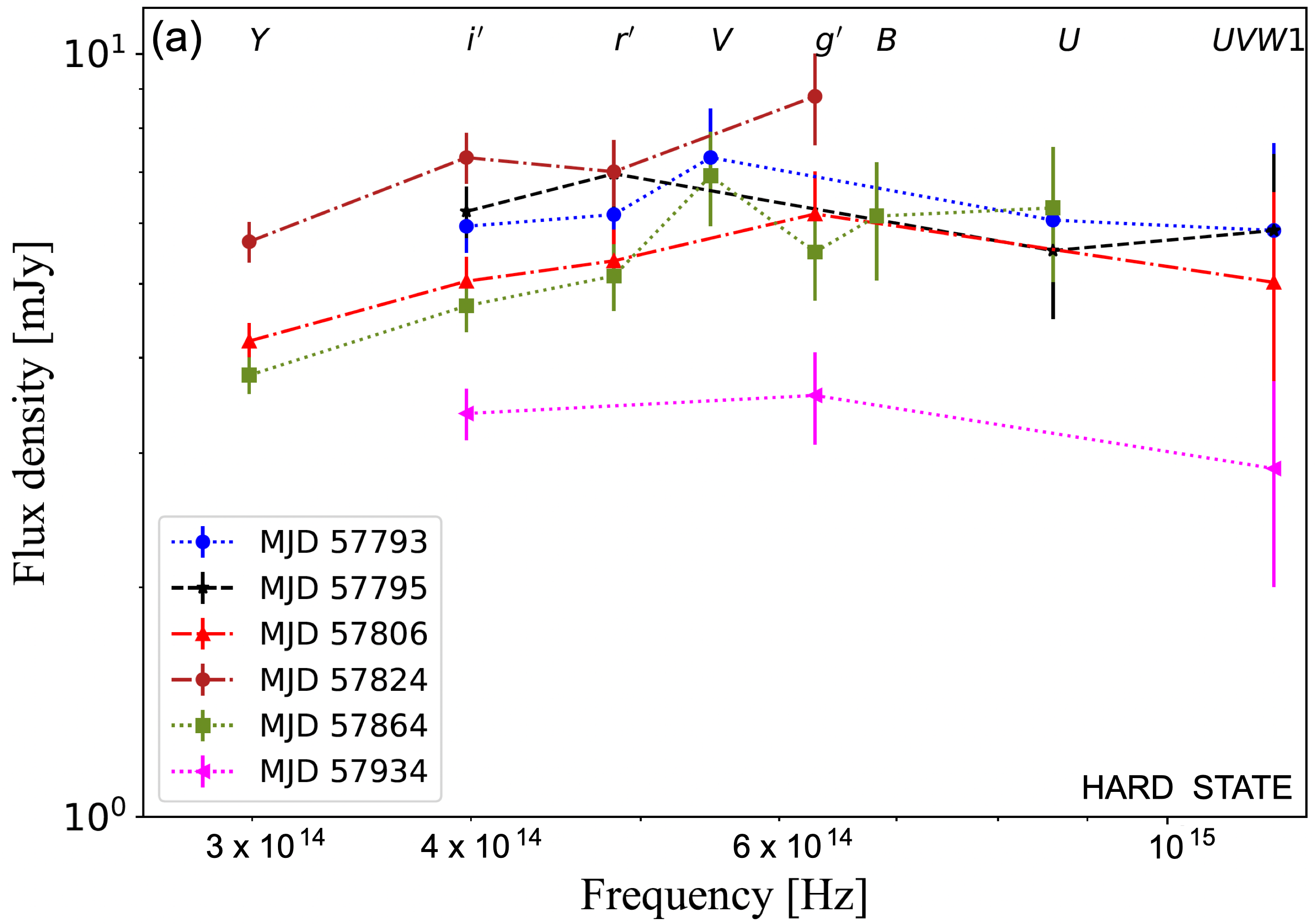}
\includegraphics[width=8.2cm,angle=0]{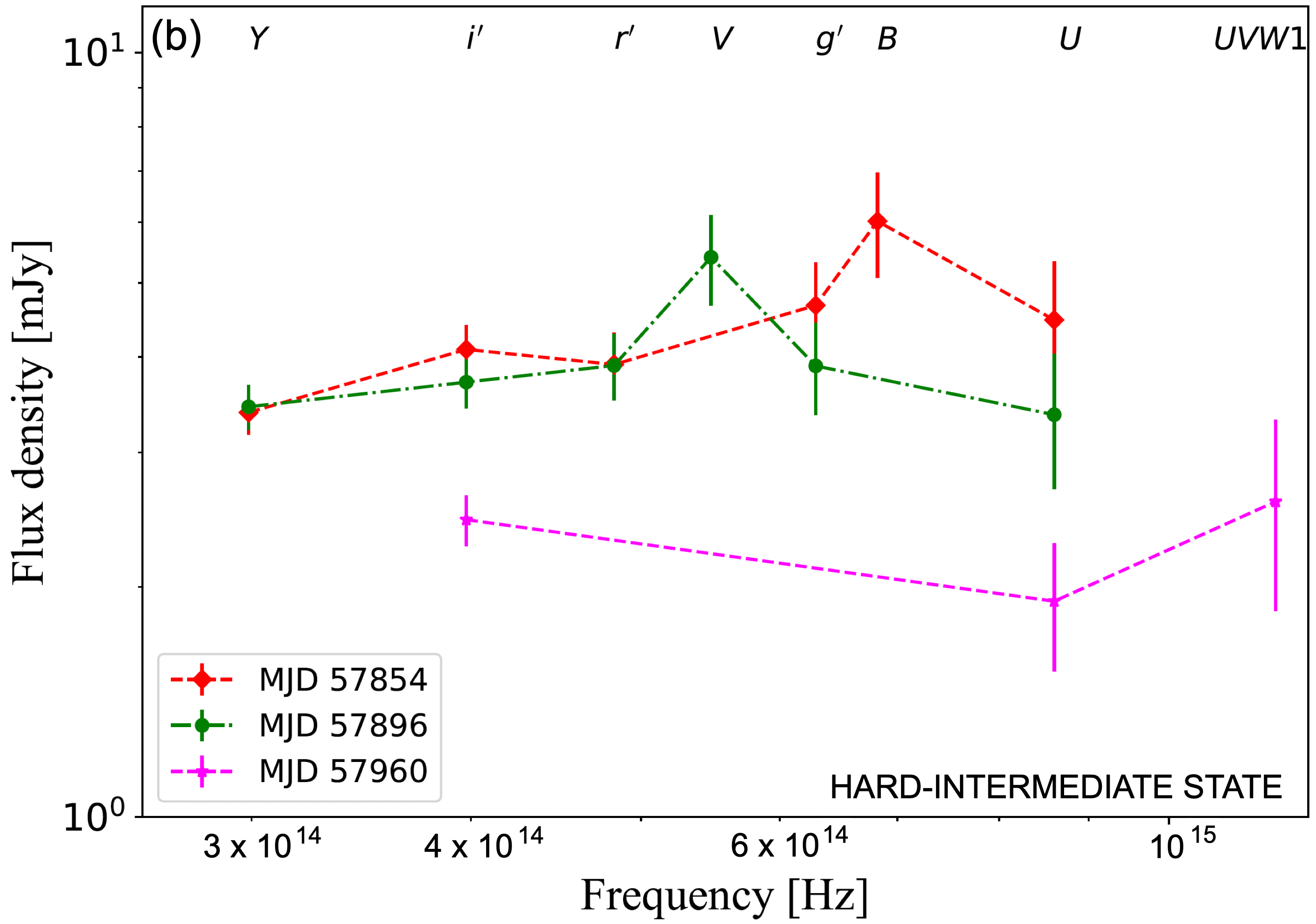}
\includegraphics[width=17.2cm,angle=0]{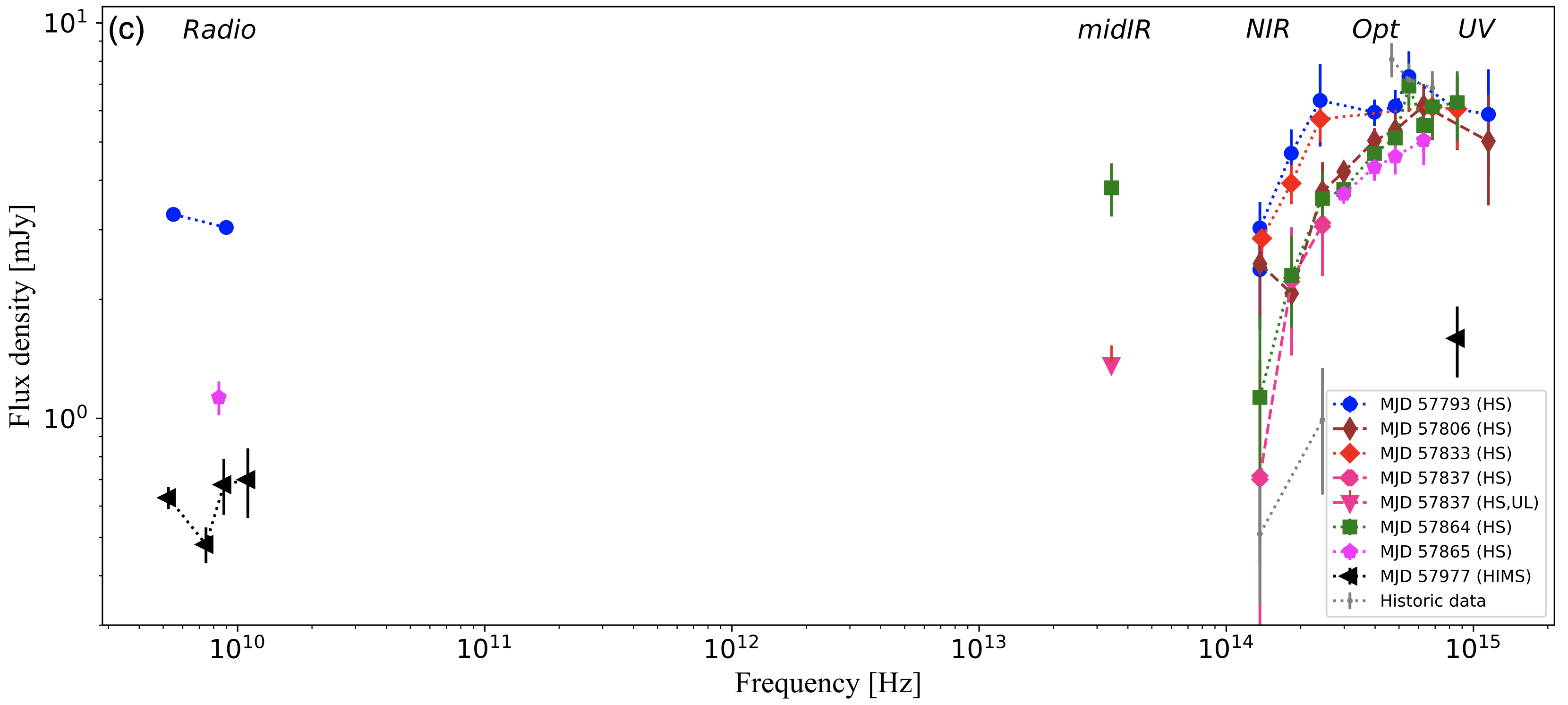}
\caption{De-reddened optical/UV spectra of GRS~1716$-$249 during \emph{(a)} the hard, and \emph{(b)} the hard-intermediate state of the outburst, when quasi-simultaneous (within 24 hours) observations were available. \emph{(c)} The de-reddened radio/mid-IR/near-IR/UV spectrum when quasi-simultaneous (within 24 hr) data were available. The mid-IR to radio spectral index measured from the VISIR mid-IR detection on MJD 57864.4 and LBA radio detection at 8.4 GHz on MJD 578865.7 is found to be $\alpha$=0.13$\pm$0.03. The mid-IR upper-limit on MJD 57837 in the J8.9 filter is plotted as an inverted triangle, to show the mid-IR variability of the source. We also plot for reference the historical optical \citep[MJD 49268,][]{della} and near-IR \citep[MJD 49541,][]{chaty} SEDs from its discovery outburst in 1993/1994 (in grey dotted lines).}
\label{fig:sed}
\end{figure*}

\begin{figure}[ht]
\center
\includegraphics[width=8.2cm,angle=0]{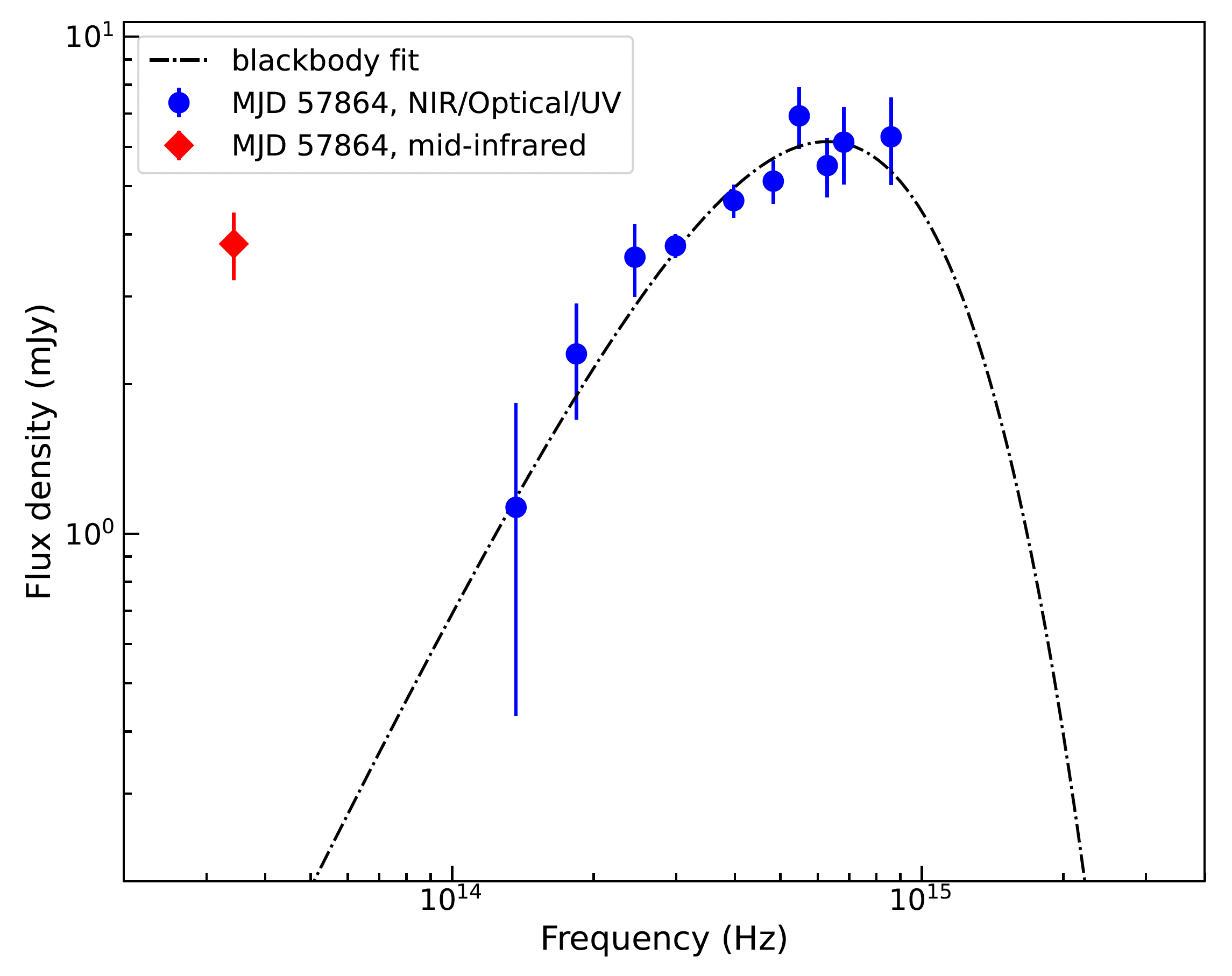}
\caption{The quasi-simultaneous, de-reddened NIR/optical/UV spectrum of GRS 1716-249 on MJD 57864 (blue circles). Superimposed is the fit of the spectra with a single-temperature black body (black line). We also overplot the quasi-simultaneous mid-IR detection of the source for comparison (red diamond). We show that the NIR/optical/UV part of the spectrum is qualitatively well represented by the single temperature black body, while the mid-IR emission is comparatively brighter probably due to an additional contribution from the jet.}
\label{fig:sedfit}
\end{figure}

\begin{figure*}[ht]
\center
\includegraphics[width=15.68cm,angle=0]{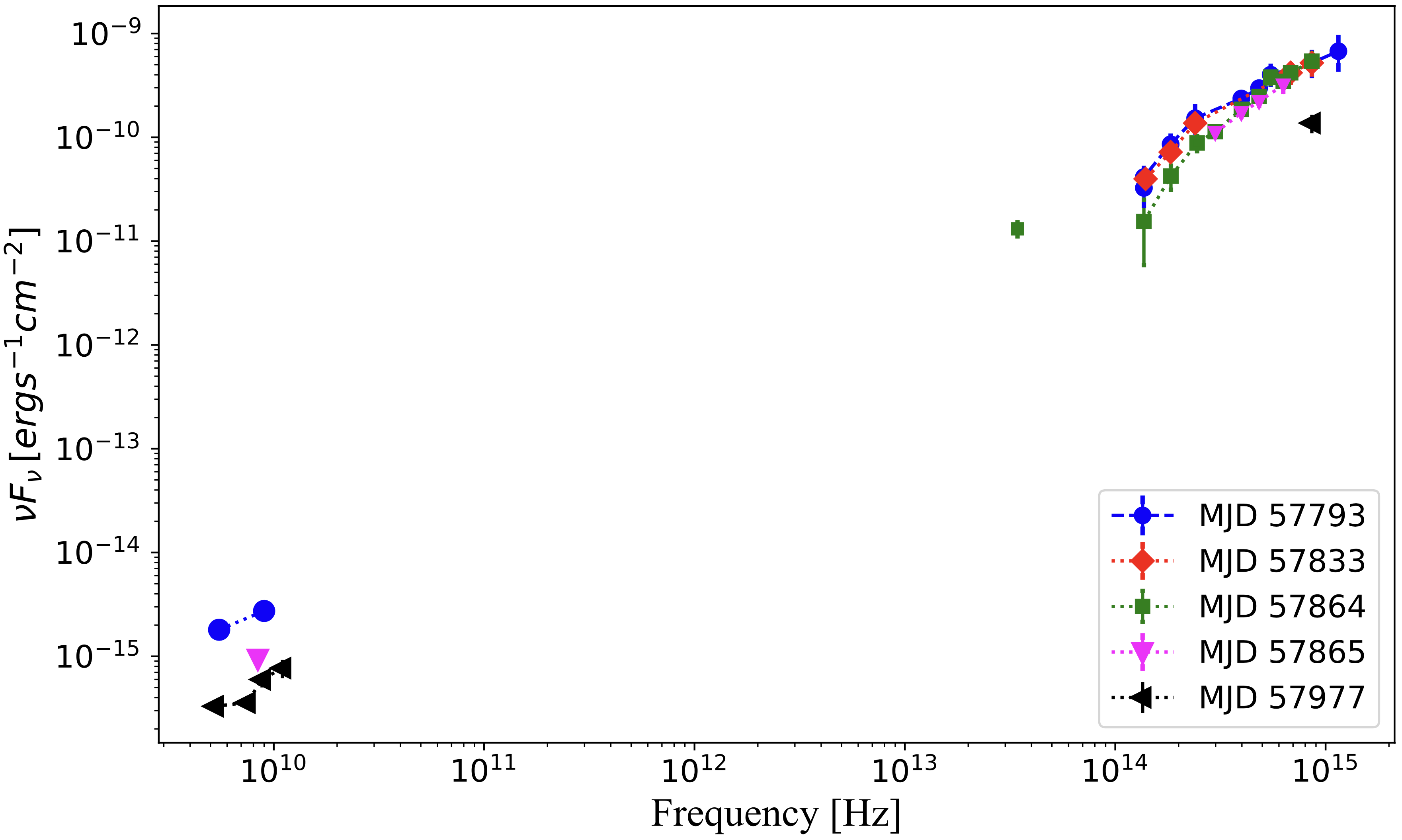}
\caption{De-reddened broadband SED of GRS~1716$-$249 on the five days during the hard state when quasi-simultaneous (within 24 hours) infrared, optical, radio and UV detections were available. The SEDs presented here are from the hard state, except on MJD 57977 (shown in black triangles) when the source was in the hard--intermediate state.}
\label{fig:broadbandsed}
\end{figure*}

\subsubsection{Infrared variability}

Fig.~\ref{fig:zoomin}b shows the zoom-in of the near-IR light curve during the peak, where the source is observed to be varying by $\sim$1 magnitude. We also observe a possible small amplitude ($\sim$0.5-0.6 mag) flare happening in $J$ and $K$ bands between MJD 57887 and 57891. However, no corresponding activity is observed in $H$-band, and the lack of time-resolved data during these days makes it difficult to study this event further. The fractional rms deviation in the infrared flux of GRS~1716$-$249 during the peak of the outburst on longer timescales (days/weeks; a frequency range of 5.8E-6 -- 8.7E-8 Hz), after removing the contribution from the blended star, is measured to be 20.69$\pm$2.34\%, 10.92$\pm$4.86\% and 34.43$\pm$4.24\% in the $J$, $H$, and $K$ bands, respectively. Hence GRS~1716$-$249 is variable in the near-IR band. Although the coverage of the outburst in the mid-IR range is scarce, the detections and the upper limits (see also Table 1 and Section 2.2.1) also point to a variable mid-IR emission, with the flux density spanning from $<1.4$ mJy to $3.2 \pm 0.6$ mJy at 8.7 $\mu$m.

A similar increase of fractional rms deviation in the flux towards longer wavelengths in the optical/IR wavelength range, is also seen in other BHXBs. For example, the rms is often 10--20\% or higher in the near-IR regime in the hard accretion states, as seen in sources like XTE~J1550--564, GX~339--4 and MAXI~J1820+070 \citep{curran2013,vincentelli2018,tetarenko2021}. In the mid-IR regime, the fractional rms increases further, with for example rms $\sim$15--22\% in MAXI~J1535--571 at a similar time resolution, which supports a jet origin to the variability on these timescales \citep{baglio18}. In XTE~J1118+480, the spectrum of the rms variability is consistent with a power law of spectral index $\alpha = -0.6$ from optically thin synchrotron radiation, spanning near-IR to X-ray \citep{hynes2003,hynes2006}. In the hard accretion state of MAXI~J1820+070, the fractional rms (in a larger integrated frequency range; $10^{-4}$--50 Hz) decreases monotonically with increasing wavelength, from tens of per cent in the optical/near-IR, to 2--8\% at radio frequencies \citep{tetarenko2021}. Other timing properties such as the frequency of the break in the power spectrum was also seen to vary smoothly with wavelength from optical to radio, with time lags between bands increasing at longer wavelengths. One interpretation is that although the fractional variability increases from optical to IR due to an increase in the jet contribution, the fractional rms drops again as it approaches radio wavelengths, because the variability in the jet dominated bands gets more smoothed out by the larger size scale of the emitting region at the longer wavelengths.

\subsection{Spectral energy distribution}

We construct the optical/UV spectra as well as the broadband SEDs of GRS~1716$-$249 in the hard (Fig.~\ref{fig:sed}a) and hard-intermediate states (Fig.~\ref{fig:sed}b), to illustrate the peculiar multi-wavelength characteristics of the source. In Fig.~\ref{fig:sed} we plot the optical/UV spectra of GRS~1716$-$249 in both the hard (Fig.~\ref{fig:sed}a) and hard-intermediate states (Fig.~\ref{fig:sed}b) of the outburst. We use quasi-simultaneous observations obtained within 24 hours, and convert the magnitudes to de-reddened fluxes as described in Section 2.1 for optical LCO magnitudes and Section 2.3 for archival Swift/UVOT observations. In both the hard-intermediate and hard states, the SEDs are found to be smooth up to the $UVW1$-band, with a shallow peak around the ${g}^{\prime }$ band. 

We use the available information to constrain the intrinsic optical/UV spectral index by fitting the function $S_{\nu} \propto \nu^{\alpha}$, where $S_{\nu}$ is the flux density, $\nu$ is the frequency and $\alpha$ is the spectral index. We obtain an average $U$-${i}^{\prime }$ spectral index of $\alpha_{U-i'}$ = $-0.1\pm$0.3 across the spectra. Generally, a negative slope $\sim -0.7$ \citep[e.g.][]{gandhi2011} is expected if there is a jet present with an optically thin synchrotron spectrum dominating the near-IR/optical regime. This value can be even more negative, as seen in cases like Swift~J1357.2$-$0933 where the quiescent optical/mid-IR SED has a power-law index of $-1.4$, arising from a weak jet \citep{shahbaz}. Although optically thick, self-absorbed synchrotron emission from a jet can produce slope $\sim -0.1$, there are only a few cases in which such emission extends to higher frequencies like optical \citep[e.g.][]{russell2013,maitra2017}. A positive slope SED is expected (with spectral index $\sim$ 1) if the optical emission is dominated by the blackbody from the outer accretion disk \citep[e.g.][]{hynes2005}. For a viscously heated disk, $\alpha \sim$ 0.3 is expected, turning to a steeper slope 0.3 $< \alpha <$ 2.0 at lower frequencies \citep[e.g.][]{fkr}. Very often, a combination of all the processes can result in an intermediate slope.

For comparison, during the hard state, the spectrum constructed for GRS~1716$-$249 on 2017 February 22 (MJD 57806), has an $\alpha_{Y-g'} \sim 0.5$, while during the hard-intermediate state on 2017 April 11 (MJD 57854), we find  $\alpha_{Y-B} \sim 0.7$ (See Fig.~\ref{fig:sed}a and \ref{fig:sed}b). This suggests the optical spectra are probably dominated by an accretion disk, but it is unlikely for the UV/optical emission to solely originate from the Rayleigh-Jeans part of a single-temperature blackbody spectrum.

\begin{figure*}[ht]
\center
\includegraphics[width=8.8cm,angle=0]{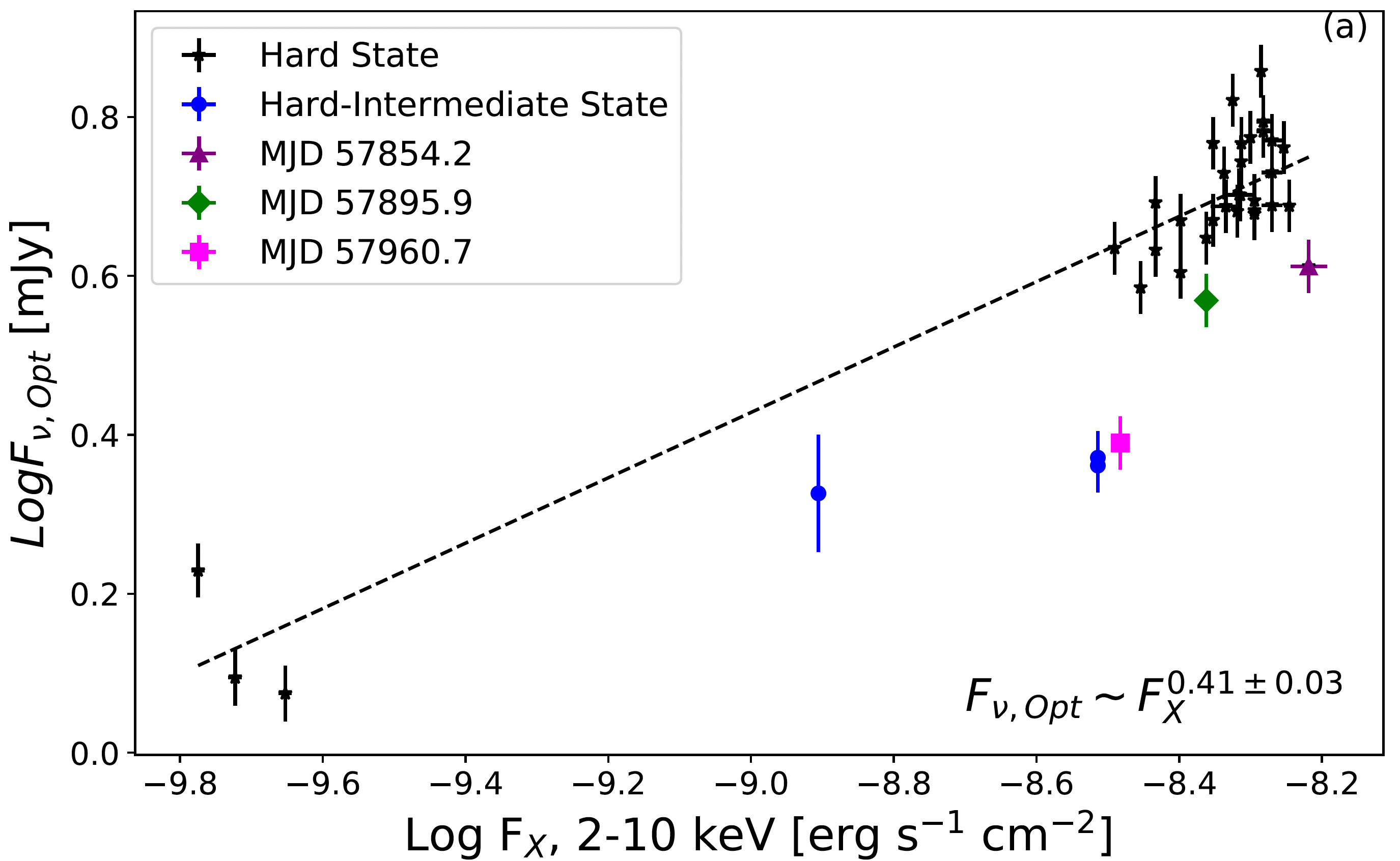}
\includegraphics[width=9.1cm,angle=0]{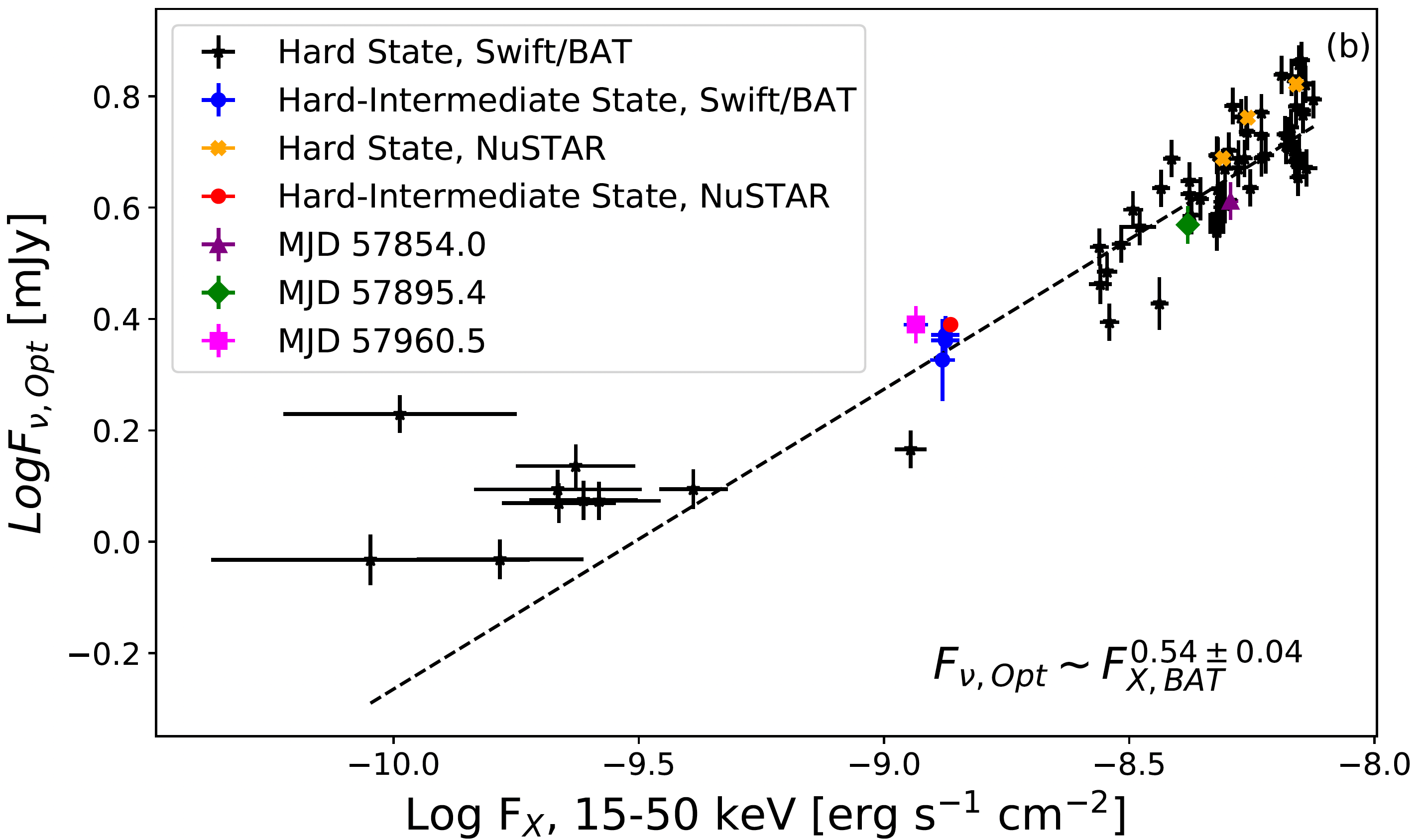}
\includegraphics[width=8.8cm,angle=0]{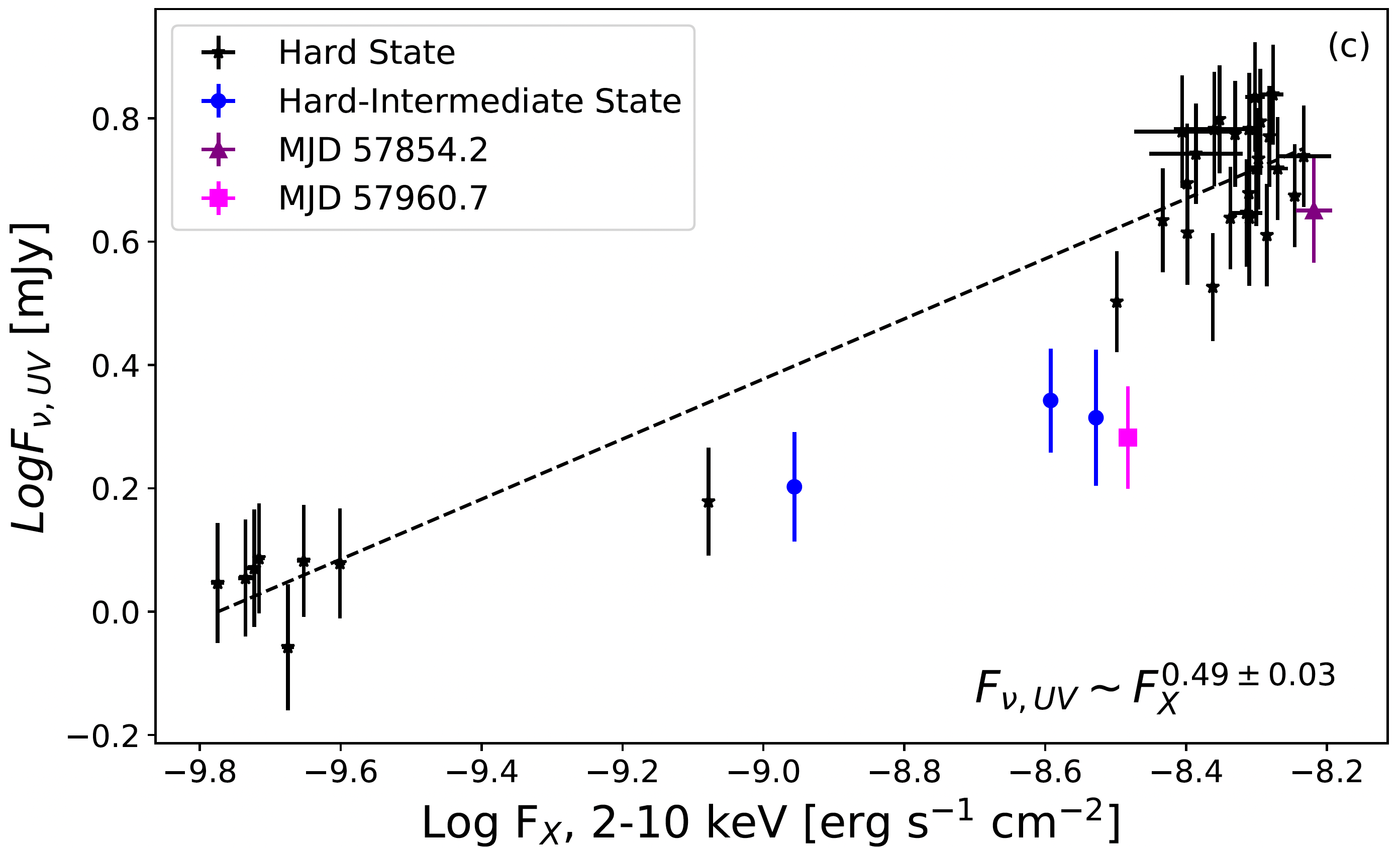}
\includegraphics[width=8.8cm,angle=0]{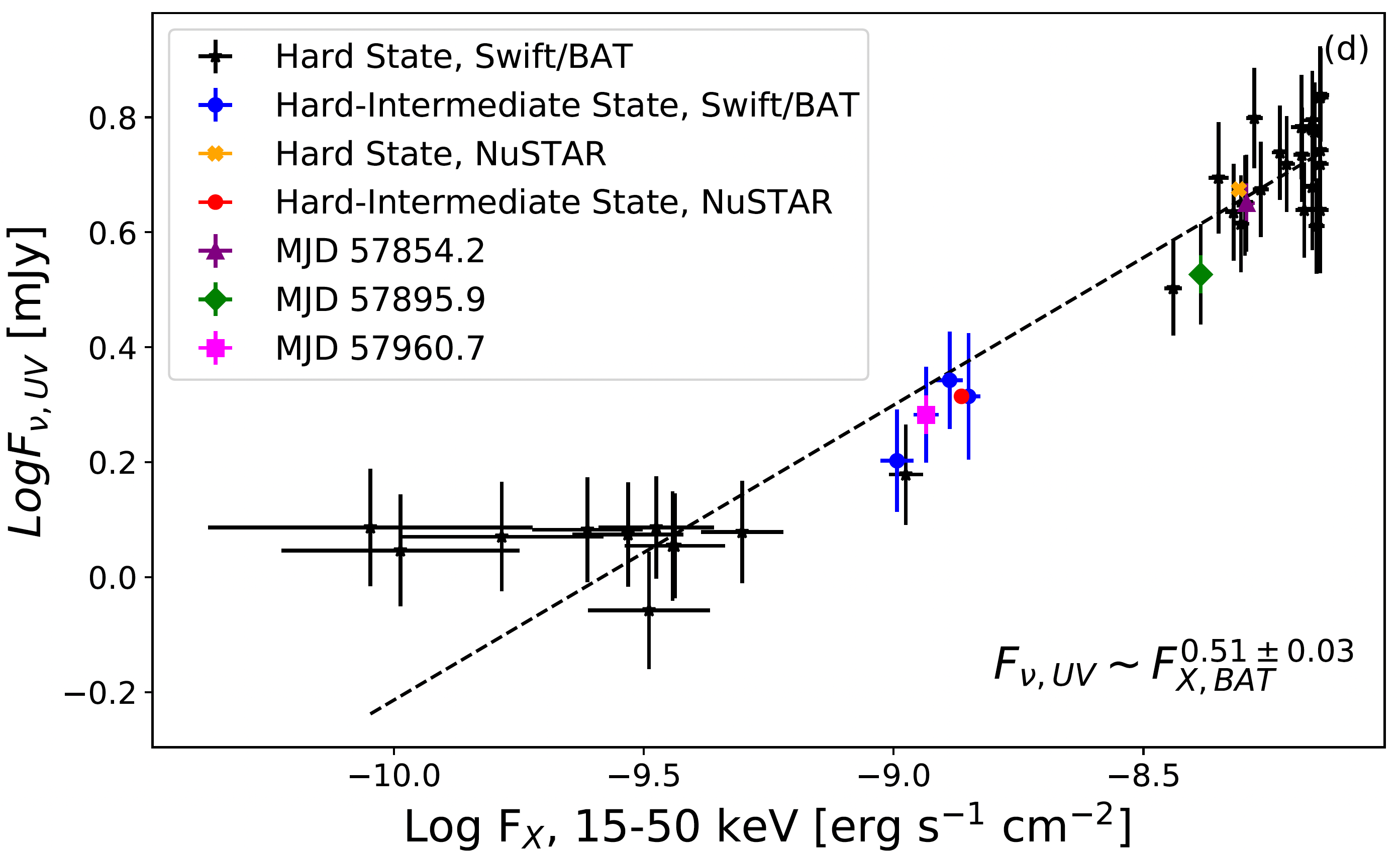}
\caption{Optical/X-ray correlation for GRS~1716$-$249. While the top two plots show the correlation with optical flux densities in ${i}^{\prime }$-band obtained from the LCO telescopes, the bottom plots use the $U$-band flux densities obtained with Swift UVOT. We use quasi-simultaneous (within 24 hours) X-rays taken either from Swift/XRT in the 2-10 keV range (for the left plots), or from Swift/BAT and NuSTAR in the 15-50 keV (in the right plots). The black dashed lines show the best-fit using the orthogonal distance regression method of least squares, and the slope is mentioned in the plot. The hard state values with Swift XRT and BAT are shown in black stars, the hard state values with NuSTAR are shown in orange crosses. The hard-intermediate state values with Swift XRT and BAT are shown in blue circles, and with NuSTAR in red circles. The softest points during the three softening events \citep{bassi} are shown in coloured symbols, where the MJD values are indicated.}
\label{fig:correlation}
\end{figure*}

To investigate the issue further and disentangle the emission processes in the optical and UV regime, we include the available IR and radio data and construct the broadband spectrum (see Fig.~\ref{fig:sed}c) and the SED (see Fig.~\ref{fig:broadbandsed}) of GRS~1716$-$249 with quasi-simultaneous (within 24 hours) optical (with LCO), near-IR (with the Mount Abu 1.2 meter telescope and REM), mid-IR (with VISIR) and radio (with ATCA, LBA and VLA) data. The broadband spectrum has a positive slope in the near-IR regime, which flattens in the optical, and gets fainter in the UV wavelengths. We fit the spectrum on MJD 57864 with a single temperature black body curve (See Fig.~\ref{fig:sedfit}), and find that the NIR/optical/UV part of the spectrum is qualitatively well represented by a black body model with a temperature $\sim$10500$\pm$200 K, while the mid-IR emission is comparatively brighter. This suggests that the overall shape of the IR/optical/UV spectra together are as expected for the outer accretion disc. The steeper slope in the NIR is the Rayleigh-Jeans limit of the blackbody with the lowest temperature. The fainter UV emission suggests that the viscous disk does not dominate in these wavelengths (as the UV emission does not keep rising with alpha $\sim$0.3). Instead, the irradiated disc most likely dominates the emission, as a peak is seen around g'-band, with UV flux densities being slightly fainter. This is similar to seen in other BHXBs where the irradiation bump peaks in the optical, with the UV slightly fainter \citep[e.g.][]{hynes2005}. The historical optical (taken in 1993) and near-IR (taken in 1994) data from the discovery outburst, plotted in Fig.~\ref{fig:sed}c for a comparison with the current outburst, show a similar brightness profile in optical, although not as flat as the recent data. In the near-IR wavelengths, although it was fainter during the historical outburst, it shows a similar shape with a positive slope.

The single mid-IR detection of the source, obtained on 2017 April 21 (MJD 57864), is significantly brighter than what is expected from the disk alone (see Fig.~\ref{fig:sedfit} and Fig.~\ref{fig:broadbandsed}), and can probably be attributed to synchrotron emission from a compact jet during the outburst. The mid-IR to radio spectral index measured from the VISIR mid-IR detection on MJD 57864.4 and the LBA radio detection at 8.4 GHz on MJD 57865.7 (a separation of 1.3 d) is found to be $\alpha$=0.13$\pm$0.03. This is slightly more positive than (but consistent with to within $< 2\sigma$), the reported radio spectral indices of $\alpha$=-0.15$\pm$0.08 and $\alpha$=-0.07$\pm$0.19, at the beginning (2017 February 9, MJD 57793.8) and close to the end of the outburst (2017 August 12, MJD 57977.3), respectively \citep{bassi}. Therefore, the radio to mid-IR spectrum is consistent with a flat or slightly inverted spectrum coming from a compact jet. We also estimate the spectral index between the mid-IR detection and quasi-simultaneous X-ray spectrum as $\alpha=-0.26$. If the mid-IR emission arises from optically thin synchrotron, then its extrapolation to X-ray is much fainter than the observed X-ray power law index, which implies that the synchrotron jet does not contribute much to the X-ray flux.

\subsection{Multi-wavelength correlations}

Another tool for disentangling the emission processes in BHXBs during outburst is multi-wavelength correlations. We study the quasi-simultaneous multi-wavelength correlation of GRS~1716$-$249, using de-reddened optical and UV fluxes as a function of the soft X-ray (2-10 keV) fluxes from Swift/XRT, and hard X-ray (15-50 keV) fluxes from Swift/BAT and NuSTAR; whenever the X-ray fluxes are obtained within a day of the optical or UV observations.

\subsubsection{Optical versus X-ray correlations}

For the optical versus soft X-ray (2--10 keV) correlation study, we use Swift/XRT flux in the 2-10 keV range for X-rays, and de-reddened optical ${i}^{\prime }$-band flux density (as ${i}^{\prime }$-band had the best coverage amongst all optical filters). We choose all the points for which we have quasi-simultaneous data (i.e., data obtained within 24 hours; see Fig.~\ref{fig:correlation}a). While the hard state values follow one single correlation, the hard-intermediate state values (shown in the plot as coloured points with non-circular symbols, where the MJD values are indicated) are generally seen to lie on the lower side of the correlation. This is in agreement with previous studies where comparatively less optical emission is observed during the state transition and soft state \citep[eg.][]{jain01,corbel2002,hb,Russell2006,Coriat2009}. Generally, this is thought to be due to a weak jet component to the optical emission, which usually fades during the transition from the hard to hard-intermediate state \citep[eg.][]{cadollebel2011,baglio18} and recovers when a BHXB returns to the hard state \citep[e.g.][]{Corbel2013,Kalemci2013,russell2013}. Another reason could be a weak disk-blackbody component which can contribute towards the X-ray luminosity in the 2-10 keV energy range during the hard-intermediate state \citep{cap,ala2020}.

The correlation is found to be significant (Pearson correlation coefficient = 0.84, p-value = $2.5 \times 10^{-10}$). The best-fit slope for the correlation in hard state using the orthogonal distance regression method of least squares is 0.41$\pm$0.03. The observed slope suggests an X-ray irradiated accretion disk \citep{van}, with possibly some contribution from the viscous disk \citep[][]{Russell2006}. But we note that the scaling relation of \cite{van} depends on an assumed geometrical configuration and is not as simple and straightforward. Recent studies have showed that the slope of the correlation can differ depending on the origin of the emission at different regimes \citep[e.g.][]{Coriat2009,be}, and irradiation from a hot dense accretion disk wind may also cause a slight distortion of the scaling relation \citep[][see also Section 4.1 for a detailed discussion]{cuneo}.

In addition, we note that there is a slight hint of the correlation flattening at the fainter end of the luminosity ranges. As an alternative explanation, we attempted to fit the data with a broken power law (keeping the break luminosity as a free parameter). The correlation obtained were found to have a steeper slope ($\sim$1.1) at the brighter end, and a shallower slope ($\sim$0.2) at the fainter end, implying that the viscous disk could play a role at the lower luminosities (the break luminosity was found to be $\sim2.5\times10^{-9}\, \rm erg\,s^{-1}\,cm^{-2}$). A prominent role of the viscous disk in the fainter part of the outburst is also hinted at by our color evolution analysis (see Section 3.5).

The hard X-ray emission (15--50 keV) of the source follows a power-law correlation with the optical flux. To check this correlation, we used hard X-ray data in the 15--50 keV range from Swift/BAT telescope and NuSTAR with de-reddened ${i}^{\prime }$-band flux density obtained from LCO (see Fig.~\ref{fig:correlation}b). The correlation with a power-law index of $\sim$0.54$\pm$0.04 is found to be significant (Pearson correlation coefficient = 0.93, p-value = $1.2 \times 10^{-13}$). The difference between hard state and hard-intermediate state here is subtle, as probably the hard X-ray flux is also fading slightly in this state compared to the hard state, such that the optical and the hard X-ray flux are both fainter, maintaining the correlation.

\subsubsection{$U$-band versus X-ray correlations}

\begin{figure*}[ht]
\center
\includegraphics[width=8.9cm,angle=0]{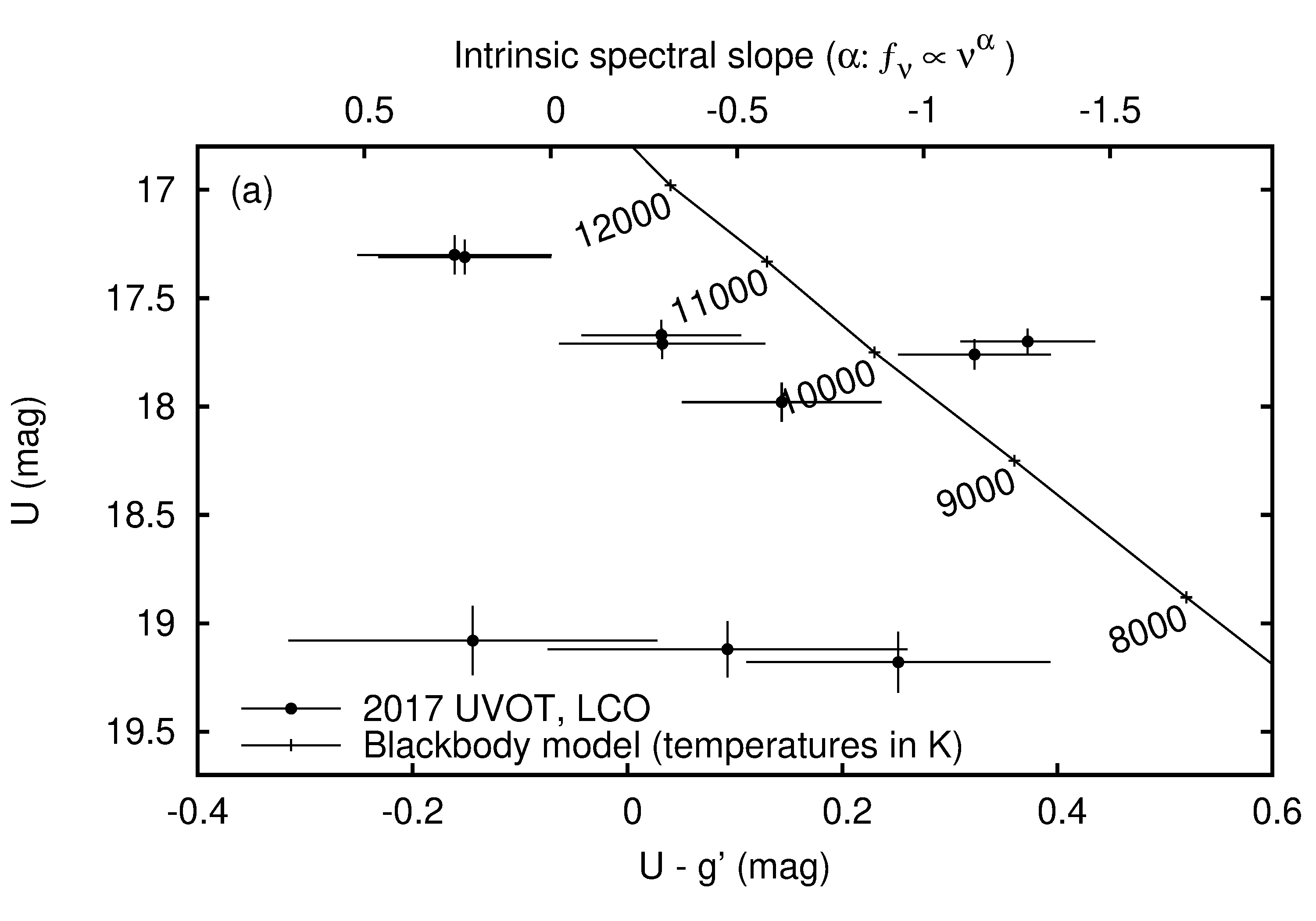}
\includegraphics[width=8.9cm,angle=0]{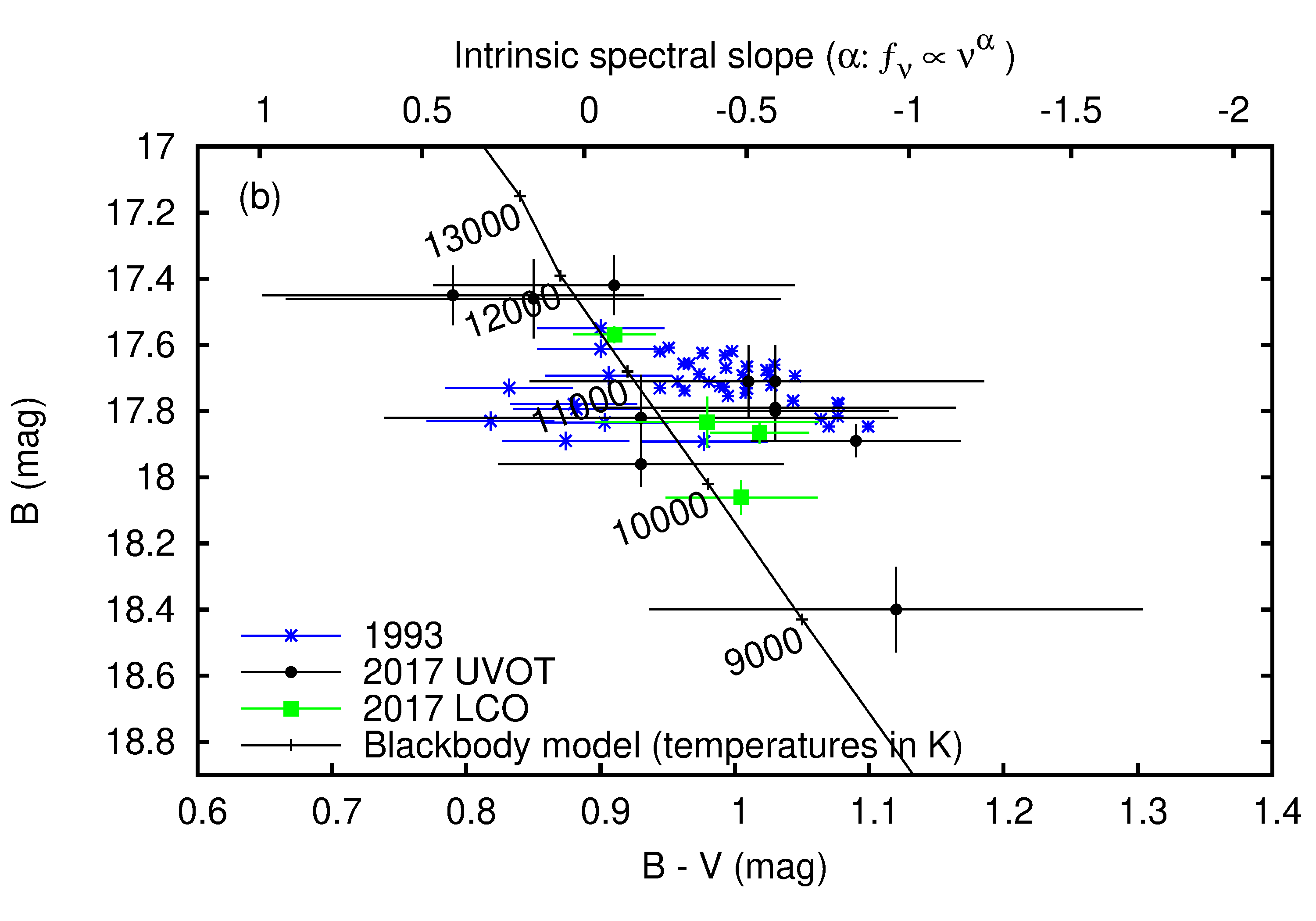} \\
\includegraphics[width=8.9cm,angle=0]{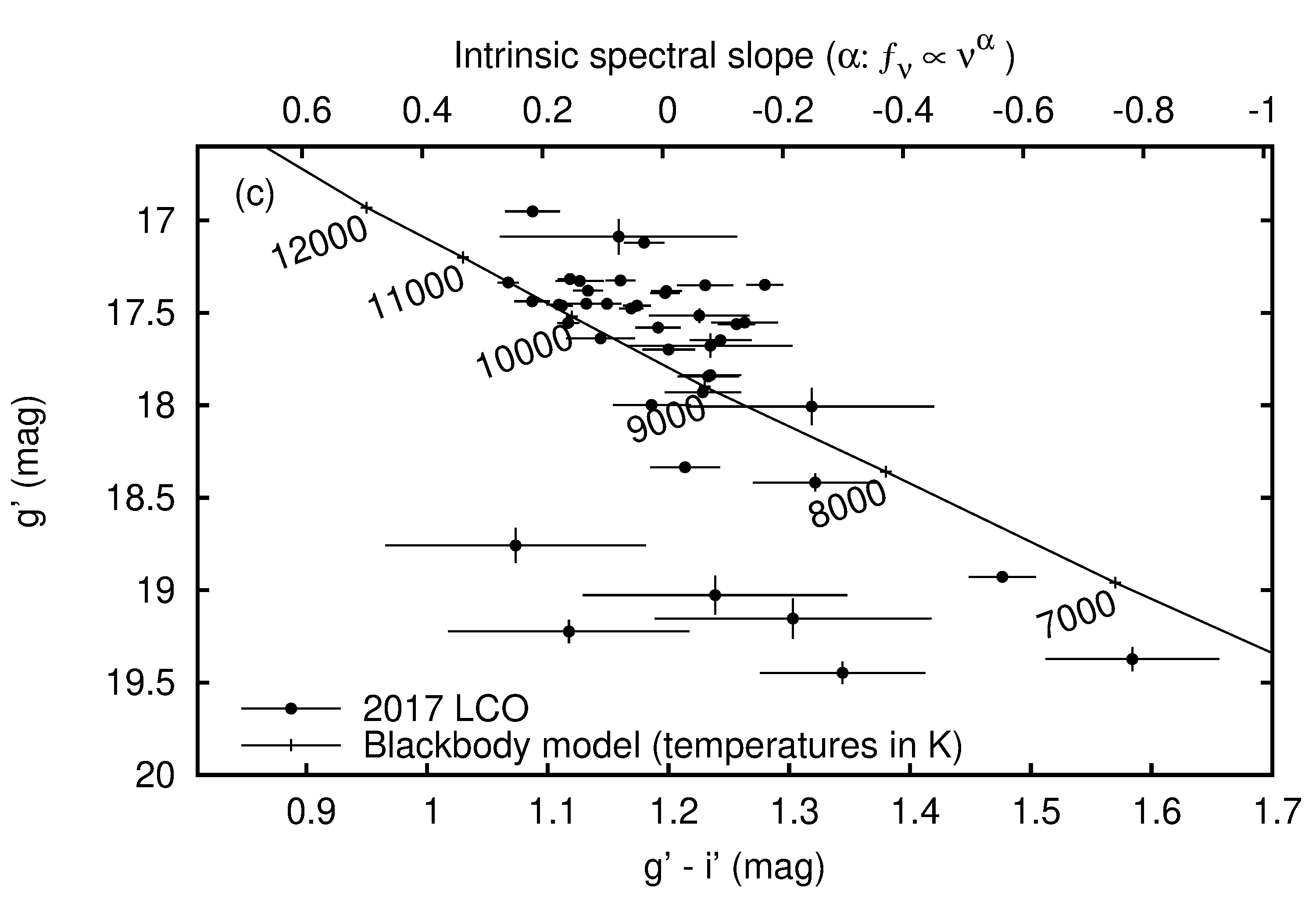}
\includegraphics[width=8.9cm,angle=0]{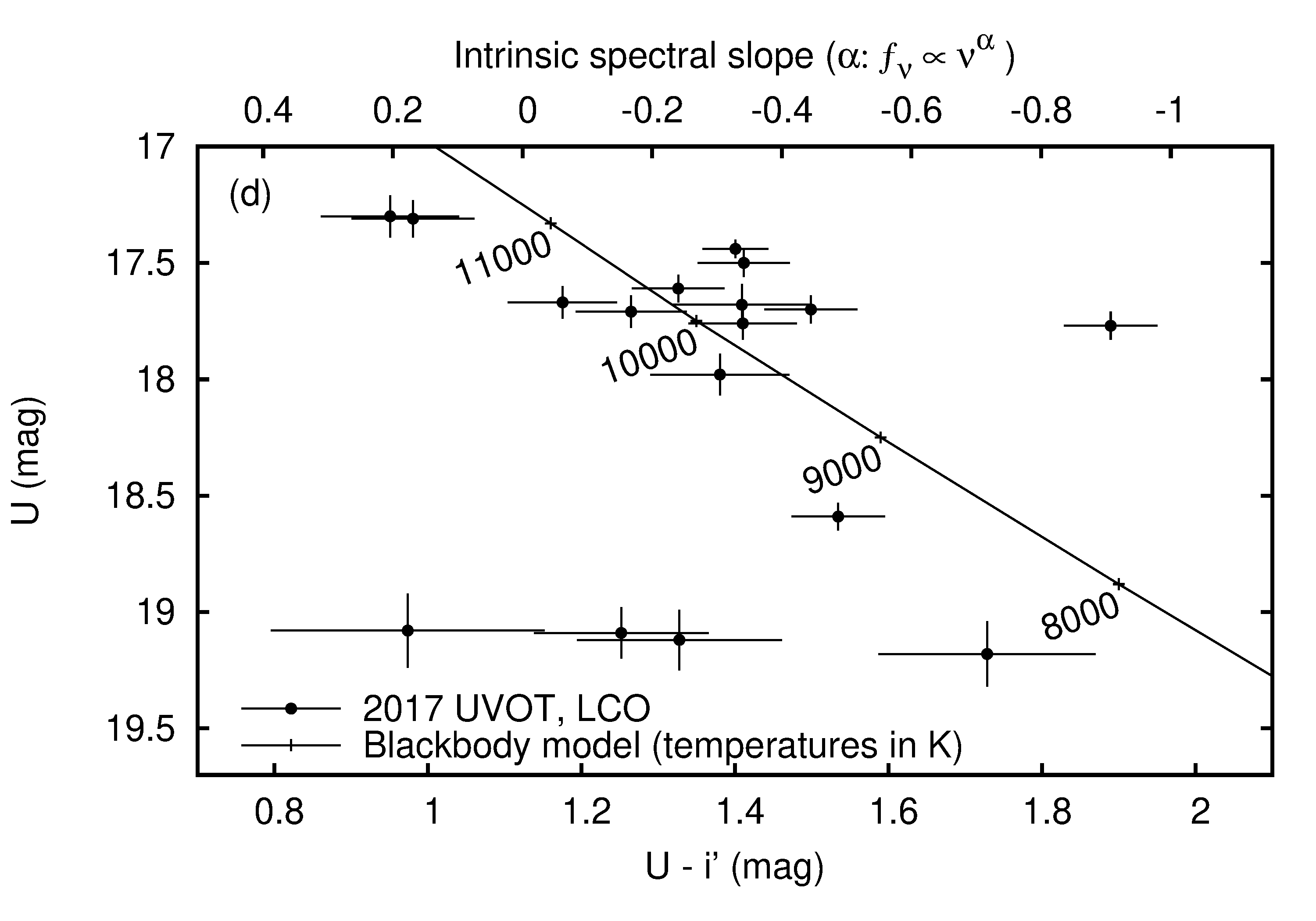}
\caption{Colour-magnitude diagrams, adopting four different filter combinations. For each combination, the bluer filter is on the y-axis. Bluer colours (greater spectral index) are to the left, redder colours (lower spectral index) are to the right. A simple model of a single temperature blackbody heating up and cooling, used to approximate emission from reprocessing on the disk, is denoted by the black line labelled with the temperature values, in each panel (see text). The $U-g^{\prime}$ CMD \emph{(a)} is the combination showing the shortest wavelengths; the reprocessing model is a poor approximation of the data, the viscously heated disk likely plays a strong role at the fainter epochs. The $B-V$ CMD \emph{(b)} also includes some data from the 1993 outburst \citep{della,masetti}; the reprocessing model approximates most of the data well (brightest epochs in both outbursts). The $g^{\prime}-i^{\prime}$ CMD \emph{(c)} is the combination with the longest wavelengths; the reprocessing model is close to the data at bright epochs but not during the outburst fade, when the viscous disk likely dominates. The $U-i^{\prime}$ CMD \emph{(d)} shows the widest wavelength range; again reprocessing can describe most of the brightest epochs, not the fainter epochs.}
\label{fig:cmd}
\end{figure*}

The $U$-band/soft X-ray correlation is plotted in Fig.~\ref{fig:correlation}c with $U$-band detections from the UVOT telescope, and simultaneous soft X-ray data from Swift/XRT in the 2-10 keV energy range. The correlation is significant (Pearson correlation coefficient = 0.94, p-value = $2.6 \times 10^{-12}$) and the power-law index of the UV/X-ray correlation is found to be $\sim$0.49$\pm$0.03, which is consistent with an irradiated accretion disk \citep{van}. Similar to the optical/X-ray correlation, the hard-intermediate state values were seen to have weaker $U$-band emission in comparison to the hard state. A correlation was also observed between the $U$-band and the hard X-ray emission (15-50 kev, from Swift/BAT telescope and NuSTAR, see Fig.~\ref{fig:correlation}d), with high significance (Pearson correlation coefficient = 0.95, p-value = $1.2 \times 10^{-15}$), and a similar slope of $\sim$0.51$\pm$0.03. There is a hint of the correlation flattening at the lower luminosity end, as also seen in the case of the optical/X-ray correlations. But due to the lack of fainter data points, and the large uncertainties associated with it, a conclusive result regarding a broken power law can not be obtained. But we note that a shallower correlation could arise due to the emergence of a viscous disk at the end of the outburst (see also Section 3.5).

\subsection{Colour-magnitude diagram}

The colour-magnitude diagrams (CMDs) are plotted in Fig.~\ref{fig:cmd}, using four different filter combinations using the ${i}^{\prime }$, ${g}^{\prime }$, $U$, $B$ and $V$ filters. We adopt the single temperature blackbody model of \cite{Maitra2008}, described in detail in \cite{Russell2011} to study the colour evolution of X-ray binaries during outbursts, which approximates the emission from the X-ray irradiated outer accretion disk. A colour change is expected due to the evolving temperature of the irradiated disk, which is assumed to have a constant emitting surface area. While at high temperatures the optical emission is expected to originate in the Rayleigh-Jeans tail of the blackbody, at lower temperatures it originates near the peak of the blackbody curve. The blackbody temperature of the model depends on the intrinsic colour and the interstellar extinction.

The normalization of the model depends on the accretion disk radius (estimated using the known orbital period, mass of the companion star and the mass of the black hole from the literature), the distance to the source, inclination angle, the disk filling factor, disk warping and the fraction of disk that is reprocessing the X-rays. As many of these parameters are not certain, we choose a value of the normalization that best describes the trend in the data. In particular, we fix the normalization using the $B$-$V$ CMD (see Fig.~\ref{fig:cmd}b) as this filter combination has the most amount of data, and is less affected by the jet emission (if present), being from the bluer wavelengths. We find that the data do not completely agree with the single temperature blackbody model, which indicates that more than one component is likely to be present. The disk temperature was roughly seen increasing from $\sim$7,000 K to $\sim$12,000 K, as expected during outbursts when hydrogen in the disk is typically ionized.

We find that the data in the $B$-$V$ CMD, which includes some data from the 1993 outburst \citep[data from][]{della,masetti}, generally follows the expected trend between colour and magnitude, with scatter of $\pm 0.1$ mag in colour. These data were from the brightest epochs in both outbursts. We adopt the same normalization that this provides to the other three filter combinations. We note that there is an uncertainty on the normalization due to the scatter in the colour, but it should not be larger than $\sim$0.1 mag in colour. Assuming that the same normalization can be applied to the other filter combinations, we can investigate deviations from the blackbody model as a function of wavelength combination. For these filter combinations, we find that the brightest epochs show data close to the blackbody model, but at lower luminosities there are significant deviations, whereby the observed colour is much bluer, in some epochs, compared to model expectations (see Fig.~\ref{fig:cmd} caption). The spectral index in these faint data points, instead of decreasing to a value of $\alpha \sim -1$ at ${g}^{\prime } > 19$ mag, diverges away from the model, to values of $\alpha = 0\textrm{--}0.3$. A spectral index of $+1/3$ is expected for the overlapping radii of a viscously heated disk \citep[eg.][]{fkr}. It may be that reprocessing on the disk becomes less important at these lower luminosities, revealing the viscously heated disk as the outburst fades.

If the viscous disk, with $\alpha = +1/3$, is responsible for the deviations from the model, one would expect this to affect the shorter wavelengths more than the longer wavelengths, since this component rises at shorter wavelengths. This seems to be the case, with a colour deviation of $\sim 0.7$--1.0 mag in the $U$-${g}^{\prime }$ and $U$-${i}^{\prime }$ CMDs, and $\sim 0.5$ mag in the ${g}^{\prime }$-${i}^{\prime }$ CMD. The companion star could start to contribute to the optical emission at low fluxes, but we consider this to be unlikely to cause the observed deviations because (a) the star would have to be rising towards the blue, requiring it to be a more massive companion than is likely in this LMXB, and (b) the fluxes during the decay are still a couple of magnitudes above the quiescent level (see below), so the star is unlikely to dominate the emission. Optical emission from the viscous disk also has a shallower relation with the X-ray flux, compared to reprocessing, and this is hinted at in Fig.~\ref{fig:correlation} whereby the correlation slopes seem to appear shallower at lower luminosities compared to higher luminosities.

\begin{figure*}[ht]
\center
\includegraphics[width=17.8cm,angle=0]{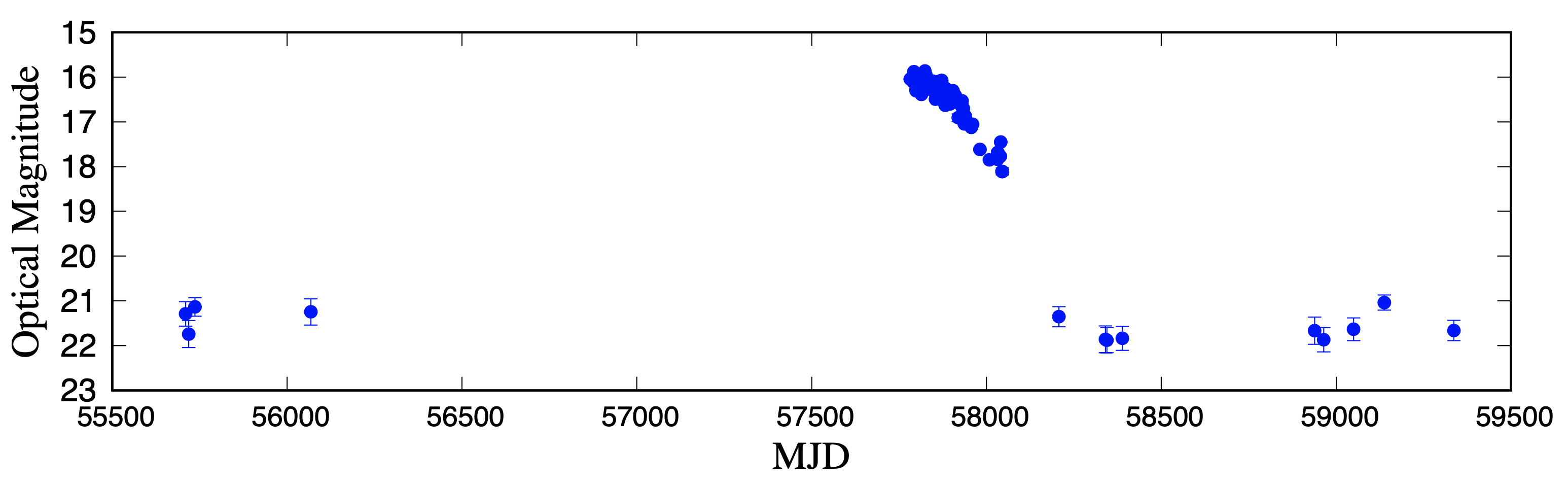}
\caption{The long-term ($\sim$10 years) light curve of GRS~1716$-$249 in the ${i}^{\prime }$-band with LCO from 2011 May (MJD 55709.39) to 2021 May (MJD 59336.68), showing the quiescent variability of the source.}
\label{fig:longterm}
\end{figure*}

In the ${g}^{\prime }$-${i}^{\prime }$ CMD (see Fig.~\ref{fig:cmd}c), some of the brightest epochs show data that deviate from the blackbody model in the opposite sense; some data points are redder than the blackbody model by up to 0.2 mag in colour. This is less prominent in the other CMD filter combinations. Since ${g}^{\prime }$-${i}^{\prime }$ is the combination with the longest wavelengths, this is likely due to an additional component that is redder than the disk component, and which only makes a contribution at high luminosities. It is also variable; some data points are close to the blackbody model and so this redder component seems to vary in time. This is therefore probably the jet making a weak contribution to the ${i}^{\prime }$-band, since we know that the jet makes a stronger contribution at longer wavelengths in the infrared (Section 3.3) and it is variable (Section 3.2).

\subsection{Long-term monitoring and quiescent magnitude}

The only report of any optical quiescent magnitudes of GRS~1716$-$249 in the literature is a single weak constraint of B$\sim$ 21.0-21.5 mag \citep{iauc}. The source was not detected with Gaia in quiescence, and it only appears in EDR3 after the new outburst data were included (see Section 2.1.2). This provides a 20.7 mag limit in the $G$-band \citep{gaia}.

We have been monitoring GRS~1716$-$249 in quiescence with LCO (mostly using the 2-m Faulkes Telescope South) for the last 15 years, since 2006 February 3 (MJD 53769; see section 2.1.1 for details). The monitoring continues past the data we report on here through 2022 February. During quiescence, all the measurements obtained with XB-NEWS are forced-photometry points centred at the position of GRS~1716$-$249. On visual inspection of the quiescent data, we find that in most of the images the target is not visible at its expected position, and the quiescent magnitudes from XB-NEWS could be contaminated by emission from a brighter source very close to ($\sim$ 2 arcseconds away from) the transient within the aperture, and a faint star 1.6 arcseconds away from the X-ray binary (see Fig. 1), making them unreliable.

To obtain a reliable quiescent optical magnitude, we select all the LCO images with good seeing ($<$ 1.6 arcseconds) and perform aperture photomertry at the source position using an aperture size of $\sim$1 arcsecond in order to exclude the flux from the nearby stars and obtain the quiescent magnitude. The finding chart in quiescence obtained from the image with the best seeing ($\sim$0.82 arcseconds) is shown in the right panel of Fig.~\ref{fig:findingchart}. We find uncontaminated detections of the source at 13 epochs during quiescence spanning a range of 10 years (2011 May - 2021 May, see Fig.~\ref{fig:longterm}). There is a slight variation during quiescence, with the ${i}^{\prime }$-band magnitudes ranging from 21.04$\pm$0.17 (MJD 59137.4) to 21.88$\pm$0.28 (MJD 58344.5). By combining all the 7 detections with seeing $<$ 1.1 arcseconds, we find the quiescent ${i}^{\prime }$-band magnitude = 21.39$\pm$0.15 mag. The position of the source in quiescence is consistent with that measured from outburst data. The quiescent magnitude does not appear to change (within errors of the variations) before versus after the 2016--2017 outburst. We also rule out there being any long mini-outbursts after the 2017 outburst.

In the near-IR wavelengths, the source is not detected with 2MASS during quiescence \citep{rout}, inferring an upper limit of 15.8 mag for the $J$ band, 15.1 mag for the $H$ band and 14.3 for the $K$ band \citep{2mass}. From archival $J$-band images of the field taken during quiescence on 1999 July 5 and 7 (MJD 51364 and 51366) with the SOFI instrument at the New Technology Telescope (NTT; La Silla, Chile), we find a 3$\sigma$ upper limit of the source of $J > 18.8$ mag, while the nearby southern star was found to have magnitudes of $J=15.38\pm0.05$ mag (see Section 2.2.2). There is a mention of probable near-IR quiescent magnitudes of GRS~1716$-$249, as \cite{chaty} detected the source in $J, H$ and $K$ bands with the 2.2-m La Silla Telescope (ESO, Chile) on 1997 July 19 (MJD 50648) when it was expected to be in quiescence. They tabulate the quiescent magnitudes of the source as $J=19.4\pm1.2$; $H=19.2\pm1.0$; $K=18.3\pm1.0$. But they note that the source was not detected on 1998 July 6 (MJD 51000), and caution that observations with more powerful telescopes are needed to confirm the quiescent magnitudes.

\section{Discussion}

Compared to the origins of X-ray or radio emission in a BHXB, the origin of the optical and near-IR emission is much less understood. Many physical processes could potentially contribute to the emission at these wavelengths, including X-ray reprocessing by the outer accretion disk \citep{Cunningham1976, Vrtilek1990}, intrinsic thermal emission from a viscously heated outer accretion disk \citep{ss,fkr}, synchrotron emission originating from a steady compact jet during the hard state \citep[e.g.][]{markoff01,jain01,corbel2002,bb04,Russell2006,Kalemci2013,saikia} and sometimes in transitional states \citep[e.g.][]{fender04,vanderHorst,Koljonen2015,Russell2020}, a hot inner flow during the hard state \citep[e.g.][]{veledina}, and the companion star during quiescence \citep[e.g.][]{casares14}. In this section, we explore the various emission processes contributing to the optical/UV fluxes of GRS~1716$-$249 using information from the methods mentioned previously, and discuss their implications on the system parameters, especially the distance to the source.

\subsection{Optical/UV/IR emission mechanism}

We study the optical/UV as well as broadband SEDs of GRS~1716$-$249 with quasi-simultaneous (within 24 hours) data, and find that they show a flat spectrum at optical/UV wavelengths (with a slight peak in the optical), with a positive slope in the near-IR regime, suggesting that the optical/UV emission mainly originates from a multi-temperature accretion disk. The optical/UV emission is near the peak of the blackbody from reprocessing. The fainter near-IR emission compared to optical is consistent with the Rayleigh-Jeans tail of the blackbody from the outer disk. The mid-IR emission on one date is comparatively brighter than what is expected from the disk alone. Such excess of emission in the IR regime is seen in many BHXBs (e.g., XTE~J1550$-$564, \citealt{jain01}; 4U~1543$-$47, \citealt{bb04}; H1743$-$322, \citealt{chaty15}; XTE~J1650$-$500, \citealt{curran}; GX~339$-$4, \citealt{corbel2002}, \citealt{homan2005}), generally associated with a compact jet. Along with being above the disk model, the mid-IR emission is also highly variable (see Table 1), and the radio to mid-IR spectrum is slightly inverted (with an index of $\alpha$=0.13$\pm$0.03), which are typical signs of jet emission from BHXBs. The broadband spectral fitting performed by \cite{rout} also shows that an irradiated accretion disk dominates the ultraviolet and optical emission. They report an IR excess compared to what is predicted by the irradiated disk model, and interpret it as due to the presence of a jet. Similarly, \cite{bassi2020} fitted their broadband SED with the irradiated disk model \textit{diskir} to describe the contribution of the accretion flow emission, which accounts for the irradiation of the outer disk and the reprocessing of the X-ray photons in the optical/UV band.

The slope of the optical/X-ray correlation also reveals the dominant emission mechanism of the accretion disk. For an X-ray reprocessing accretion disk the slope of the correlation is theoretically expected to be $\sim$0.5 \citep{van}. But we note that the theoretical value can be slightly different if there are extra contributions coming from additional emission components like irradiation from a disk wind, and can have a much larger range of slopes depending on which wavelength is used and whether the optical emission is coming from the Rayleigh-Jeans tail (RJ) or closer to the peak of the blackbody disk \citep{be,shahbaz2015,Coriat2009}. On the other hand, for a viscously-heated disk the slope of the correlation is expected to have a wavelength-dependent value $\sim$0.3 \citep{Russell2006}, and for an optically thick jet the expected slope is $\sim$0.5-0.7 \citep{Corbel2003,Russell2006}. For GRS~1716$-$249, the best-fit power-law correlations indicate the optical/X-ray slope to be 0.41$\pm$0.03 (see Section 3.4). This value is consistent with the X-ray irradiated accretion disk \citep{van}, with additional contribution from the viscous disk, which could lower the value of the fitted slope from the theoretical value of $\sim$ 0.5 for irradiation \citep{Russell2006}. An X-ray irradiated accretion disk with optical emission coming from the peak of the blackbody, is also favoured by recent studies that find that the expected slope in the hard state can range from 0.13 (optical flux at RJ tail) to 0.33 (flux in the multicolour disk blackbody) for a viscously heated disc, and from 0.14 (RJ tail) to 0.67 (disk) for X-ray reprocessing with an isothermal disk \citep[for a detailed calculation, see][]{be,Coriat2009}. For cases like GRS~1716$-$249, where the outer disk temperature rises to $\sim$10000 K in outburst (see Fig.~\ref{fig:cmd}), the optical flux is found at the spectral transition between the RJ tail and the multicolour blackbody \citep{Russell2006}, and hence the optical/X-ray slope of 0.41$\pm$0.03 is consistent with the scenario of X-ray reprocessing. Similar values of power-law correlations have also been seen in other X-ray binaries like XTE~J1817--330 \citep[0.47$\pm$ 0.03,][]{rykoff}, GX~339--4 \citep[0.44$\pm$ 0.01,][]{Coriat2009}, GS~1354--64 \citep[$\sim$0.4-0.5,][]{Koljonen2016}. On the other hand, many sources like Swift~J1357.2--0933 \citep{armas2013}, Swift~J1910.2--0546 \citep{saikia_submitted}, SAX~J1808.4--3658 \citep{patruno} and Cen~X-4 \citep{Baglio_submitted} show a significantly shallower correlation ($\sim$0.1-0.3). A slightly steeper correlation ($\sim$0.56) is seen for V404~Cyg \citep{Bernardini2016,hynes2019,oates2019}, probably arising from contamination in optical fluxes from jet contribution. From our multi-wavelength correlation and spectral energy distribution analysis, we can rule out a significant optical emission component arising from a jet, in GRS~1716$-$249.

This is also supported by our variability studies. Generally, sources with strong optical/IR variability on short (seconds-minute) timescales are known to have a strong jet contribution, and the variability is stronger at longer wavelengths where the disk makes a smaller contribution \citep{gandhi2009,gandhi2010,baglio18,tetarenko2021}. Disk variability is driven by changes in the mass accretion rate, which happen on the viscous timescale (days--weeks) for the viscously heated disk, and shorter (minute) timescales for reprocessing on the disk surface, if the X-rays have strong variability (with the reprocessing being smeared). The lack of strong variability in our optical data on short (minute) timescales, along with the presence of correlated variability on longer (days) timescales, suggests that the disk is producing the optical emission, and the contribution of the synchrotron jet emission at optical wavelengths is low in GRS~1716$-$249. The emission at near-IR wavelengths is dominated by the accretion disk, with a weak and variable jet component contributing towards the $K$-band in a few epochs. At mid-IR wavelengths, we find evidence for a highly variable jet component as suggested by the variable emission and the mid-IR to radio spectral index.

\begin{figure*}[ht]
\center
\includegraphics[width=8.95cm,angle=0]{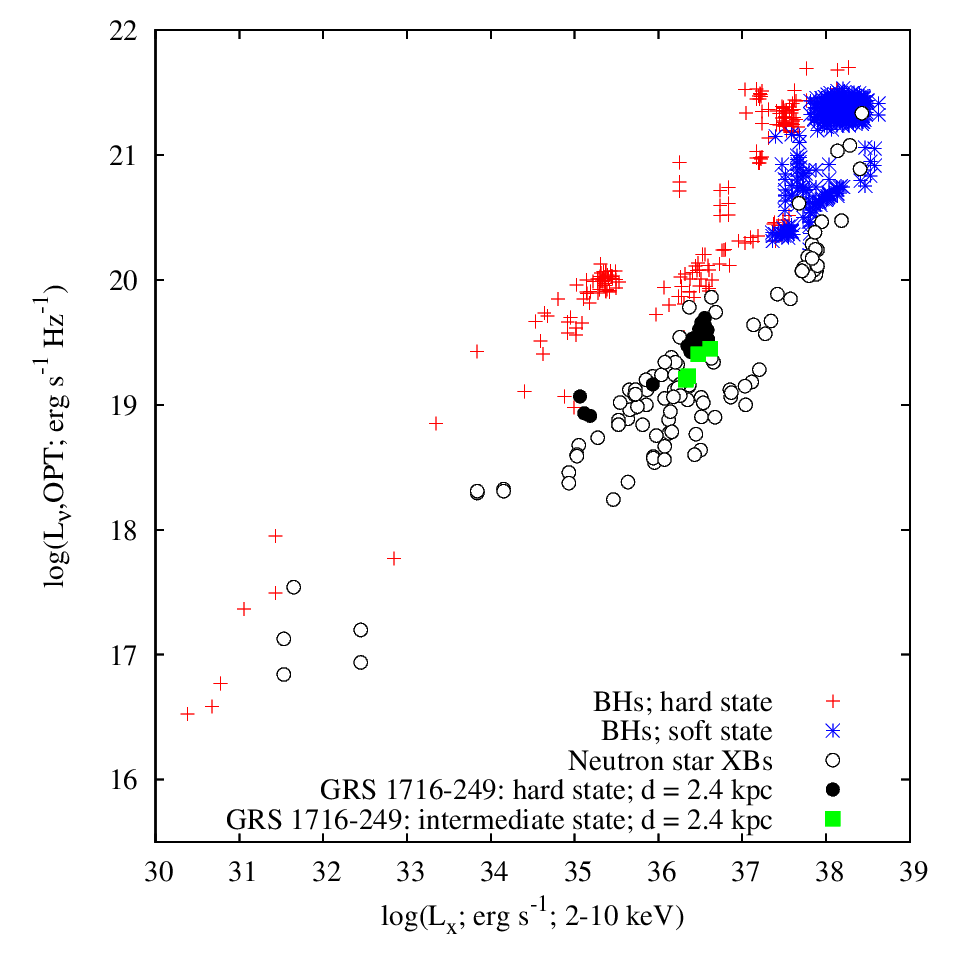}
\includegraphics[width=8.95cm,angle=0]{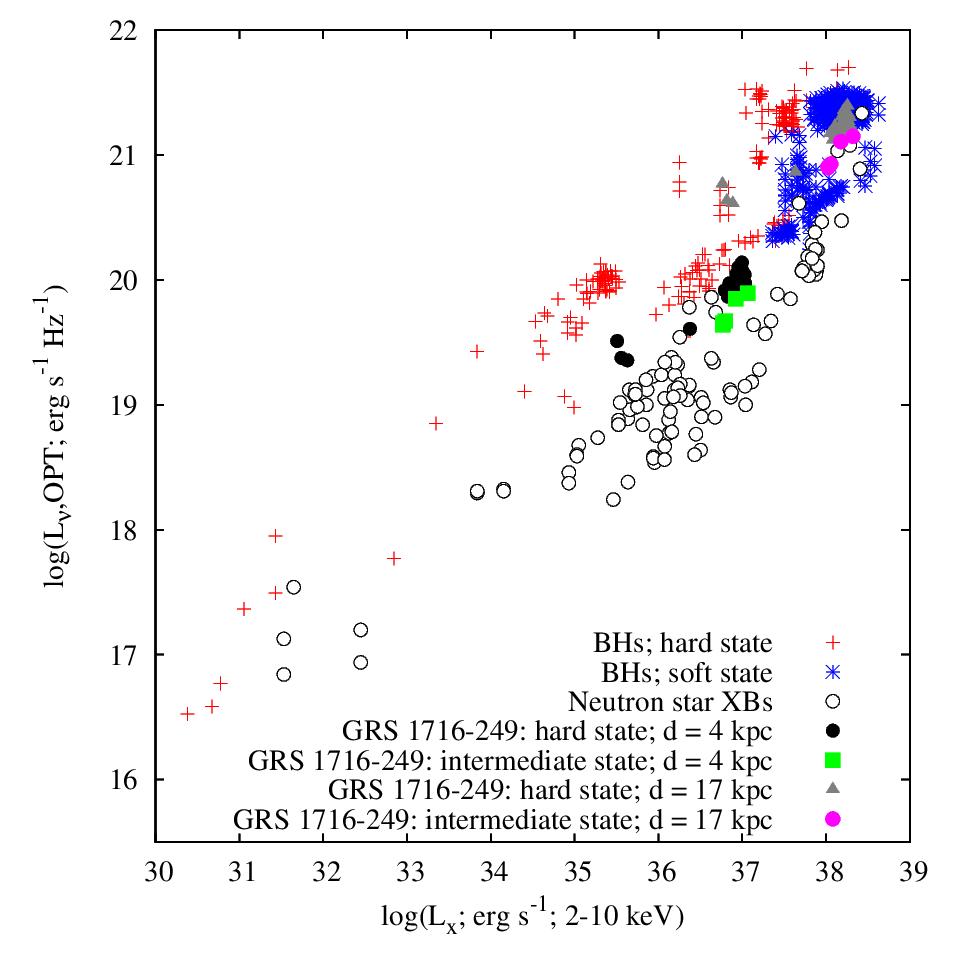}
\caption{Optical/X-ray correlation for GRS~1716$-$249 and samples of black hole and neutron star LMXBs, assuming the distances of 2.4 kpc (left; literature value), and the upper and lower bounds of the range that is empirically consistent with other BHXBs; 4--17 kpc (right).}
\label{fig:globalcorrelation}
\end{figure*}

In addition to this, we investigated the color evolution of the source during its outburst. Our CMD analysis shows that the observed optical data mostly agrees with the single temperature blackbody model (at least at higher luminosities), with a scatter of $\pm 0.1$ mag in colour. This agreement supports the finding that the optical emission is originating mainly from a disk with varying temperature. The disk temperature varied between $\sim$7,000 K to $\sim$12,000 K, which is optimal for ionizing hydrogen in the disk. At the brighter epochs in the CMD filter combination with the longest wavelengths (${g}^{\prime }$-${i}^{\prime }$), we found that the data are slightly redder and brighter than what is expected from the disk model, with possibly some contribution coming from a jet.  At the fainter epochs, we found significant deviations of the data from the reprocessing model, where the observed colour was much bluer, shifting the spectral index from $\alpha = 0$ to +0.3 \cite[which is expected in a viscously heated disk,][]{fkr}. 

It is also worth considering the possibility of optical or IR emission from the hot flow. In this scenario, synchrotron emission from overlapping components of the hot flow contribute to the optical emission \citep{veledina}. We have found that the optical spectrum is well described by a multi-temperaure disk, with the RJ tail in the IR (Section 3.3). The irradiation peak is detected, and there is low short term variability (Section 3.2). These characteristics, along with the behaviour in the optical/X-ray correlation and the CMDs, strongly favor a disk origin. In addition, \cite{cuneo} detected variable, double peaked emission lines from hydrogen and helium, in the optical spectrum. These lines originate in the rotating accretion disk, and P-Cygni profiles were also detected from a disc wind. The hot flow model predicts strong short timescale optical variability (stronger in the optical compared to the IR because variations are amplified closer to the black hole, and the IR synchrotron emission in the hot flow orginates at larger radii) and a flat optical spectrum with a downturn at longer wavelengths \citep{veledina}. We find stronger, high amplitude variability in the IR compared to optical, with the mid-IR flux density being higher than the optical on one date. The spectrum, emission lines, and variability properties are therefore not consistent with expectations from the hot flow. The hot flow model predicts a lower flux in the IR because the synchrotron-emitting region is physically limited by the inner edge of the accretion disc. So, while the optical is dominated by the disc, the mid-IR must be dominated by the jet, not the hot flow.

\subsection{Constraints on the system parameters}

We also conduct a comparative study of the quasi-simultaneous optical/X-ray emission of GRS~1716$-$249 against a large sample of black hole and neutron star LMXBs in Fig.~\ref{fig:globalcorrelation}, with data taken from \cite{Russell2006,Russell2007}. Both these classes of LMXBs are known to show different correlations, with the neutron star LMXBs being around 20 times optically fainter than black hole LMXBs \citep{Russell2006}, for reasons discussed in \cite{Bernardini2016}. Assuming the previously estimated distance of 2.4$\pm$0.4 kpc \citep{della} is correct, GRS~1716$-$249 is found to be much more optically faint (or X-ray bright) compared to other BHXB samples (see the left panel in Fig.~\ref{fig:globalcorrelation}); in fact, at this distance GRS~1716$-$249 agrees more with the neutron star track in the global optical/X-ray correlation plot. 

\subsubsection{BH nature of GRS~1716$-$249}

A BH nature of the source was first inferred by \cite{masetti}, who derived a lower limit for the compact object mass of $>$4.9 M$_{\odot}$ from the superhump period of 14.7 hrs. Superhumps generally appear in disks with viscous-dominated emission, where the luminosity variations are caused by viscous dissipation associated with tidal deformation of the disk when it reaches the 3:1 resonance radius. We have shown in Section 4.1 with several lines of reasoning that the disk emission in GRS~1716$-$249 during outburst is dominated by X-ray irradiation. Such systems can have orbital modulations due to irradiation, rather than (or in addition to) superhumps \citep[see the discussion in][]{haswell}, especially when part of the optical variability comes from the irradiated face of the donor star. The superhump variability can be dominant at high orbital inclinations when the donor star to BH mass ratio is low and the donor star is shielded from irradiation, but optical modulation can be expected when the ratio is higher \citep[see the discussion in][]{torres}. As it is not clearly known whether the optical variability reported by \cite{masetti} was a superhump or an irradiation effect, we cannot use it as a reliable constraint to the BH mass.

\cite{masetti} also noted that a massive primary is expected from the very long decay time of the X-ray light curve. Later, \cite{tao2019} studied three quasi-simultaneous NuSTAR and Swift datasets of the system in its hard intermediate state, and assuming a distance of 2.4 kpc, constrained the upper limit for the compact object mass to be $<$8.0 M$_{\odot}$, at a 90\% confidence level. \cite{chatterjee} also used X-ray spectral analysis of the source during outburst to suggest a BH nature of the compact object. They fitted the X-ray spectra of GRS~1716$-$249 with the physical two-component advective flow (TCAF) model, keeping the mass of the primary as a free parameter, and constrained the mass of the compact object in the range of 4.5--5.9 M$_{\odot}$, but the values obtained are highly model-dependent and very unlikely to be a realistic range. The lack of Type I bursts during the outburst despite the presence of hydrogen \citep[as suggested by the H$\alpha$ lines,][]{cuneo} also provide strong evidence against a NS accretor. Moreover, \cite{tao2019} show that good quality NuSTAR X-ray spectra of the source in the intermediate states can be fitted by BH models. In addition to the previous evidence, the X-ray timing properties of GRS~1716$-$249 also suggest that the compact object of the system is a BH. \citet{chatterjee} report different power density spectra (PDS) of GRS~1716$-$249 in their Fig.~4, all of which show a strong decline from $\sim$3 Hz to 10 Hz. This behaviour is more typical of BH systems since the PDS of BHXBs show a strong decline at frequencies above 10--50 Hz \citep[][]{Sunyaev2000}. NS systems, on the contrary, can show variability up to 500--1000 Hz. The lack of X-ray pulsations and kilo-herz QPOs in the PDS (typical signatures of neutron star systems), the presence of type-C and type-B QPOs in the PDS of GRS~1716$-$249 \citep[][]{chatterjee}, and the strong decline of the power spectra below 10 Hz all reinforce the identification of the compact object as a BH.

If GRS~1716$-$249 is indeed a black hole, then the discrepancy shown by GRS~1716$-$249 with respect to other BHXBs in the global optical/X-ray correlation space could have two possible explanations. Either the source is intrinsically much more optically faint than what has been observed in other BHXBs at a given X-ray luminosity, or it is located much further away than was previously thought.

\subsubsection{Distance to GRS~1716$-$249}

The original distance estimate of 2.4$\pm$0.4 kpc is based on a comparison of the source to other X-ray binaries with data from a few decades before \citep{della}. They argued that the lower limit on the distance is expected to be $\sim$2 kpc from the equivalent width of the NaD absorption lines. To constrain the upper limit, the peak optical brightness was compared to other BHXB outbursts known at the time \citep{della}. Since then, a distance of 2.4$\pm$0.4 kpc has been used by various studies concerning GRS~1716$-$249. Later \cite{hynes2005} notes that one should be cautious about using such method to constrain the upper limit on the distance. It has now become clear that BHXB outbursts can peak at different luminosities, from close to the Eddington limit, down to $\sim 10^{36}$ erg s$^{-1}$ or less \cite[e.g. the Very Faint X-ray Binaries, or mini-outbursts;][]{heinke,zhang2019}. Moreover, the historic peak optical brightness of LMXBs during outburst used by \cite{della} was based on a compilation that neither corrected for orbital period, nor performed sorting of neutron stars versus black holes \citep{1981}. In light of all these arguments we do not consider the formerly estimated upper limit of 2.8 kpc as a reliable constraint.

\begin{figure}[t]
\center
\includegraphics[width=8.5cm,angle=0]{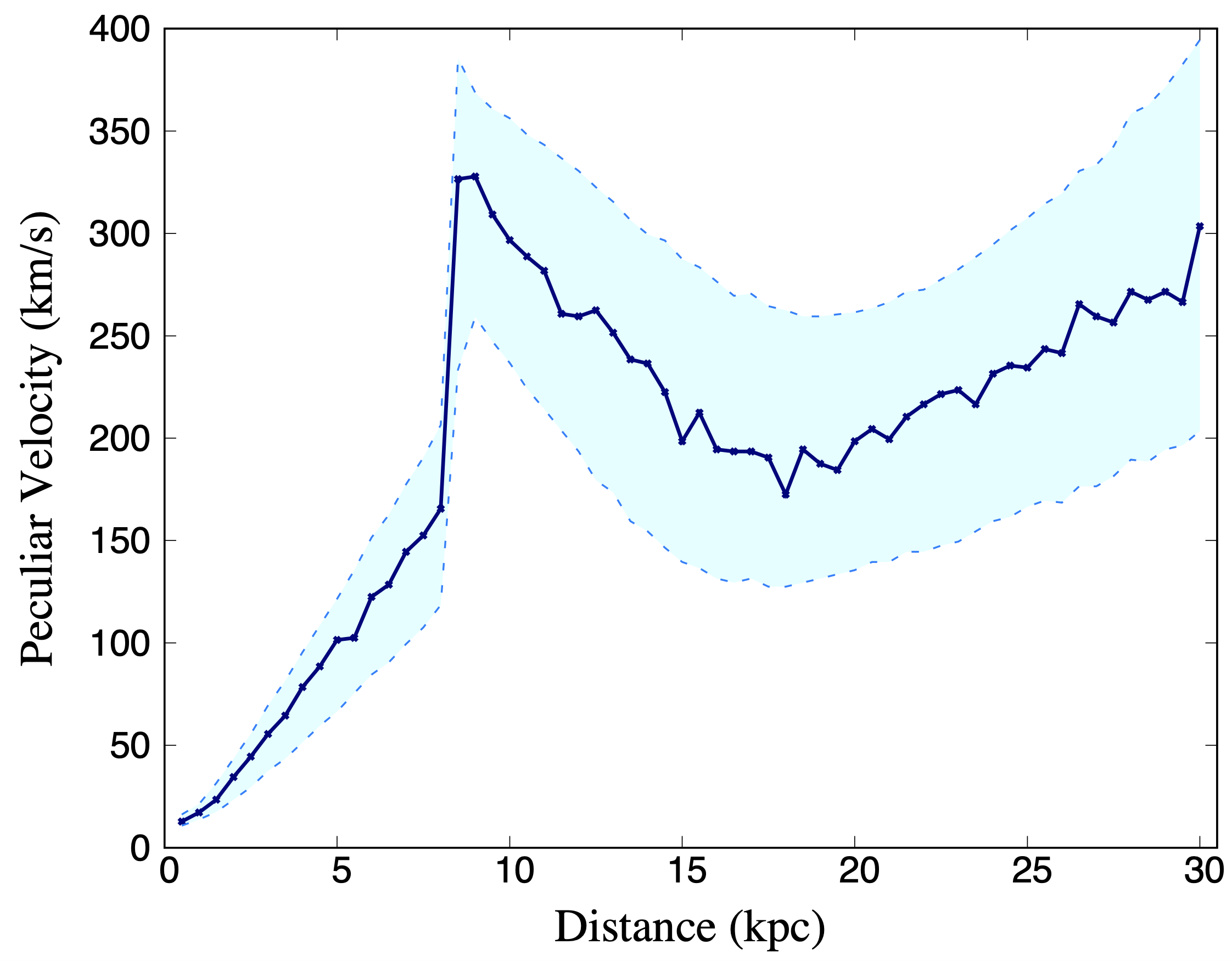}
\caption{Expected peculiar velocity ($\upsilon_{pec}$) of GRS~1716$-$249 for a range of possible distances over 0.5–30 kpc. The shaded region bound by the two dashed lines represent the $1 \sigma$ scatter. The kinematics of the system favour distances of $\lesssim$8 kpc, as for BH systems the natal kick probability distribution ranges only up to $\sim$150 km/s \citep{mm2020}.}
\label{fig:vpec}
\end{figure}

\cite{masetti} had discovered optical modulations in the source with a prominent period of $\sim$ 14.7 hrs, and found that the secondary star in the system should be substantially brighter than claimed by \cite{della}. To explain this discrepancy, they suggested that either the distance of 2.4$\pm$0.4 kpc has been underestimated, or the secondary is a slightly evolved late-type star.

In addition to the previous arguments, we also find that a higher value of distance is expected from the state transition luminosity distribution of the source \citep[e.g.][]{macca,Kalemci2013,macca2}. It has been observed that BHXBs transit from the soft state to the hard state at luminosities between 0.3-3\% percent of the Eddington luminosity \citep{Kalemci2013}, with a mean value of 1.9$\pm$0.2\% \citep{macca}. The state transition luminosity has been used to estimate the distances to many BHXB sources \citep[e.g.][]{homan2006,miller2012}. Although GRS~1716$-$249 did not go to a soft state, we use the luminosity during transition from the final hard/intermediate state to the hard state (MJD 57978, Swift/XRT flux $9.25\times10^{-10}\, \rm erg\,s^{-1}\,cm^{-2}$ at 2-10 keV) to estimate the distance. Assuming a BH of mass 7M$_{\odot}$, 1.9$\pm$0.2\% Eddington luminosity and a bolometric correction factor of 2 relative to the Swift/XRT band, we obtain a probable distance of 8.7$\pm$0.5 kpc for the source. For a more conservative range of 0.3-3\% percent Eddington luminosity \citep{Kalemci2013}, the distance range is found to be 3.46--10.94 kpc.

Moreover, for a distance of 2.4 kpc, the inner disk radius depending on the inclination angle is r$_{\rm in} \sim$ 15 km \citep[see Fig.~6 of][]{bassi}, which is very unusual for a BH disk spectrum; while a more plausible value of r$_{\rm in}>$ 50 km is obtained for distances $d>$ 8 kpc. An underestimated distance could also explain the discrepancy we see for this source with respect to other BHXBs in the optical/X-ray correlation plots (see Fig.~\ref{fig:globalcorrelation}). From our global correlation comparison, we find that for a distance of 4 kpc and less the data are more consistent with being a neutron star, and for distances more than 4 kpc the data are more consistent with a BH.

We also place a conservative upper limit on the distance as 17 kpc from the global optical/X-ray corelation plot (see Fig.~\ref{fig:globalcorrelation}), as for a greater distance, the source would be the most X-ray luminous BHXB, probably exceeding the Eddington limit (depending on the black hole mass). The proper motion estimate of the source is $\sim 4.65\pm1.12$ mas/year, and the potential kick velocity (after removing Galactic rotation) is $\sim$70-100 km/s for a distance $d=$2.4 kpc \citep{atri}. We performed a simulation using all the standard assumptions of \cite{gandhi2019} and the measured proper motions assuming a radial velocity of -10 km/s \citep[for further details, see][]{atri}, and found that at any distance higher than 6 kpc, the space velocity of the source starts to exceed 100 km/s. At the Galactic centre distance, the source peculiar velocity increases to $\sim$150 km/s (see Fig.~\ref{fig:vpec}). On the other side of the Galaxy, however, median peculiar velocities are predicted to be between $\sim$190-330 km/s. Such high velocities are not expected in BH systems, where the natal kick probability distribution ranges up to $\sim$150 km/s with a root-mean-square kick of $\sim$60 km/s \citep{mm2020}. So the kinematics of the system favour distances of $\lesssim$8 kpc, suggesting that it is significantly closer than 17 kpc. We also note that the source had a failed-transition outburst and did not show a transition to the soft state \citep{bassi}. Generally, the failed-transition outbursts reach lower peak X-ray luminosities than full outbursts \citep{tt,alabarta}. For example, in the case of one of the best studied BHXB GX~339$-$4, the luminosity at which the hard-to-soft state transition occurs during a full outburst is $\sim 0.11 L_{\rm Edd}$, and the luminosity during failed-transition outbursts are always equal to or lower than this value \citep{tt}. Assuming a similar behaviour of $\sim$10\% $L_{\rm Edd}$ during the peak flux in the case of GRS~1716$-$249, a conservative mass of 7M$_{\odot}$ results in a distance of 5.0 kpc, and adds additional support to a closer distance. From all the arguments stated above, we place a conservative upper limit of 17 kpc for the system, although our lines of evidence suggest a much lower value ($\sim$8 kpc).

\begin{figure}[t]
\center
\includegraphics[width=8.4cm,angle=0]{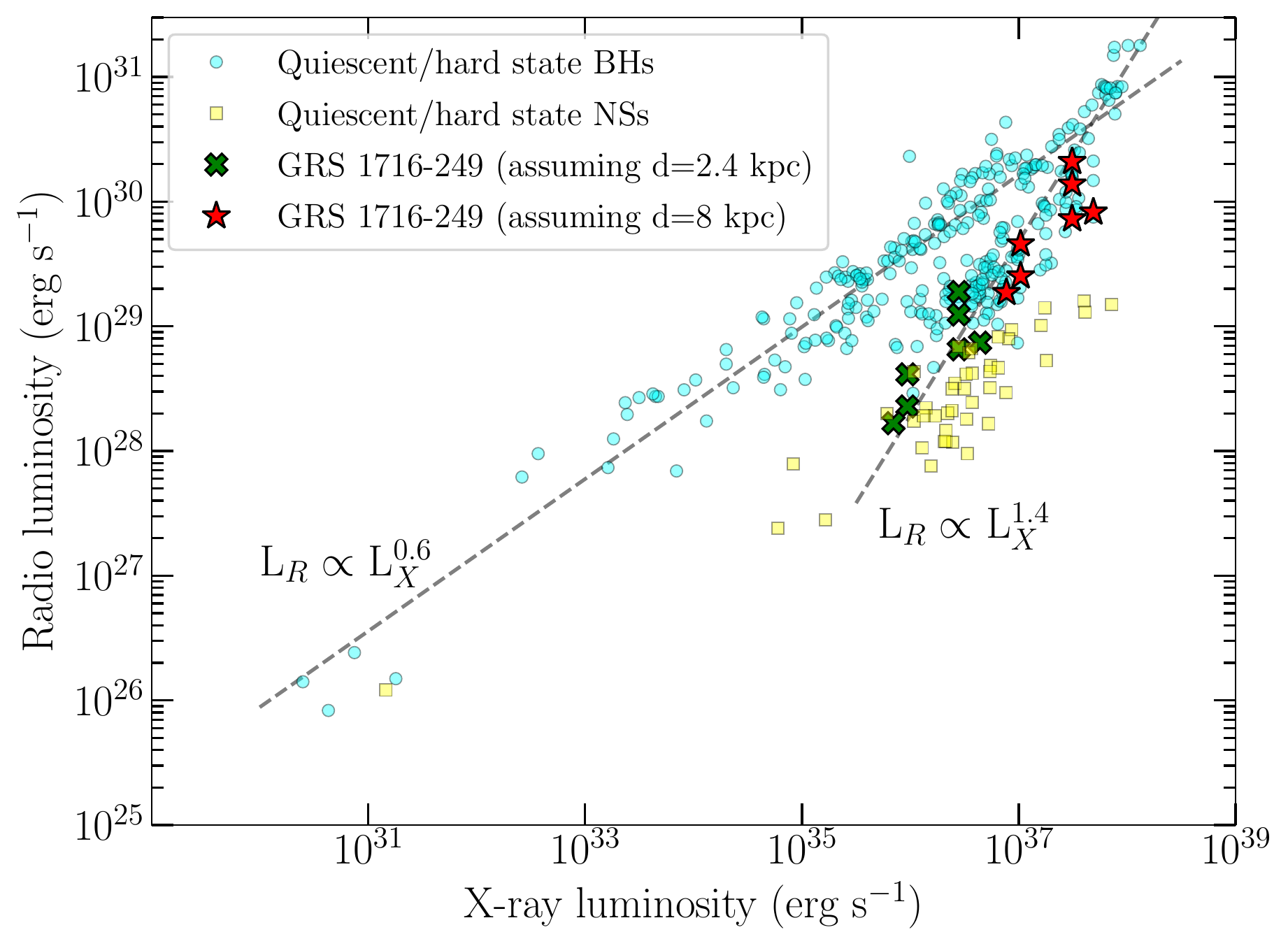}
\caption{Radio/X-ray luminosity correlation plot of GRS~1716$-$249 with quasi-simultaneous data from \cite{bassi}, and a large sample of black hole and neutron star LMXBs with data made available by \cite{bahr}. The X-ray luminosities are in the 1-10 keV energy range, while the radio luminosities are taken in $\sim$5 and 9 GHz. The green crosses depict the position of the source assuming the distance of 2.4 kpc (literature value), and the red stars represent the position when a greater distance (in this case, 8 kpc) is assumed.}
\label{fig:lxlr}
\end{figure}

From the global correlation plot and the list of reasoning mentioned, we constrain the distance of GRS~1716$-$249 to likely be in the range of 4--17 kpc (see the right panel in Fig.~\ref{fig:globalcorrelation}), with a most likely range of $\sim 4$--8 kpc. This improved distance estimate will have implications for models of GRS~1716$-$249 that depend on its distance \citep[e.g.][]{bassi2020,zhang2021,chatterjee}; and affect the parameters of GRS~1716$-$249, like the inferred masses, spins and inclination angles, which depends critically on the assumption of $d=$2.4 kpc \citep{tao2019}. For example, the same spectral modelling assuming a distance of $d=$8 kpc results in M$_{\rm BH}=24.8^{+1.7}_{-10.2}$ M$_{\odot}$ (compared to M$_{\rm BH}=7.6^{+0.8}_{-2.7}$ M$_{\odot}$ obtained assuming $d=$2.4 kpc), which shows that the parameters of the system are clearly model dependent and sensitive to the distance (Lian Tao, private communication). So distances beyond 8 kpc would make it the most massive stellar-mass BH known in our Galaxy \citep[Cyg X-1 currently holds the record with 21.2$\pm$2.2 M$_{\odot}$,][]{miller2021}, which is unlikely. This implies that a distance above 8 kpc is highly implausible and again argues for an upper limit that is substantially lower than 17 kpc.

A greater distance than the literature value also changes the position of the source in the radio/X-ray luminosity correlation plot \citep{bassi}. Generally, BHXBs follow two different tracks in the radio/X-ray luminosity correlation plot (where $L_{R} \propto L_{X}^{\beta}$) - the ‘standard’ track with a power law index $\beta \sim$0.5-0.7 \citep[e.g.][]{Corbel2003,Corbel2013,gallo2018}, and the much steeper ‘outlier’ track with $\beta \geq$1 \citep[e.g.][]{Corbel2004,Coriat2011,gallo2012}, although the existence of two separate tracks has been questioned statistically \citep[e.g.][]{gallo2014,gallo2018}. Newer studies have also shown that the two tracks are not well-defined and there is evidence for standard track sources to get steeper at high X-ray luminosity \citep{karridave}, and outlier track sources to get shallower and ultimately rejoin the standard track at low X-ray luminosities \citep[see e.g.][]{Coriat2011,c2021}. Although the underlying physics behind the two tracks is not completely clear, it has been suggested that the dichotomy could originate either from the structure of the inner accretion flow or from different physical properties of the jets resulting in different levels of radio emission \citep{Coriat2011}. NSXBs, especially in the hard state, also showed a similar correlation with fainter radio emission compared to BHs and a steeper slope \citep[with $\beta \sim$1.4, e.g.][]{ref1,ref2}, although there is strong evidence of different classes of NSXBs showing different behaviour in the radio/X-ray correlation plane \citep[see e.g.][]{tudor,ns}. We find that a greater distance shifts GRS~1716$-$249 from the lower luminosity part of the outlier track (luminosities where mostly NSs are observed and the BHs seem to shift to the standard track) to the higher luminosities where the majority of the BHXB sample following the outlier track lies (see Fig.~\ref{fig:lxlr}). A greater distance also implies that the source was potentially formed in the bulge, and hence its proper motion is not necessarily representative of its natal kick, since the bulge itself has a large velocity dispersion and scale height \citep{atri}.

\section{Conclusion}

The 2016--2017 outburst of the BHXB GRS~1716$-$249 (or GRO J1719$-$24) is well-studied at X-ray and radio wavelengths. In this work, we investigate the optical, near-IR, mid-IR and UV wavelength monitoring data of GRS~1716$-$249 in outburst using LCO, REM, VLT (VISIR) and \textit{Swift's} UVOT, and compare them with the multi-wavelength archival data from Gaia, Mount Abu 1.2 meter telescope, \textit{Swift} XRT, NuSTAR, MAXI, ATCA, VLA and LBA. We also report the long-term ($\sim$ 10 years) optical light curve of the source using LCO and find that the quiescent ${i}^{\prime }$-band magnitude is 21.39$\pm$0.15 mag.

We find that the optical and UV emission of the source in outburst is mainly originating from a multi-temperature accretion disk, with X-ray reprocessing dominating at high luminosities, and with some contribution at the fainter end from the viscously heated disk. Although the near-IR emission is dominated by the emission from the accretion disk, it has a weak contribution from the variable jet in a few epochs in the $K$-band. The mid-IR and radio emission of the source are dominated by the synchrotron emission from a compact jet. In the hard state, the optical/UV emission of the source is correlated with both the soft and hard X-ray emission. The power-law coefficient of the correlation is consistent with the optical emission coming from an X-ray irradiated accretion disk with possibly some additional contribution from the viscous disk, as a hint of a shallower coefficient at low luminosities. This is also supported by the spectral energy distributions, variability studies and color-magnitude diagrams of the source during the outburst.

Finally, we discuss how the previous estimates of system parameters of the source (especially its mass and distance) are based on various assumptions, and cannot be completely trusted. From the global optical/X-ray correlation study in comparison with other black hole and neutron star X-ray binaries, and several other lines of reasoning, we show that GRS~1716$-$249 is much further away than what has previously been assumed, with a probable distance within the range 4--17 kpc, and a most likely range of $\sim 4$--8 kpc.

\acknowledgments

The authors thank the anonymous referee for useful comments and suggestions. The authors also thank Lian Tao for the revised black hole mass estimation using the updated distance. DMR and DMB acknowledge the support of the NYU Abu Dhabi Research Enhancement Fund under grant RE124. JJ acknowledges the support of the Leverhulme Trust, the Isaac Newton Trust and St. Edmund's college, University of Cambridge. RS acknowledges grant number 12073029 from the Natural National Science Foundation of China (NSFC). SM is thankful for support by Dutch Research Council (NWO) VICI award, grant Nr. 639.043.513. TMB acknowledges financial contribution from the agreement ASI- INAF n.2017-14-H.0 and from PRIN-INAF 2019 N.15. TDR acknowledges financial contribution from the agreement ASI-INAF n.2017-14-H.0. TS acknowledges financial support from the Spanish Ministry of Science, Innovation and Universities (MICIU) under grant PID2020-114822GB-I00. GRS is supported by NSERC Discovery Grants RGPIN-2016-06569 and RGPIN-2021-0400. KIIK acknowledges funding from the European Research Council (ERC) under the European Union’s Horizon 2020 research and innovation programme (grant agreement No. 101002352) and from the Academy of Finland projects 320045 and 320085. This work uses data from the Faulkes Telescope Project, which is an education partner of Las Cumbres Observatory (LCO). The Faulkes Telescopes are maintained and operated by LCO. This work also uses observations made with the REM Telescope, INAF Chile. It is also based on observations collected at the European Southern Observatory under ESO programmes 098.D-0893 and 099.D-0884 (PI : D. Russell). Support for this work was provided by NASA through the NASA Hubble Fellowship grant HST-HF2-51494.001 awarded by the Space Telescope Science Institute, which is operated by the Association of Universities for Research in Astronomy, Inc., for NASA, under contract NAS5-26555. This work also makes use of data supplied by the UK \emph{Swift} Science Data Centre at the University of Leicester, and the MAXI data provided by RIKEN, JAXA and the MAXI team.

\clearpage
\restartappendixnumbering
\appendix

\section{Appendix information}

\startlongtable
\begin{deluxetable*}{|lll|lll|lll|}
\tablecaption{Faulkes/LCO optical detections (AB magnitudes) of GRS~1716$-$249 during the 2016-2017 outburst.}
\tablehead{ \multicolumn{3}{c}{${i}^{\prime }$-band}  \vline                         & \multicolumn{3}{c}{${g}^{\prime }$-band}     \vline  & \multicolumn{3}{c}{${r}^{\prime }$-band} \\ \hline
\colhead{MJD}                                 & \colhead{magnitude} & \colhead{error}  \vline & \colhead{MJD}       & \colhead{magnitude} & \colhead{error}  \vline & \colhead{MJD}         & \colhead{magnitude} & \colhead{error}}
\startdata
57781.76280                          & 16.046   & 0.008 &             &           &        &             &           &        \\ 
57790.37413                         & 16.107   & 0.009 &             &           &        & 57790.37282 & 16.563   & 0.010 \\ 
57792.11238                         & 15.881   & 0.011 &             &           &        & 57792.11104 & 16.365   & 0.010 \\ 
57794.34359                         & 16.089   & 0.017 &             &           &        & 57794.34227 & 16.553   & 0.016 \\ 
57794.65647                         & 16.086   & 0.007 &             &           &        &             &           &        \\ 
57795.72785                         & 16.039   & 0.015 &             &           &        & 57795.72654 & 16.419   & 0.040 \\ 
57798.35717                         & 16.306   & 0.012 &             &           &        & 57798.35586 & 16.725   & 0.014 \\ 
57798.64443                         & 16.278   & 0.005  &             &           &        &             &           &        \\ 
57807.36263                         & 16.269   & 0.005 & 57807.36557 & 17.336   & 0.007 & 57807.36849 & 16.704   & 0.005 \\ 
57807.69145                         & 16.316   & 0.006 & 57807.69308 & 17.451   & 0.017 & 57807.69460  & 16.704    & 0.007  \\ 
57808.72058                         & 16.329   & 0.005  & 57808.73102 & 17.461   & 0.010 & 57808.73395 & 16.815   & 0.008  \\ 
57808.72806                         & 16.287   & 0.006 &             &           &        &             &           &        \\ 
57811.03827                         & 16.165   & 0.006 & 57811.04126 & 17.325   & 0.011 & 57811.04423 & 16.660   & 0.006  \\ 
57814.03003                         & 16.389   & 0.011 & 57814.03302 & 17.580   & 0.015  & 57814.03905 & 16.895   & 0.008 \\ 
57816.02459                         & 16.347    & 0.006 & 57816.02759 & 17.456   & 0.009 & 57816.03362 & 16.766   & 0.006 \\ 
57818.01929                         & 16.247   & 0.007 & 57818.02229 & 17.380   & 0.010 & 57818.02832 & 16.751   & 0.007  \\ 
57820.01393                         & 16.196   & 0.007 & 57820.01691 & 17.393   & 0.010 & 57820.02297 & 16.779   & 0.006 \\ 
57822.65082                         & 15.929   & 0.019  &             &           &        & 57822.65975 & 16.442   & 0.011 \\ 
57824.00293                         & 15.863   & 0.008 & 57824.00592 & 16.951   & 0.021 & 57824.01197 & 16.412   & 0.009 \\ 
57826.25255                         & 15.940     & 0.060 & 57826.25548 & 17.120   & 0.016 & 57826.26145 & 16.370   & 0.007 \\ 
57828.00147                         & 16.201    & 0.009 & 57828.00446 & 17.328   & 0.018 & 57828.01048 & 16.713   & 0.010 \\ 
57838.64016                         & 16.268   & 0.009 & 57838.64180  & 17.552   & 0.026 & 57838.64500   & 16.772   & 0.014 \\ 
57845.62141                         & 16.088   & 0.003 &             &           &        &             &           &        \\ 
57845.97092                         & 16.303   & 0.007  & 57845.97392 & 17.452   & 0.010 &             &           &        \\ 
57846.00164                         & 16.199   & 0.006  & 57846.00463 & 17.317   & 0.008 & 57846.01070  & 16.696   & 0.006 \\ 
57849.21821                         & 16.184   & 0.006 & 57849.22115 & 17.382   & 0.012 & 57849.22709 & 16.728   & 0.007 \\ 
57850.60261                         & 16.120   & 0.008 & 57850.60557 & 17.351   & 0.022 & 57850.61154 & 16.603   & 0.009 \\ 
57852.49929                         & 16.284   & 0.006 &             &           &        &             &           &        \\ 
57854.04275                         & 16.495   & 0.014 & 57854.04514 & 17.638   & 0.025  & 57854.05076 & 17.046   & 0.022 \\ 
57856.79728                         & 16.271   & 0.009 &             &           &        &             &           &        \\ 
57863.46871                         & 16.100   & 0.004 &             &           &        &             &           &        \\ 
57864.27177                         & 16.349   & 0.005 & 57864.26883 & 17.461   & 0.007 & 57864.27766 & 16.754   & 0.005 \\ 
57865.27192                         & 16.437   & 0.006 & 57865.26898 & 17.554   & 0.007 & 57865.27784 & 16.872   & 0.005 \\ 
57867.27192                         & 16.307   & 0.005 & 57867.26897 & 17.476   & 0.009 & 57867.27783 & 16.782    & 0.006 \\ 
57871.81080                          & 16.073   & 0.007 & 57871.80948 & 17.348   & 0.014 &             &           &        \\ 
57874.73148                         & 16.504   & 0.005 &             &           &        &             &           &        \\ 
57874.80092                         & 16.499   & 0.010 & 57874.79960  & 17.699   & 0.021 &             &           &        \\ 
57878.68604                         & 16.518   & 0.005 &             &           &        &             &           &        \\ 
57878.77927                         & 16.350   & 0.008 & 57878.78106 & 17.437    & 0.012 &             &           &        \\ 
57882.75055 & 16.246   & 0.013 & 57882.75196 & 17.515    & 0.040 & 57882.75494 & 16.724   & 0.018 \\ 
57883.63992                         & 16.297   & 0.009 &             &           &        &             &           &        \\ 
57883.72140                          & 16.445   & 0.018 & 57883.72281 & 17.678   & 0.066 & 57883.72579 & 16.884   & 0.031 \\ 
57889.71699                         & 16.423   & 0.049 &             &           &        &             &           &        \\ 
57895.43395                         & 16.600   & 0.011 & 57895.43536 & 17.836   & 0.023 & 57895.43833 & 17.050   & 0.014 \\ 
57896.70059                         & 16.381   & 0.010   & 57896.70235 & 17.648   & 0.024 & 57896.70641 & 16.892   & 0.015 \\ 
57897.46580                          & 16.464   & 0.005  &             &           &        &             &           &        \\ 
57904.49230                          & 16.304   & 0.008 & 57904.49406 & 17.561    & 0.013 & 57904.49704 & 16.761   & 0.009 \\ 
57907.45315                         & 16.562   & 0.006 &             &           &        &             &           &        \\ 
57910.77233                         & 16.412   & 0.009 &             &           &        &             &           &        \\ 
57919.69822                         & 16.905   & 0.085 &             &           &        &             &           &        \\ 
57929.48416                         & 16.612   & 0.010 & 57929.48535 & 17.844   & 0.023 &             &           &        \\ 
57930.47042                         & 16.535   & 0.007 &             &           &        &             &           &        \\ 
57930.62906                         & 16.670   & 0.017 & 57930.63024 & 18.007   & 0.100 &             &           &        \\ 
57932.39149                         & 16.813   & 0.012 & 57932.39266 & 17.998   & 0.030 &             &           &        \\ 
57934.10396                         & 16.701   & 0.014 & 57934.10530  & 17.930   & 0.028 &             &           &        \\ 
57937.29423                         & 17.043   & 0.010 &             &           &        &             &           &        \\ 
57939.53233                         & 16.873   & 0.030 &             &           &        &             &           &        \\ 
57956.59849                         & 17.121   & 0.018 & 57956.60008 & 18.335   & 0.023 & 57956.60167 & 17.640   & 0.019 \\ 
57956.62583                         & 17.073   & 0.015 & 57956.62725 & 18.418   & 0.049 & 57956.62854 & 17.552   & 0.025 \\ 
57960.53634                         & 17.055   & 0.010 &             &           &        &             &           &        \\ 
57981.30615                         & 17.617   & 0.013 &             &           &        &             &           &        \\ 
58008.43489                         & 17.847   & 0.020 &             &           &        &             &           &        \\ 
                                    &           &        & 58029.04037 & 19.041   & 0.047  &             &           &        \\ 
58031.37832                         & 17.838    & 0.029 &             &           &        &             &           &        \\ 
58032.41498                         & 17.685   & 0.051 &             &           &        &             &           &        \\ 
58033.41110                          & 17.803   & 0.035  &             &           &        &             &           &        \\ 
58035.37890                          & 17.792   & 0.030 &             &           &        &             &           &        \\ 
58039.39175                         & 17.770   & 0.030 & 58039.39461 & 19.373   & 0.065 &             &           &        \\ 
58041.00927                         & 17.456   & 0.019 & 58041.01225 & 18.928   & 0.020 &             &           &        \\ 
58044.39425                         & 18.108   & 0.031 & 58044.39704 & 19.448   & 0.061 &             &           &        \\ 
58045.74268                         & 18.106   & 0.079 & 58045.74565 & 19.224   & 0.062 &             &           &        \\ 
\enddata
\end{deluxetable*}


\startlongtable
\begin{deluxetable*}{|lll|lll|lll|}
\tablecaption{REM near-IR detections of GRS~1716$-$249 during the 2016-2017 outburst. The Vega magnitudes before the subtraction of the contribution from the nearby star is reported here. If the source is not detected, the $3\sigma$ upper limit (UL) is reported.}
\tablehead{\multicolumn{3}{c}{$J$-band}                            & \multicolumn{3}{c}{$H$-band}     & \multicolumn{3}{c}{$K$-band}  \\
\hline
\colhead{MJD}                                 & \colhead{magnitude} & \colhead{error}  & \colhead{MJD}         & \colhead{magnitude} & \colhead{error}  & \colhead{MJD}         & \colhead{magnitude} & \colhead{error}}
\startdata
57792.32435  & 14.190   & 0.108 & 57792.32820 & 13.552   & 0.103 &    57792.33073         & 13.034          & 0.176       \\ 
57793.32431   &  14.270   	&0.121	& 57793.32817  &  13.616       &  0.122	& 57793.33068  &  13.462 	&  0.207    \\
57794.32440  &  14.079  &0.074	& 57794.32832   &  13.586       &  0.113	&57794.33087  &  13.262     &   0.173   \\
57795.33416   &  14.372 	&0.113	& 57795.33802  &  13.726      &  0.102	&57795.34053  &  13.562 	&  0.211    \\
57796.37340  &  14.287 	&0.107	&57796.37726   &  13.826      &  0.129	& 57796.37977  &  13.528 	&  0.198    \\
57798.30813  &  14.392  	&0.128	& 57798.30461  &  13.786      &  0.142	& 57798.30076  &  14.101       &   0.298   \\
             &     &     & 57799.31212   &  13.873      &  0.112	& 57799.31462  &  13.600       &   0.240   \\
57800.30831  &  14.299 	&0.094	& 57800.31216  &  13.633      &  0.105	& 57800.31468  &  13.123 	&  0.176    \\
57801.30816  &  14.133	&0.072	& 57801.31200  &  13.717      &  0.092	& 57801.31452  &  13.011  	&  0.143	   \\
57802.31356  &  14.443 	&0.132	& 57802.31742   &  13.839     &   0.151	&57802.31990  &  13.626       &   0.205   \\
57803.31346  &  14.155	&0.128	& 57803.31733  &  13.613      &  0.122	& 57803.31984  &  13.035       &   0.151    \\
57805.26872  &  14.242 	&0.112	& 57805.26521  &  13.597      &  0.097	& 57805.26137  &  13.001  	&  0.174	   \\
57807.35268  &  14.254  	&0.096	& 57807.35655  &  13.817       &  0.103	& 57807.35907  &  13.218       &   0.158   \\
57809.28189  &  14.569	& 0.105	& 57809.27835  &  13.864     &   0.097	& 57809.27451   &  13.674       &   0.254   \\
57812.32570  &  14.245 	&0.110	& 57812.32958  &  13.695     &   0.134	& 57812.33209  &  12.903  	&  0.228    \\
57813.32666  &  14.602 	&0.109	& 57813.33053  &  14.051      &  0.129	& 57813.33304  &  13.137 	&  0.152    \\
57815.23916  &  14.159  	&0.101	& 57815.23564  &  13.911       &  0.146	& 57815.23180  &  12.987	&   0.126   \\
             &     &     & 57816.26729  &  13.784      &  0.143	& 57816.26982  &  13.382	&   0.174   \\
57817.28778  &  14.144	&0.069	& 57817.29164  &  13.751      &  0.124	& 57817.29417  &  13.104 	&   0.174   \\
57818.28780    &  14.246	&0.092	& 57818.29173  &  13.806      &  0.127	& 57818.29424  &  13.584	&   0.229   \\
57819.28782  &  14.129	&0.077	& 57819.29169   &  13.701      &  0.116	& 57819.29419  &  13.490  	&   0.226    \\
57820.29095  &  14.404  	&0.100	& 57820.29483  &  13.753      &  0.134	& 57820.29736  &  13.298	&   0.154   \\
57821.29092  &  14.323	& 0.113	& 57821.29478  &  13.700      &  0.112	& 57821.29728  &  13.441 	&   0.232   \\
57822.30718  &  14.093	& 0.067	& 57822.31108  &  13.601       &  0.097	& 57822.31358  &  13.396	&   0.164   \\
57823.32352  &  14.347	& 0.098	& 57823.31999  &  13.695     &   0.136	& 57823.31615  &  13.296 	&   0.144   \\
57824.32348  &  14.139 	&0.085	& 57824.31995  &  13.642     &   0.115	& 57824.31613  &  12.981	&   0.116  \\
57835.26807  &  14.247	& 0.088	& 57835.27195  &  13.676      &  0.120	& 57835.27452  &  13.469	&   0.180   \\
57836.27332  &  14.381  	&0.124	& 57836.27719  &  13.777      &  0.139	& 57836.27970    &  13.799	&   0.207   \\
57840.28813  &  14.130	& 0.078	& 57840.28460  &  13.575      &  0.076	& 57840.28075  &  13.454	&   0.161   \\
57841.28813  &  14.207	& 0.132	& 57841.28462  &  13.757     &   0.108	& 57841.28082  &  13.515 	&   0.145   \\
57842.28812   &  14.318	& 0.132	& 57842.28461  &  13.979     &   0.181	& 57842.28084  &  13.486	&   0.185   \\
57843.37491  &  14.141	& 0.097	& 57843.37876  &  13.614      &  0.103& & & \\
& & & & & & 57844.38131  &  13.407	&   0.232   \\
57847.22599  &  14.083 	&0.057	& 57847.22991  &  13.594      &  0.081	& 57847.23242  &  13.169	&   0.137   \\
57849.22592  &  14.230 	&0.086	& 57849.22983   &  13.582      &  0.090	& 57849.23236  &  13.506	&   0.169   \\
57851.24054   &  14.440 	&0.087	& 57851.24440  &  13.962     &   0.112	& 57851.24692  &  13.610	&   0.171   \\
& & & 57854.2229  &  13.680      &  0.071	& 57854.22542  &  13.279	&   0.136   \\
57856.21885   &  14.267 	&0.091	& 57856.22273   &  13.943     &   0.119	&57856.22526  &  13.800	&   0.197   \\
57862.17564  &  14.269 	&0.101	& 57862.17210  &  13.922       &  0.103	& 57862.16824  &  13.614  &   0.204	   \\
57864.17594  &  14.284 	&0.082	& 57864.17243  &  13.767     &   0.100	& 57864.16858  &  13.628  &   0.192   \\
57866.18714  &  14.307 	&0.082	& 57866.18362  &  13.807     &   0.101	& 57866.17983  &  13.329  &   0.163   \\
57868.26877  &  14.306 	&0.075	& 57868.27263  &  13.615     &   0.096	&57868.27513  &  13.262 	&   0.137   \\
& & & 57870.27550   &  13.454       &  0.107	& 57870.27801  &  13.000  &   0.107  \\
57873.07972  &  14.378	& 0.079	& 57873.07621  &  13.995     &   0.137	& 57873.07240  &  13.311 	&   0.097   \\
57875.08102  &  14.700 	&0.195	& & & & 57875.07377   &  13.397 	&   0.163  \\
57877.13749   &  14.200	& 0.109	& 57877.13396  &  13.750      &  0.103	& 57877.13012  &  13.419	&   0.128   \\
57879.16786  &  14.313	& 0.104	& 57879.16433  &  13.605      &  0.073	& 57879.16049  &  13.427	&   0.145   \\
57882.13166   &  14.255	& 0.064	& 57882.12812  &  13.826       &  0.090	& 57882.12427  &  13.282	&   0.102   \\
57887.22384  &  13.873	& 0.266	& & & & & & \\
57888.25770  &  14.236	& 0.092	& 57888.26169  &  13.464      &  0.093	& 57888.26419  &  13.465	&   0.166   \\
57889.27466  &  14.583	& 0.127	& 57889.27853  &  13.697      &  0.114	& 57889.28102  &  12.682	&   0.164  \\
57890.27447  &  13.678	& 0.066	& & & & 57890.28084  &  12.772	&   0.093   \\
57894.11317  &  14.363	& 0.093	& 57894.10966  &  13.889      &  0.126	& 57894.10581   &  13.278	&   0.101   \\
57895.11313   &  14.320	& 0.084	& 57895.10962  &  13.838       &  0.110	& 57895.10578  &  13.810	&   0.183   \\
57901.02803  &  14.343	& 0.108	& 57901.02453  &  13.686      &  0.139	& 57901.02068   &  13.318 	&   0.124   \\
57902.05513  &  14.335  	&0.091	& 57902.05157  &  13.808       &  0.122	& 57902.04774  &  13.726	&   0.148    \\
57904.00570  &  14.414 	&0.097	& 57904.00218  &  13.655     &   0.097	& & & \\
57905.06105  &  14.642  	&0.152	& 57905.05752  &  13.792      &  0.118	& 57905.05370  &  13.307	&   0.127  \\
57906.06101  &  14.298 	&0.061	& 57906.05749   &  13.648&   0.065	& 57906.05365  &  13.585	&   0.177   \\
57907.06098  &  14.688 	&0.137	& 57907.05748  &  13.798       &  0.104	& 57907.05363  &  13.446	&   0.158    \\
57908.07134  &  14.302 	&0.105	& 57908.06780    &  13.759      &  0.182	& 57908.06397  &  13.272	&   0.207   \\
57909.07118   &  14.404 	&0.134	& 57909.06766  &  13.775     &   0.174	& 57909.06381  &  13.502	&   0.168   \\
57910.07129  &  14.292 	&0.131	& 57910.06778  &  13.745     &   0.088	& 57910.06393  &  13.609	&   0.160   \\
57911.07136  &  14.307  	&0.094	& 57911.06784   &  13.694&   0.124	& 57911.06399  &  13.214 	&   0.113  \\
57913.03877  &  14.175	& 0.092	& 57913.03525  &  13.711     &   0.143	& 57913.03135  &  13.390	&   0.139   \\
57915.05323  &  14.320	& 0.1195	& 57915.04971  &  13.616     &   0.104	& 57915.04584  &  13.730	&   0.176   \\
57916.05831  &  14.516 	&0.107	& 57916.05479  &  14.055      &  0.130& 57916.05095  &  13.552	&   0.145   \\
57917.06366  &  14.461	& 0.116	& 57917.06013   &  13.888     &   0.112	& 57917.05629  &  13.406 	&   0.135   \\
57918.06363  &  14.461	& 0.116	& 57918.06012  &  13.750      &  0.111	& & & \\
57922.04947  &  14.613  	&0.127	& 57922.04595  &  13.887     &   0.157	& 57922.04211  &  13.494 & 0.159	   \\
57924.15149  &  14.307     &0.052	& 57924.15530  &  13.665     &   0.075	& 57924.15786  &  13.198	&   0.129   \\
57925.23085  &  14.328 	&0.090	& 57925.23477  &  14.045     &   0.121	& 57925.23726  &  13.432	&   0.144   \\
57933.12670  &  14.100	 	&0.222	& & & & & & \\
57934.12671  &  14.568  	&0.113& 57934.13060  &  13.948      &  0.103	& 57934.13311   &  13.564	&   0.182   \\
57936.12677  &  14.318  	&0.156	& 57936.13060  &  14.159      &  0.180	& & & \\
57937.14052  &  14.425       &0.131	& 57937.13702  &  13.948&   0.156	& 57937.13322  &  13.817	&   0.184   \\
57938.30936  &  14.453  	&0.106	& 57938.30584  &  13.820&   0.117	& 57938.30203  &  13.735	&   0.166   \\
57949.33352  &  13.510 	&0.136	& 57949.33735   &  13.048&   UL		& 57949.33980 & 12.613	&   UL		   \\
57951.09202  &  14.409  	&0.0815	& 57951.09588  &  14.209     &   0.170	& 57951.09841  &  13.755	&   0.236   \\
57954.23148  &  14.267     &0.102	& 57954.23537  &  13.714       &  0.140	& 57954.23789  &  13.132 	&   0.142   \\
57955.23142  &  14.241   	&0.102	& 57955.23530    &  13.741      &  0.140	& 57955.23779  &  13.273 	&   0.221   \\
57956.23142  &  14.368   	&0.116	& 57956.23534  &  13.872       &  0.128	& 57956.23785  &  13.737	&   0.182   \\
57958.26998  &  14.480	&   0.129& 57958.26647  &  14.104     &   0.180	& 57958.26263  &  13.693  &   0.195   \\
57959.31483   &  14.342  	&0.082	& 57959.31134  &  14.032&   0.130	& 57959.30750    &  13.640  &   0.173   \\
57962.05381  &  14.313  & 0.145	& 57962.05775  &  13.856&   0.146	&57962.06027  &  13.403 	&   0.168	   \\
57964.22435  &  14.578  	&0.099	& 57964.22824  &  13.685&   0.100	& 57964.23077  &  13.800		&   0.221   \\
& & & & & & 57967.02934  &  13.494	&   0.144   \\
57969.96398  &  14.581	& 0.120	& 57969.96786   &  14.109     &   0.132	& 57969.97039  &  13.193	&   0.136   \\
& & & & & & 57971.04542  &  13.458 	&   0.197   \\
57972.04106   &  13.162	& 0.050	& 57972.04494  &  15.766       &   UL		& 57972.04740  &  13.395 	&   0.323   \\
57973.24798  &  14.892	& 0.157	& 57973.25183  &  14.429       &   0.181	& 57973.25434  &  14.100	&   0.232   \\ 
57974.24868   &  14.665  	&0.078	& 57974.25258  &  13.836&   0.113	& 57974.25510    &  13.860	&   0.222   \\
57982.22021  &  14.582  	&0.076	& 57982.22410  &  13.725       &   0.115	& 57982.22663  &  13.699	&   0.211   \\
57983.24220  &  14.961  	&0.129	& 57983.24606  &  14.473&   0.230	& 57983.24860  &  14.321  	&    UL		   \\
57984.24218  &  15.228	& 0.220	& 57984.24604   &  13.968&   0.160	& 57984.24853   &  13.734 	&     UL	   \\
57987.21943  &  14.858	& 0.126	& 57987.22329  &  14.048       &   0.133	& 57987.22578  &  13.916	&   0.293   \\
57990.09610  &  13.988  & 0.075	& 57990.09999  &  13.340&   0.083	& 57990.10249  &  13.431	&   0.201   \\
57991.22900     &  15.209  & UL		& & & & 57991.23530   & 14.023 	&   UL		   \\
57994.21801  &  14.930	&   0.139& 57994.22188  &  14.082       &   0.160	& 57994.22438  & 13.794 	&  0.179   \\
57998.21709  &  14.981        & UL		& 57998.22096  &  14.139      &   UL		& & & \\
58000.15082 &    14.856 	&0.121	& 58000.15469  &   14.359    &    0.165 & 58000.15717  &   13.678	&  0.159    \\
58001.15070 &    15.253  	&0.207	& 58001.15456  &   14.207     &    0.135 & 58001.15705  &  13.673 	&  0.156	   \\
58002.15071  &    14.783 	&0.209	& 58002.15457  &   14.781    &    UL		& 58002.15709  &  13.083	&  0.259   \\
58003.15487 &    14.695	&   0.082& 58003.15874   &   14.168     &    0.126 & 58003.16120  &  13.668	&  0.179    \\
58004.15478  &    14.899 	&0.117	& 58004.15863  &   14.377     &    0.158	& 58004.16112  &  13.720	&  0.140    \\
58007.12125  &   14.785  	&0.103	& 58007.12511  &   14.358      &    0.154 & 58007.12761  &  13.978	&  0.192    \\
58008.12116   &   14.795 	&0.102	& 58008.12502  &   14.138    &    0.168 & 58008.12752  & 13.987 	&  0.241	   \\
58010.12534  &   14.823	&  0.076 & 58010.12920  &   14.298    &    0.130 & 58010.13173  & 13.847  	&  0.150    \\
58011.12521  &   14.957	&  0.120 & 58011.12904  &   14.371    &    0.130 & 58011.13155  & 13.974  	&  0.187   \\
58012.12525  &   14.837  	&0.091	& 58012.12911  &   14.194&   0.142	& 58012.13163  & 13.772  	&  0.144	   \\
58017.08652  &   14.884	&  0.136 & 58017.09039  &   14.302      &   0.146	& 58017.09290  & 13.339  	&  0.137    \\
58019.06545  &   14.613	&  0.059 & 58019.06929  &   14.008      &   0.115	& 58019.07177  & 13.613  	&  0.172   \\
58021.05757  &   14.563 	&0.130	& 58021.06152  &   14.072&   0.125	& 58021.06401  & 13.726	&  0.230\\
58022.05786  &   14.886	&  0.127	& 58022.06174  &   14.232&   0.138	& 58022.06424  & 13.345 	&  0.171    \\
58024.08948  &   14.672	&   0.127& 58024.09339  &   14.433      &  0.258	& 58024.09590  &	12.944	&  0.372	   \\
58025.08981  &   14.914 	&0.137	& 58025.09367  &   14.390     &  0.197	& & & \\
\enddata
\end{deluxetable*}


\begin{thebibliography}{99}
\bibliographystyle{aasjournal}
\bibitem[Alabarta et al.(2020)]{ala2020} Alabarta K. et al., 2020, MNRAS, 497, 3896
\bibitem[Alabarta et al.(2021)]{alabarta} Alabarta K., Altamirano D., Mendez M., Cuneo V. A., Vincentelli F. M., Castro-Segura N., Garcia F., Luff B. \& Veledina A., 2021, MNRAS, 507, 5507
\bibitem[Armas Padilla \& Munoz-Darias(2017)]{armas} Armas Padilla M. \& Munoz-Darias T., 2017, The Astronomer's Telegram, 10236
\bibitem[Armas Padilla et al.(2013)]{armas2013} Armas Padilla M., Degenaar N., Russell D.~M., Wijnands R., 2013, MNRAS, 428, 3083
\bibitem[Atri et al.(2019)]{atri} Atri P., Miller-Jones J. C. A., Bahramian A., Plotkin R. M., Jonker P. G. et al., 2019, MNRAS, 489, 3, 3116
\bibitem[Baglio et al.(2018)]{baglio18} Baglio M. C., Russell D. M., Casella P., et al., 2018, ApJ, 867, 114
\bibitem[Baglio et al.(2022)]{Baglio_submitted} Baglio M. C., Saikia P., Russell D. M., et al., 2022, ApJ, 930, 20
\bibitem[Bahramian et al.(2015)]{bahr15} Bahramian A., Heinke C. O., Degenaar N., Chomiuk L., Wijnands R., Strader J., Ho W. C. G., Pooley D., 2015, MNRAS, 452, 4, 3475, https://doi.org/10.1093/mnras/stv1585
\bibitem[Bahramian et al.(2018)]{bahr} Bahramian A., et al., 2018, https://doi.org/10.5281/zenodo.1252036
\bibitem[Ballet et al.(1993)]{ballet93} Ballet J., Denis M., Gilfanov M., Sunyaev R., Harmon B. A., Zhang S. N., Paciesas W. S., Fishman G. J., 1993, IAU Circ., 5874
\bibitem[Bassi et al.(2017)]{bassi17} Bassi T., Del Santo M., Motta S. E., 2017, The Astronomer's Telegram, 10371
\bibitem[Bassi et al.(2019)]{bassi} Bassi T., Del Santo M., Dai A., et al., 2019, MNRAS, 482, 1587
\bibitem[Bassi et al.(2020)]{bassi2020} Bassi T., Malzac J., Del Santo M. et al., 2020, MNRAS, 494, 1, 571
\bibitem[Belloni et al.(2002)]{bb02} Belloni T. M., Colombo A. P., Homan J., Campana S., van der Klis M., 2002, A\&A, 390, 199
\bibitem[Belloni(2010)]{b10} Belloni T. M., 2010, in Belloni T., ed., Lecture Notes in Physics, Berlin Springer Verlag Vol. 794, Lecture Notes in Physics, Berlin Springer Verlag. p. 53 (arXiv:0909.2474)
\bibitem[Bernardini et al.(2016)]{Bernardini2016} Bernardini F., Russell D. M., Kolojonen K. I. I. et al., 2016, ApJ, 826, 149
\bibitem[Bharali et al.(2019)]{bharali} Bharali P., Chandra S., Chauhan J., Garcia J. A., Roy J., Boettcher M. \& Boruah K., 2019, MNRAS, 487, 3150
\bibitem[Blandford \& Konigl(1979)]{bk} Blandford R. D. \& Konigl A., 1979, ApJ, 232, 34
\bibitem[Bramich \& Freudling(2012)]{dan} Bramich D. M. \& Freudling W., 2012, MNRAS, 424, 1584
\bibitem[Brown et al.(2016)]{gaia} Brown A. G. A., Vallenari A., Prusti T., de Bruijne J. H. J., et al., 2016, A\&A, 595, A2
\bibitem[Buxton \& Bailyn(2004)]{bb04} Buxton M. M. \& Bailyn C. D., 2004, ApJ, 615, 880
\bibitem[Cadolle Bel et al.(2011)]{cadollebel2011} Cadolle Bel M., Rodriguez J., D'Avanzo P., et al., 2011, A\&A, 534, A119 
\bibitem[Capitanio et al.(2009)]{cap} Capitanio F., Belloni T., Del Santo M., Ubertini P., 2009, MNRAS, 398, 1194
\bibitem[Cardelli et al.(1989)]{cardelli} Cardelli J. A., Clayton G. C. \& Mathis J. S., 1989, ApJ, 345, 245
\bibitem[Carotenuto et al.(2021)]{c2021} Carotenuto F., Corbel. S, Tremou E., Russell T. D., Tzioumis A. et al., 2021, MNRAS, 504, 444
\bibitem[Casares et al.(2009)]{154} Casares J., Orosz J. A., Zurita C., Shahbaz T., et al., 2009, ApJS, 181, 238
\bibitem[Casares \& Jonker(2014)]{casares14} Casares J., Jonker P. G., 2014, SSRv, 183, 223
\bibitem[Chatterjee et al.(2021)]{chatterjee} Chatterjee K., Debnath D., Chatterjee D., Jana A., Nath S. K., Bhowmick R., Chakrabarti S. K., 2021, Ap\&SS, 366, 63
\bibitem[Chaty et al.(2002)]{chaty} Chaty S., Mirabel I. F., Goldoni P., Mereghetti S., Duc P. A., Marti J., Mignani R. P., 2002, MNRAS, 331, 4, 1065
\bibitem[Chaty et al.(2015)]{chaty15} Chaty S., Munoz Arjonilla A. J. \& Dubus G., 2015, A\&A, 577, A101
\bibitem[Chevalie et al.(1989)]{cheva} Chevalier C., Ilovaisky S. A., van Paradijs J., Pedersen H., van der Klis M., 1989, A\&A, 210, 114
\bibitem[Corbel et al.(2000)]{Corbel2000} Corbel S., Fender R. P., Tzioumis A. K., et al. 2000, A\&A, 359, 251
\bibitem[Corbel \& Fender(2002)]{corbel2002} Corbel S., Fender R. P., 2002, ApJ, 573, L35 
\bibitem[Corbel et al.(2003)]{Corbel2003} Corbel S., Nowak M. A., Fender R. P., Tzioumis A. K., Markoff S., 2003, A\&A, 400, 1007
\bibitem[Corbel et al.(2004)]{Corbel2004} Corbel S., Fender R. P., Tomsick J. A., Tzioumis A. K., Tingay S., 2004, ApJ, 617, 1272
\bibitem[Corbel et al.(2013)]{Corbel2013} Corbel S., Coriat M., Brocksopp C., Tzioumis A. K., Fender R. P., Tomsick J. A., Buxton M. M., Bailyn C. D., 2013, MNRAS, 428, 2500
\bibitem[Coriat et al.(2009)]{Coriat2009} Coriat M., Corbel S., Buxton M. M., Bailyn C. D., Tomsick J. A., Kording E., Kalemci E., 2009, MNRAS, 400, 123
\bibitem[Coriat et al.(2011)]{Coriat2011} Coriat M., Corbel S., Prat L., Miller-Jones J. C. A., Cseh D.,  Tzioumis A. K., Brocksopp C., Rodriguez J.,  R. P. Fender R. P. \& Sivakoff G. R., 2011, MNRAS, 414, 677
\bibitem[Corral-Santana et al.(2016)]{bc} Corral-Santana J.~M., Casares J., Mu{\~n}oz-Darias T., Bauer F.~E., Mart{\'\i}nez-Pais I.~G., Russell D.~M., 2016, A\&A, 587, A61
\bibitem[Cunningham(1976)]{Cunningham1976} Cunningham C., 1976, ApJ, 208, 534
\bibitem[Cuneo et al.(2020)]{cuneo} Cuneo V. A., Munoz-Darias T., Sanchez-Sierras J., Jimenez-Ibarra F., Armas Padilla M. et al., 2020, MNRAS, 498, 1, 25
\bibitem[Curran \& Chaty(2013)]{curran2013} Curran P. A., Chaty S., 2013, A\&A, 557, A45
\bibitem[Curran et al.(2012)]{curran} Curran, P. A., Chaty, S., \& Zurita Heras J. A.  2012, A\&A, 547, A41
\bibitem[Del Santo et al.(2017)]{delsanto} Del Santo M., et al., 2017, The Astronomer's Telegram, 10069
\bibitem[della Valle et al.(1993)]{iauc} della Valle M., Mirabel I. F., Cordier B., Bonibaker J., Stirpe G., Rodriguez L. F., 1993, IAU Circ., 5876
\bibitem[della Valle et al.(1994)]{della} della Valle M., Mirabel I. F., Rodriguez L. F., 1994, A\&A, 290, 803
\bibitem[Done et al.(2007)]{done} Done C., Gierlinski M., Kubota A., 2007, A\&AR, 15, 1
\bibitem[Evans et al.(2007)]{evans1} Evans P. A., et al., 2007, A\&A, 469, 379
\bibitem[Evans et al.(2009)]{evans2} Evans P. A., et al., 2009, MNRAS, 397, 1177
\bibitem[Falcke et al.(2004)]{falcke}
Falcke H., Koerding E. \& Markoff S., 2004, A\&A, 414, 895
\bibitem[Fender et al.(1999)]{Fender1999} Fender R. P. et al., 1999, ApJ, 519, L165 
\bibitem[Fender \& Hendry(2000)]{ref1} Fender R. P. \& Hendry M. A., 2000, MNRAS, 317, 1
\bibitem[Fender \& Kuulkers(2001)]{ref2} Fender R. P. \& Kuulkers E., 2001, MNRAS, 324, 923
\bibitem[Fender, Belloni \& Gallo(2004)]{fender04} Fender R. P., Belloni T. M., Gallo E., 2004, MNRAS, 355, 1105
\bibitem[Fitzpatrick(1999)]{fitz} Fitzpatrick E. L. 1999, PASP, 111, 63
\bibitem[Foight et al.(2016)]{foight16} Foight D. R., Guver T., Ozel F., \& Slane P. O., 2016, ApJ, 826, 66
\bibitem[Frank et al.(2002)]{fkr} Frank J., King A., Raine D. J., 2002, Accretion Power in Astrophysics, 3rd edn. Cambridge Univ. Press, Cambridge
\bibitem[Gallo, Miller \& Fender(2012)]{gallo2012} Gallo E., Miller B. P., \& Fender R., 2012, MNRAS, 423, 590
\bibitem[Gallo et al.(2014)]{gallo2014} Gallo E., et al., 2014, MNRAS, 445, 290
\bibitem[Gallo, Degenaar \& van den Eijnden(2018)]{gallo2018} Gallo E., Degenaar N., van den Eijnden J., 2018, MNRAS, 478, L132
\bibitem[Gandhi(2009)]{gandhi2009}Gandhi P., ApJ, 2009, 697, L167
\bibitem[Gandhi et al.(2010)]{gandhi2010}Gandhi P., Dhillon V. S., Durant M., et al., 2010, MNRAS, 407, 2166
\bibitem[Gandhi et al.(2011)]{gandhi2011}Gandhi P., Blain A. W., Russell D. M., Casella P., Malzac J., et al., ApJ, 2011, 740, L13
\bibitem[Gandhi et al.(2016)]{gandhi2016}Gandhi P., Littlefair S. P., Hardy L. K., et al., 2016, MNRAS, 459, 554
\bibitem[Gandhi et al.(2019)]{gandhi2019} Gandhi P., Rao A., Johnson M. A. C., Paice J. A., Maccarone T. J., 2019, MNRAS, 485, 2642
\bibitem[Goodwin et al.(2020)]{goodwin2020} Goodwin A. J., Russell D. M., Galloway D. K., et al., 2020, MNRAS, 498, 3429
\bibitem[Guver et al.(2009)]{av1} Guver T. \& Ozel F., 2009, MNRAS, 400, 2050
\bibitem[Harmon et al.(1993)]{harmon93} Harmon B. A., Fishman G. J., Paciesas W. S., Zhang S. N., 1993, IAU Circ., 5900
\bibitem[Haswell et al.(2001)]{haswell} Haswell C. et al., 2001, MNRAS, 321, 475
\bibitem[Heinke et al.(2015)]{heinke} Heinke C.~O., Bahramian A., Degenaar N., Wijnands R., 2015, MNRAS, 447, 3034
\bibitem[Hjellming \& Johnston (1988)]{hjellming88} Hjellming R. M. \& Johnston K. J., 1988, ApJ, 328, 600
\bibitem[Hjellming et al.(1996)]{hjellming96} Hjellming, R. M. et al. 1996, ApJ, 470, L105
\bibitem[Homan et al.(2001)]{homan2001} Homan J., Wijnands R., van der Klis M., Belloni T., van Paradijs J., KleinWolt M., Fender R., Mendez M., 2001, ApJS, 132, 377
\bibitem[Homan \& Belloni(2005)]{hb} Homan J. \& Belloni T., 2005, Ap\&SS, 300, 107
\bibitem[Homan et al.(2005)]{homan2005} Homan J., Buxton M., Markoff S., Bailyn C. D., Nespoli E., Belloni T., 2005, ApJ, 624, 295
\bibitem[Homan et al.(2006)]{homan2006} Homan J., Wijnands R., Kong A., Miller J. M., Rossi S., Belloni T., Lewin W. H. G., 2006, MNRAS, 366, 235
\bibitem[Hynes et al.(1998)]{hynes1998} Hynes R. I., O'Brien K., Horne K., Chen W., Haswell C. A., 1998, MNRAS, 299, L37
\bibitem[Hynes et al.(2003)]{hynes2003} Hynes R. I., Haswell C. A., Cui W., et al., 2003, MNRAS, 345, 292
\bibitem[Hynes(2005)]{hynes2005} Hynes R. I., 2005, ApJ, 623, 1026
\bibitem[Hynes et al.(2006)]{hynes2006} Hynes R. I., Robinson E. L., Pearson K. J., et al., 2006, ApJ, 651, 401 
\bibitem[Hynes et al.(2009)]{hynes2009} Hynes R. I., O'Brien, K., Mullally F., Ashcraft T., 2009, MNRAS, 399, 281
\bibitem[Hynes et al.(2019)]{hynes2019} Hynes R. I., Robinson E. L., Terndrup D. M., et al., 2019, MNRAS, 487, 1, 60
\bibitem[Ingram \& Motta(2019)]{ingram} Ingram A. R. \& Motta S. E. 2019, New Astronomy Reviews, 85, 101524
\bibitem[Jain et al.(2001)]{jain01} Jain R. K.,Bailyn C. D.,Orosz J. A.,McClintock J. E. \& Remillard R. A., 2001, ApJ, 554, L181
\bibitem[Jiang et al.(2017)]{jiang} Jiang J., FUrst F, Walton D. J., Parker M. L. \& Fabian A. C., 2020, MNRAS, 492, 2, 1947
\bibitem[Joshi et al.(2017)]{joshi17} Joshi V., Vadwale S., Ganesh S., Aarthy E., 2017, The Astronomer's Telegram, 10196
\bibitem[Kalemci et al.(2013)]{Kalemci2013} Kalemci E., Din{\c{c}}er T., Tomsick J. A., Buxton M. M., Bailyn C. D., Chun Y. Y., 2013, ApJ, 779, 95
\bibitem[Koljonen et al.(2015)]{Koljonen2015} Koljonen K. I. I., Russell D. M., Fern{\'a}ndez-Ontiveros J. A., et al., 2015, ApJ, 814, 139
\bibitem[Koljonen et al.(2016)]{Koljonen2016} Koljonen K. I. I., Russell D. M., Corral-Santana J. M. et al., 2016, MNRAS, 460, 942
\bibitem[Koljonen et al.(2018)]{Koljonen2018} Koljonen, K. I. I., Maccarone, T., McCollough, M. L., et al. 2018, A\&A, 612, A27
\bibitem[Koljonen \& Russell(2019)]{karridave} Koljonen K. I. I. \& Russell D. M., ApJ, 871, 26, 8, 10.3847/1538-4357/aaf38f
\bibitem[Krimm et al.(2013)]{bat} Krimm H. A., et al., 2013, ApJSS, 209, 14
\bibitem[Lagage et al.(2004)]{Lagage2004} Lagage P.O., Pel J., Authier M., et al., 2004, Messenger 117, 12
\bibitem[Lewis(2018)]{Lewis2018} Lewis, F., 2018, Robotic Telescope, Student Research and Education Proceedings, 1, 237
\bibitem[Lewis et al.(2008)]{Lewis2008} Lewis, F., Russell, D. M., Fender, R. P., Roche, P., Clark, J. S. 2008, AIP Conf. Proc, 1010, 204-206
\bibitem[Maccarone(2003)]{macca} Maccarone T. J., 2003, A\&A, 409, 697
\bibitem[Maitra et al.(2017)]{maitra2017} Maitra D., Scarpaci J. F., Grinberg V., et al. 2017, ApJ, 851, 148
\bibitem[Maitra \& Bailyn (2008)]{Maitra2008} Maitra D. \& Bailyn C. D., 2008, ApJ, 688, 537
\bibitem[Markoff, Falcke \& Fender (2001)]{markoff01} Markoff S., Falcke H. \& Fender R., 2001, A\&A, 372, L25
\bibitem[Masetti et al.(1996)]{masetti} Masetti N., Bianchini A., Bonibaker J., della Valle M., Vio R., 1996, A\&A, 314, 123
\bibitem[Mashumitsu et al.(2016)]{mashu} Masumitsu T., et al., 2016, The Astronomer's Telegram, 9895
\bibitem[Matsuoka et al.(2009)]{maxi} Matsuoka M., et al., 2009, PASJ, 61, 999
\bibitem[Mathis (1990)]{matthis} Mathis, 1990, ARA\&A, 28, 37
\bibitem[McCully et al.(2018)]{banzai} McCully C., Volgenau N. H., Harbeck D.-R., et al., 2018 Proc. SPIE 10707 107070K
\bibitem[Merloni et al.(2003)]{merloni} Merloni A., Heinz S. \& di Matteo T., 2003, MNRAS, 345, 1057
\bibitem[Miller et al.(2017)]{milleratel} Miller J. M. et al., 2017, The Astronomer's Telegram, 10296
\bibitem[Miller-Jones et al.(2012)]{miller2012} Miller-Jones J. C. A. et al., 2012, MNRAS, 421, 468
\bibitem[Miller-Jones et al.(2021)]{miller2021} Miller-Jones J. C. A., et al. 2021, Science, 371, 6533, 1046
\bibitem[Miyamoto et al.(1995)]{my} Miyamoto S., Kitamoto S., Hayashida K. \& Egoshi W, 1995, ApJ, 442, 1, L13
\bibitem[Mandel \& Muller (2020)]{mm2020} Mandel I. \& Muller B., 2020, MNRAS 499, 3, 3214
\bibitem[Negoro et al.(2016)]{negoro} Negoro H., et al., 2016, The Astronomer's Telegram, 9876
\bibitem[Oates et al.(2019)]{oates2019} Oates S. R., Motta S., Beardmore A. P., et al., 2019, MNRAS, 488, 4, 4843
\bibitem[O'Brien et al.(2002)]{obrien2002} O'Brien K., Horne K., Hynes R. I., Chen W., Haswell C. A., Still M. D., 2002, MNRAS, 334, 426
\bibitem[Paice et al.(2019)]{paice2019} Paice J. A., Gandhi P., Charles P. A., et al., 2019, MNRAS, 488, 512
\bibitem[Patruno et al.(2016)]{patruno} Patruno A., Maitra D., Curran P.~A., D'Angelo C., Fridriksson J.~K., Russell D.~M., Middleton M., et al., 2016, ApJ, 817, 100
\bibitem[Pirbhoy et al.(2020)]{Pirbhoy2020} Pirbhoy S. F., Baglio M. C., Russell D.M., Bramich D. M., Saikia P., Yazeedi A. A., Lewis F., 2020, The Astronomer's Telegram, 13451
\bibitem[Revnivtsev et al.(1998)]{rr98} Revnivtsev M., et al., 1998, A\&A, 331, 557
\bibitem[Ross \& Fabian(2007)]{ross} Ross R. R. \& Fabian A. C., 2007, MNRAS, 381, 1697
\bibitem[Rout et al.(2021)]{rout} Rout S. K.,  Vadawale S. V.,  Aarthy E.,  Ganesh S.,  Joshi V., Roy, J.,  Misra R. \&  Yadav J. S., 2021, J Astrophys Astron 42, 39
\bibitem[Russell et al.(2006)]{Russell2006} Russell D. M., Fender R. P., Hynes R. I., Brocksopp C., Homan J., Jonker P. G., Buxton M. M., 2006, MNRAS, 371, 1334
\bibitem[Russell et al.(2007)]{Russell2007} Russell D. M., Fender R. P., Jonker P. G., 2007, MNRAS, 379, 1108
\bibitem[Russell et al.(2011)]{Russell2011} Russell D. M., Maitra D., Dunn R. J. H. \& Fender R. P., 2011, MNRAS, 416,2311
\bibitem[Russell et al.(2011b)]{Russell2011b} Russell D. M., Miller-Jones J. C. A., Maccarone T. J., Yang Y. J., Fender R. P., Lewis F., 2011, ApJ, 739, L19
\bibitem[Russell et al.(2013)]{russell2013} Russell D. M. et al., 2013, MNRAS, 429, 815
\bibitem[Russell et al.(2019)]{Russell2019} Russell D. M., Bramich D. M., Lewis F., et al. 2019, Astronomische Nachrichten, 340, 278
\bibitem[Russell et al.(2019b)]{r2019}
Russell T. D., Tetarenko A. J., Miller-Jones J. C. A., Sivakoff G. R., Parikh A. S. et al., 2019, ApJ, 883, 198
\bibitem[Russell et al.(2020)]{Russell2020} Russell, D. M., Casella P., Kalemci E., Vahdat Motlagh A., Saikia P., Pirbhoy S. F., Maitra, D., 2020, MNRAS, 495, 182
\bibitem[Rykoff et al.(2007)]{rykoff} Rykoff E.~S., Miller J.~M., Steeghs D., Torres M.~A.~P., 2007, ApJ, 666, 1129
\bibitem[Saikia et al.(2015)]{saikia15} Saikia P., Koerding E. \& Falcke H., 2015, MNRAS, 450, 2317
\bibitem[Saikia et al.(2018)]{saikia18} Saikia P., Koerding E., Coppejans D. L., Falcke H., Williams D., Baldi R. D., Mchardy I., Beswick R., 2018, A\&A, 616, A152
\bibitem[Saikia et al.(2019)]{saikia} Saikia P., Russell D. M., Bramich D. M., Miller-Jones J. C. A. , Baglio M. C. \& Degenaar N., 2019, ApJ, 887, 21
\bibitem[Saikia et al.(2022)]{saikia_submitted} Saikia P., Russell D. M., et al., 2022, submitted
\bibitem[Shahbaz et al.(2013)]{shahbaz} Shahbaz T., Russell D. M., Zurita C., Casares J., Corral-Santana J. M., Dhillon V. S., Marsh T. R., 2013, MNRAS, 434, 2696
\bibitem[Shahbaz et al.(2015)]{shahbaz2015} Shahbaz T., Linares M., Nevado S. P., Rodríguez-Gil P., Casares J., Dhillon V. S., Marsh T. R., Littlefair S., Leckngam A., Poshyachinda S., 2015, MNRAS, 453, 4, 3461
\bibitem[Shakura \& Sunyaev(1973)]{ss} Shakura N. \& Sunyaev R. A., 1973, A\&A, 24, 337
\bibitem[Skrutskie(2006)]{2mass} Skrutskie, M. ,F. 2006, AJ, 131, 1163
\bibitem[Stetson(1990)]{Stetson} Stetson, P. B. 1990, PASP, 102, 932
\bibitem[Sunyaev \& Titarchuk (1980)]{sunyaev} Sunyaev R. A. \& Titarchuk L. G., 1980, A\&A, 86, 121
\bibitem[Sunyaev \& Revnivtsev (2000)]{Sunyaev2000} Sunyaev R. A. \& Revnivtsev, M., 2000, AAP, 358, 617
\bibitem[Tananbaum et al.(1972)]{oldjet} Tananbaum H., Gursky H., Kellogg E., Giacconi R. \& Jones C., 1972, ApJ, 177, L5
\bibitem[Tao et al.(2019)]{tao2019} Tao L., Tomsick J. A., Qu J., Zhang S., Zhang S., Bu Q., 2019, ApJ, 887, 184
\bibitem[Tetarenko et al.(2016)]{tt} Tetarenko B. E., Sivakoff G. R., Heinke C. O., Gladstone J. C., 2016, ApJS, 222, 15
\bibitem[Tetarenko et al.(2020)]{be} Tetarenko B. E., Dubus G., Marcel G., Done C. \& Clavel M., 2020, MNRAS, 495, 4, 3666
\bibitem[Tetarenko et al.(2021)]{tetarenko2021} Tetarenko A. J., Casella P., Miller-Jones J. C. A., et al., 2021, MNRAS, 504, 3862
\bibitem[Thorne \& Price(1975)]{thorne} Thorne K. S. \& Price R. H., 1975, ApJ, 195, L101
\bibitem[Tonry et al.(2018)]{tonry} Tonry et al. 2018, ApJ, 867, 105
\bibitem[Torres et al.(2021)]{torres} Torres M. A. P., Jonker P. G., Casares J., Miller-Jones J. C. A. \& Steeghs D., 2021, MNRAS, 501, 2174,
\bibitem[Tudor et al.(2017)]{tudor} Tudor V., Miller-Jones J. C. A., Patruno A., et al., 2017, MNRAS 470, 324, doi:10.1093/mnras/stx1168
\bibitem[Vahdat Motlagh et al.(2019)]{macca2} Vahdat Motlagh A., Kalemci E., Maccarone T. J., 2019, MNRAS, 485, 2, 2744
\bibitem[van den Eijnden et al.(2021)]{ns} van den Eijnden J., Degenaar N., Russell T. D., Wijnands R., Bahramian A. et al., 2021, MNRAS, 507, 3899, https://doi.org/10.1093/mnras/stab1995
\bibitem[van der Hooft et al.(1996)]{vander} van der Hooft F., Kouveliotou C., van Paradijs J., Rubin B. C. et al., 1996, A\&AS, 120, 141
\bibitem[van der Horst et al.(2013)]{vanderHorst} van der Horst A. J., et al., 2013, MNRAS, 436, 2625
\bibitem[van Paradijs(1981)]{1981} van Paradijs J., 1981, A\&A, 103, 140
\bibitem[van Paradijs \& McClintock(1994)]{van} van Paradijs J., McClintock J. E., 1994, A\&A, 290, 133
\bibitem[Vaughan et al.(2003)]{vaughan} Vaughan S., Edelson R., Warwick R.S. et al., 2003, MNRAS, 345, 1271
\bibitem[Veledina, Poutanen \& Vurm(2013)]{veledina} Veledina A., Poutanen J., Vurm I., 2013, MNRAS, 430, 3196
\bibitem[Vincentelli et al.(2018)]{vincentelli2018} Vincentelli F. M., Casella P., Maccarone T. J., et al., 2018, MNRAS, 477, 4524 
\bibitem[Vrtilek et al.(1990)]{Vrtilek1990} Vrtilek S. D., Raymond J. C., Garcia M. R., Verbunt F., Hasinger G., Kurster M., 1990, A\&A, 235, 162
\bibitem[Watson (2011)]{av2} Watson D., 2O11, A\&A 533, A16
\bibitem[Willingale et al.(2013)]{av3} Willingale R., Starling L. C., Beardmore A. P., Tanvir N. R., OBrien P. T., 2013, MNRAS, 431, 1, 394
\bibitem[Zhang et al.(2019)]{zhang2019} Zhang G.-B., Bernardini F., Russell D.~M., Gelfand J.~D., Lasota J.-P., Qasim A.~A., AlMannaei A., et al., 2019, ApJ, 876, 5
\bibitem[Zhang et al.(2021)]{zhang2021} Zhang Z., Liu H., Abdikamalov A. B., Ayzenberg D., Bambi C. and Zhou M., submitted, 2021, arXiv:2106.03086 
\end{thebibliography}
\end{document}